\documentclass{aa}
\bibpunct{(}{)}{;}{a}{}{,} 
\usepackage[english]{babel}
\usepackage{multirow}
\usepackage{graphicx}
\usepackage{txfonts}

\setcitestyle{notesep={ }}
\usepackage{hyperref}
\hypersetup{colorlinks=true}
\hypersetup{citecolor=blue}
\hypersetup{linkcolor=blue}
\hypersetup{urlcolor=blue}
\newcommand{\ninecolumns}[9]{ #1 & #2 & #3 & #4 & #9 & #5 & #6 & #7\\}
\newcommand{\eightcolumns}[8]{ #1 & #2 & #3 & #4 & #5 & #6 & #7 & #8 \\}
\newcommand{\sevencolumns}[7]{ #1 & #2 & #4 & #3 & #5 & #6 & #7 \\}

\begin{document}

   \title{The CARMENES search for exoplanets around M dwarfs}

   \subtitle{Nine new double-line spectroscopic binary stars}

   \author{D. Baroch \inst{1,2},
          J. C. Morales \inst{1,2},
          I. Ribas \inst{1,2},
          L. Tal-Or \inst{3,4},
          M. Zechmeister \inst{3},
          A. Reiners \inst{3},
          J. A. Caballero\inst{5},
          A.~Quirrenbach \inst{6},
          P.~J. Amado \inst{7},
          S.~Dreizler \inst{3},
          S.~Lalitha \inst{3},
          S.~V. Jeffers \inst{3},
          M.~Lafarga \inst{1,2},
          V.~J.~S. B\'ejar \inst{8,9},
          J.~Colom\'e \inst{1,2},
          M.~Cort\'es-Contreras \inst{7,5},
          E.~D\'iez-Alonso \inst{5},
          D.~Galad\'i-Enr\'iquez \inst{10},
          E.~W. Guenther \inst{11},
          H.-J. Hagen \inst{12},
          T.~Henning \inst{13},
          E.~Herrero \inst{1,2},
          M.~K\"urster \inst{13},
          D.~Montes \inst{14},
          E.~Nagel \inst{12},
          V.~M. Passegger \inst{12},
          M.~Perger \inst{1,2},
          A.~Rosich \inst{1,2},
          A.~Schweitzer \inst{12},
          W.~Seifert \inst{6}.
          }

   \institute{Institut de Ci\`encies de l'Espai (ICE, CSIC),
              Campus UAB, C/ de Can Magrans s/n, E-08193 Cerdanyola del Vall\`es, Spain
         \and
              Institut d'Estudis Espacials de Catalunya (IEEC),
              C/ Gran Capit\`a 2-4, E-08034 Barcelona, Spain
         \and
              Institut f\"ur Astrophysik, Georg-August-Universit\"at,
              Friedrich-Hund-Platz 1, D-37077 G\"ottingen, Germany
         \and
              School of Geosciences, Raymond and Beverly Sackler Faculty of Exact Sciences, Tel Aviv University, Tel Aviv, 6997801, Israel
         \and
              Departamento de Astrof\'isica y Ciencias de la Atm\'osfera, Facultad de Ciencias F\'isicas,
              Universidad Complutense de Madrid, E-28040 Madrid, Spain
         \and
              Landessternwarte, Zentrum f\"ur Astronomie der Universt\"at Heidelberg,
               K\"onigstuhl 12, D-69117 Heidelberg, Germany
          \and
              Centro de Astrobiolog\'ia (CSIC-INTA), ESAC,
              Camino Bajo del Castillo s/n, E-28692 Villanueva de la Ca\~nada, Madrid, Spain
         \and
              Instituto de Astrof\'isica de Canarias,
              V\'ia L\'actea s/n, 38205 La Laguna, Tenerife, Spain
              \and
              Departamento de Astrof\'isica, Universidad de La Laguna,
              38026 La Laguna, Tenerife, Spain
                       \and
              Centro Astron\'onomico Hispano Alem\'an, Observatorio de Calar Alto, Sierra de los Filabres, E-04550 Gergal, Spain
         \and
              Th\"uringer Landesstenwarte Tautenburg,
              Sternwarte 5, 07778 Tautenburg, Germany
         \and
              Hamburger Sternwarte,
              Gojenbergsweg 112, D-21029 Hamburg, Germany
         \and
              Max-Planck-Institut f\"ur Astronomie,
              K\"onigstuhl 17, 69117 Heidelberg, Germany
         \and
              Instituto de Astrof\'isica de Andaluc\'ia (IAA-CSIC),
              Glorieta de la Astronom\'ia s/n, E-18008 Granada, Spain
             }

   \titlerunning{CARMENES new double-line spectroscopic binaries}

   \authorrunning{Baroch et al.}

   \date{\today}

  \abstract
   {The CARMENES spectrograph is surveying $\sim$300 M dwarf stars in search for exoplanets. Among the target stars, spectroscopic binary systems have been discovered, which can be used to measure fundamental properties of stars.}
   {Using spectroscopic observations we determine the orbital and physical properties of nine new double-line spectroscopic binary systems by analysing their radial velocity curves.}
   {We use two-dimensional cross-correlation techniques to derive the radial velocities of the targets, which are then employed to determine the orbital properties. Photometric data from the literature are also analysed to search for possible eclipses and to measure stellar variability, which can yield rotation periods.}
   {Out of the 342 selected stars for CARMENES survey, 9 have been found to be double-line spectroscopic binaries, with periods ranging from 1.13 to $\sim$8000 days and orbits with eccentricities up to 0.54. We provide empirical orbital properties and minimum masses for the sample of spectroscopic binaries. Absolute masses are also estimated from mass-luminosity calibrations, ranging between $\sim$0.1~M$_{\odot}$ and $\sim$0.6~M$_{\odot}$.}
   {These new binary systems increase the number of double-line M dwarf binary systems with known orbital parameters by 15\%, and they have lower mass ratios on average.}


   \maketitle
%

\section{Introduction}
\label{sec:introduction}
Binary systems are essential for the study of stellar structure and evolution. Depending on their nature, they can yield fundamental properties such as the masses, radii, and luminosities of the components independently from calibrations and stellar models and with very high precision. This enables critical comparisons with stellar model predictions and the determination of empirical calibrations that can be used for single stars \citep[see][for a review]{Torres2010}. Due to the increasing interest in the discovery of exoplanets, several instruments were developed to spectroscopically survey a large number of stars. In addition to planetary objects, these projects can also reveal new binary systems that are interesting on their own, because they can be used to constrain the stellar structure and evolution models and to improve the multiplicity statistics of late-type stars \citep{Halbwachs2003,Mazeh2003}.

This is the case of the CARMENES survey \citep[][]{Quirrenbach2016}. This survey monitors about 300 M dwarf stars to uncover exoplanets in their habitable zones. Targets were selected from available M dwarf catalogues and photometric surveys, and were also carefully studied to discard unsuitable targets such as visual double systems, known spectroscopic binaries and very faint stars \citep[see e.g.][, for more details]{AlonsoFloriano2015,CortesContreras2017,Jeffers2017}. The CARMENES collaboration has already announced its first planet detections \citep{Reiners2018,Sarkis2018,Trifonov2018}. In addition to the new planets, several spectroscopic binary systems were identified with the first observations of the sample and they were followed-up to characterize them.

The binary systems discovered with CARMENES are especially interesting because the number of known M dwarf binary systems is still scarce  \citep[see e.g. the Ninth Catalogue of Spectroscopic Binary Orbits, hereafter SB9\footnote{\texttt{http://sb9.astro.ulb.ac.be/}} -- ][]{Pourbaix2004}. The distribution of mass ratios and orbital elements may help to understand the formation and evolution of low-mass stars, brown dwarfs or giant planets in M dwarf stellar systems. Besides, they are also valuable for constraining the properties of M dwarfs, which still show some discrepancies with stellar model predictions \citep[see e.g.][]{Morales2010,Feiden2013,Feiden2014}.

In this paper we present nine new double-line spectroscopic binary (SB2) systems discovered in the CARMENES survey. Orbital properties were derived for all of them, yielding their mass ratios and periods for the first time. In Sect.~\ref{sec:observations} we describe the observations for each system. In Sect.~\ref{subsec:rv-analysis}, the radial velocity analysis of each system is shown. Photometric light curves gathered from the literature and public databases are compiled and discussed in Sect.~\ref{sec:phot-analysis}. Finally, the results are discussed in Sect.~\ref{sec:discussion} and our conclusions are presented in Sect.~\ref{sec:conclusions}. Figures of the radial velocity data and the photometric periodogram analysis are compiled in the Appendix.

\section{Spectroscopic data}
\label{sec:observations}

\subsection{Spectroscopic observations}
\label{subsec:rv_obs}

High-resolution spectroscopic observations of the targets were taken with the visual (VIS) and near infrared (NIR) channels of the CARMENES spectrograph from January 2016 to March 2018, covering a wavelength range from 5200 {\AA} to 9600 {\AA} with a measured resolving power of R=94,600 in the VIS, and from 9600 {\AA} to 17100 {\AA} with a measured resolving power of R=80,400 in the NIR \citep{Quirrenbach2016}. For few nights when the NIR channel was not available, only VIS spectra is used. From the over 300 studied stars \citep{Reiners2017}, we have so far identified nine SB2 systems. Between 10 and $\sim$20 observations were taken sampling the orbital phases of short period systems.

Table \ref{tab:basicprop} lists some basic information for each target. Six publicly available additional HARPS-N \citep{Cosentino2012} observations were found for one of the systems (Ross 59) and also included in our analysis.

\begin{table*}[t]
\centering
\caption{Main properties and observing log of the spectroscopic binaries studied in this work.\tablefootmark{a} }
\label{tab:basicprop}
\resizebox{\textwidth}{!}{
\begin{tabular}[t]{ll cc c c c c c c}
\hline
\hline
\noalign{\smallskip}
Name & Karmn & \multicolumn{2}{c}{N$_{\rm obs}$} & $\Delta$t & Sp. & Ref.\tablefootmark{b} & $\pi$ & $G$ & $K_s$ \\
 & & VIS & NIR & [d] & type &  & [mas] & [mag] & [mag] \\
\noalign{\smallskip}
 \hline
 \noalign{\smallskip}
EZ Psc & J00162$+$198W & 10 & 10 & 528.6 & M4.0V+ & AF15 & 65.72$\pm$0.10& 10.8968$\pm$0.0021 &7.083$\pm$0.023\\
GJ 1029 & J01056$+$284 & 15 & 15 & 443.8 & M5.0 V+ & PMSU & 79.84$\pm$0.34 & 12.9701$\pm$0.0019 & 8.550$\pm$0.020\\
Ross 59 & J05532$+$242 & 22\tablefootmark{c} & 16 & 1893.7\tablefootmark{c} & M1.5 V+ &  PMSU &  51.5$\pm$4.6 & 9.9176$\pm$0.0040 & 6.633$\pm$0.021\\
NLTT 23956 & J10182$-$204 & 14 & 14 & 359.0 & M4.5 V+ & Ria06 & 39.483$\pm$0.087& 12.3261$\pm$0.0025 & 8.145$\pm$0.023\\
GJ 3612 & J10354$+$694 & 21 & 14 & 686.0 & M3.5 V+ & PMSU & 77.34$\pm$0.29 & 10.7133$\pm$0.0019 & 7.161$\pm$0.020\\
GJ 1182 & J14155$+$046 & 21 & 18 & 428.8 & M5.0 V+ & PMSU & 71.11$\pm$0.39 & 12.6739$\pm$0.0020& 8.618$\pm$0.025\\
UU UMi & J15412$+$759& 19 & 18 & 727.9 & M3.0 V+ & PMSU & 68.3$\pm$1.5 & 11.0493$\pm$0.0020 & 7.442$\pm$0.023\\
LP 395-8 & J20198$+$229 & 14 & 12 & 500.7 & M3.0 V+ & Lep13 & 34.081$\pm$0.074 & 10.9990$\pm$0.0024 & 7.283$\pm$0.018\\
GJ 810A & J20556$-$140N & 18 & 18 & 556.6 & M4.0 V+ & PMSU & 77.02$\pm$0.40 & 11.0453$\pm$0.0024 & 7.365$\pm$0.026\\
\noalign{\smallskip}
\hline
\end{tabular}
}
\tablefoot{
\tablefoottext{a}{$K_s$ magnitudes are from the 2MASS survey \cite{2006AJ....131.1163S}. Parallaxes and $G$ magnitudes are from the Gaia Data Release 2 \citep{2016A&A...595A...1G}, except for Ross 59, which comes from \cite{1995gcts.book.....V}.}
\tablefoottext{b}{AF15: \cite{AlonsoFloriano2015}; APASS: \cite{2015AAS...22533616H}; Lep13: \cite{Lepine2013}; PMSU: \cite{Hawley1996}; Ria06: \cite{Riaz2006}; UCAC4: \cite{2013AJ....145...44Z}.}
\tablefoottext{c}{Including six additional observations from HARPS-N.}
}
\end{table*}

\subsection{Radial velocity determination}
\label{subsec:rv_curves}
The candidate spectroscopic binary systems were identified by large variations in their radial velocities, which are routinely calculated by the CARMENES SERVAL pipeline \citep{Caballero2016b, Zechmeister2017}. This algorithm is based on least-squares fitting of the spectra, providing very accurate radial velocities for single stars \citep[see][]{Anglada2012}. However, it does not yield the velocity of secondary components in binary systems. For that reason, radial velocities of both components were also determined using {\sc todmor} \citep{Zucker2003}, a modern implementation of the two-dimensional cross-correlation technique {\sc todcor} \citep{Zucker1994} for multi-order spectra.

To derive the radial velocities of each component of the binary system, we used PHOENIX stellar models \citep{Husser2013} as templates for the calculation of the cross-correlation functions (CCFs). Using {\sc todmor}, we explored a grid of values for the effective temperatures, flux ratios in the observed wavelength band, and spectral-line broadening to fit all the spectra of each target. Spectral orders with low signal-to-noise ratio or telluric contamination were discarded. To obtain the final radial velocity curves of each system in a consistent way, we selected as the template for each system the one that produces the highest CCF peak for spectra with radial velocities obtained close to quadratures. Orbital phases close to conjunction, where the radial velocities of the components can not be disentangled due to rotational broadening and unfavorable flux ratio, were not considered in this analysis. For this reason, we discarded 6 spectra for GJ 1029, 2 for Ross 59, 3 for GJ 1182 and 7 for GJ 810 A in both the VIS and NIR channels. In the case of the VIS spectra, all the parameters could be reliably optimized. However, the effective temperatures resulting from the optimization process in the NIR spectra always led to unrealistic values at the edge of the grid. Therefore, for the NIR we adopted the values derived from the VIS data and the optimization was performed on the flux ratio and broadening, which can differ in the NIR channel because of the different wavelength coverage and resolving power.

Only in the case of UU~UMi we used observed spectra as templates with TODMOR, which performed better than synthetic spectra due to the small radial velocity difference between the components and the long period of this system. The co-added spectra of an M2.5 star (Gl~436) and an M5 star (GJ~1253) obtained with CARMENES, were used for the primary and secondary components, respectively. Consequently, the systemic radial velocity derived for UU~UMi depends on those of the templates; therefore its uncertainty might be larger. The use of observed spectra do not significantly change the parameters for the other binaries analysed in this work.

Table \ref{tab:TODMOR} shows the optimized parameters of the templates for each target, except for UU~UMi, for which effective temperatures are obtained from \cite{Passegger2018}. These are the parameters used to obtain the radial velocities of the systems with {\sc todmor}, which are provided, together with their uncertainties, in Table \ref{tab:rvs} of Appendix \ref{sec:appdata}.

\begin{table*}[t]
\centering
\caption{Spectral properties of the templates used to derive radial velocities with {\sc todmor} in the VIS and NIR channel spectra. Uncertainties indicate the step size used in the grid of models.}
\label{tab:TODMOR}

\begin{tabular}[t]{lcccccc}
\hline
\hline
\noalign{\smallskip}
\multirow{3}{*}{Name} & $T_{\rm eff,1}$ & $T_{\rm eff,2}$ & \multicolumn{2}{c}{Spectral-line broadening} & \multicolumn{2}{c}{$L_2/L_1$} \\
 & \multirow{ 2}{*}{[K]} & \multirow{ 2}{*}{[K]} & VIS & NIR & \multirow{ 2}{*}{VIS} & \multirow{ 2}{*}{NIR}\\
 & & & [km s$^{-1}$] & [km s$^{-1}$] & & \\
\noalign{\smallskip}
 \hline
 \noalign{\smallskip}
EZ Psc & 3300$\pm$100 & 2900$\pm$100 & 5.5$\pm$0.1 & 7.1$\pm$0.1 & 0.11$\pm$0.01 & 0.12$\pm$0.01\\
GJ 1029 & 3100$\pm$100 & 2900$\pm$100 & 4.1$\pm$0.1 & 3.6$\pm$0.1 & 0.24$\pm$0.01 & 0.32$\pm$0.01\\
Ross 59 & 3900$\pm$100 & 3600$\pm$100 & 1.6$\pm$0.1 & 4.9$\pm$0.1 & 0.10$\pm$0.01 & 0.11$\pm$0.01\\
NLTT 23956 & 3100$\pm$100 & 3000$\pm$100 & 6.0$\pm$0.1 & 5.9$\pm$0.1 & 0.29$\pm$0.01 & 0.37$\pm$0.01\\
GJ 3612 &  3600$\pm$100 & 3300$\pm$100 & 2.2$\pm$0.1 & 4.6$\pm$0.1 & 0.21$\pm$0.01 & 0.25$\pm$0.01\\
GJ 1182 & 3300$\pm$100 & 2900$\pm$100 & 2.4$\pm$0.1 & 4.0$\pm$0.1 & 0.20$\pm$0.01 & 0.29$\pm$0.01\\
UU UMi\tablefootmark{a} & 3500$\pm$51 & 3300$\pm$51 & 2.6$\pm$0.1 & 6.0$\pm$0.1 & 0.22$\pm$0.01 & 0.27$\pm$0.01\\
LP 395-8 & 3600$\pm$100 & 3300$\pm$100 & 15.0$\pm$0.1 & 18.1$\pm$0.1 & 0.14$\pm$0.01 & 0.11$\pm$0.01\\
GJ 810A & 3400$\pm$100 & 3300$\pm$100 & 4.3$\pm$0.1 & 4.9$\pm$0.1 & 0.61$\pm$0.01 & 0.47$\pm$0.01\\
\noalign{\smallskip}
\hline
\end{tabular}
\tablefoot{
\tablefoottext{a}{Real templates are used instead of synthetic. Effective temperatures from \cite{Passegger2018}}}
\end{table*}

\section{Data analysis}
\label{sec:analysis}

\subsection{Radial velocity analysis}
\label{subsec:rv-analysis}

The orbital parameters of each target were derived using the SBOP \citep{Etzel1985} code, which fits the seven parameters of a Keplerian orbit simultaneously to both components: the period ($P_{\rm orb}$), the time of periastron passage ($T$), the eccentricity ($e$) and argument of the periastron ($\omega$), the radial velocity semi-amplitudes of each component of the system ($K_1$ and $K_2$ for the primary and secondary components, respectively), and the barycentric radial velocity of the system ($\gamma$). An initial estimate of the periods was obtained from a Lomb-Scargle periodogram analysis \citep{Scargle1982} of the radial velocities, and used as input parameter for SBOP.

Although NIR CARMENES measurements have lower precision than those from the VIS channel \citep{Tal-Or2018}, we fitted radial velocities from both channels simultaneously for consistency and considered the respective uncertainties. We also allowed for an adjustable radial velocity jitter ($ Jit_{\rm VIS/NIR,1/2}$) in the fit, as defined by \cite{Baluev2009}, different for each channel and component. This jitter term represents unaccounted error sources in the estimation of the uncertainties of the measurements. The results show that the jitter parameter of the NIR channel radial velocity of the primary component is always above twice that of the VIS channel, except for LP 395-8. However, in this case, the dispersion of the VIS channel may be affected by the large residual of the observation close to conjunction at orbital phase~$\sim$0.8 that does not have a NIR counterpart. Final parameters and uncertainties were computed running the Markov Chain Monte Carlo (MCMC) sampler \texttt{emcee} \citep{Foreman-Mackey2013} with a model based on SBOP with additional jitter terms. Parameters uncertainties were derived from the 68.3\% credibility interval of the resulting posterior parameter distribution.

The fitted orbital parameters of all targets and their computed physical elements are given in Tables \ref{tab:orbital} and \ref{tab:physical}, respectively. Figure~\ref{fig:plot-rv} shows the radial velocity fits of all systems. We found 3 systems in close orbits with periods between 1 and 6 days (EZ Psc, NLTT 23956, and LP 395-8), 3 systems with intermediate periods between 70 and 160 days (GJ 1029, GJ 3612, and GJ 1182), and 3 systems with periods longer than about 2 years, for which further measurements are needed to better constrain the parameters (Ross 59, UU UMi, and GJ 810A). All systems show eccentric orbits, with smaller eccentricity in the case of short period binaries, except UU UMi. For this system, due to its long period and the short orbital phase sampled with CARMENES, a circular orbit was assumed in the present work.

\begin{table*}[t]
\centering
\caption{Radial velocity parameters fitted for each binary system.}
\label{tab:orbital}
\resizebox{\textwidth}{!}{
\begin{tabular}{lccccccccccccc}
\hline\hline
\noalign{\smallskip}

\eightcolumns{Name}{$P_{\rm orb}$}{$T$}{$e$}{$\omega$}{$K_1$}{$K_2 $}{$\gamma$ & $\sigma_1$ & $\sigma_2$ & $Jit_{\rm VIS,1}$ & $Jit_{\rm NIR,1}$ & $Jit_{\rm VIS,2}$ & $Jit_{\rm NIR,2}$}
\eightcolumns{}{[d]}{[JD-2457000]}{}{[$\deg$]}{[km s$^{-1}$]}{[km s$^{-1}$]}{[km s$^{-1}$] & [km s$^{-1}$] & [km s$^{-1}$] & [km s$^{-1}$] & [km s$^{-1}$] & [km s$^{-1}$] & [km s$^{-1}$]}
\noalign{\smallskip}
\hline
\noalign{\smallskip}
\eightcolumns{EZ Psc}{3.956523}{709.24}{0.00220}{28}{27.639}{78.56}{$-$0.501 & 0.154 & 0.611 & 0.054 & 0.43 & 0.17 & 0.51}
\eightcolumns{}{$^{+0.000071}_{-0.000092}$}{$^{+0.25}_{-0.18}$}{$^{+0.00096}_{-0.00090}$}{$^{+23}_{-16}$}{$^{+0.049}_{-0.054}$}{$^{+0.23}_{-0.24}$}{$^{+0.031}_{-0.042}$ & & & $^{+0.049}_{-0.037}$ & $^{+0.22}_{-0.21}$ & $^{+0.92}_{-0.80}$ & $^{+0.29}_{-0.22}$}
\noalign{\vskip 1mm}
\eightcolumns{GJ 1029}{95.69}{857.31}{0.3859}{209.2}{6.799}{9.59}{$-$11.338 & 0.485 & 0.369 & 0.033 & 0.33 & 0.49 & 0.23}
\eightcolumns{}{$^{+0.12}_{-0.13}$}{$^{+0.45}_{-0.44}$}{$^{+0.0047}_{-0.0048}$}{$^{+1.5}_{-1.5}$}{$^{+0.041}_{-0.045}$}{$^{+0.10}_{-0.10}$}{$^{+0.032}_{-0.032}$&&&$^{+0.037}_{-0.022}$&$^{+0.12}_{-0.10}$ & $^{+0.14}_{-0.10}$ & $^{+0.14}_{-0.13}$}
\noalign{\vskip 1mm}

  \eightcolumns{Ross 59}{721.4}{1006.5}{0.5089}{109.2}{2.911}{8.75}{27.496 & 0.110 & 0.322 & 0.051 & 0.200 & 0.73 & 0.54}
    \eightcolumns{}{$^{+2.2}_{-2.0}$}{$^{+1.6}_{-1.5}$}{$^{+0.0089}_{-0.0086}$}{$^{+1.7}_{-1.7}$}{$^{+0.038}_{-0.039}$}{$^{+0.13}_{-0.13}$}{$^{+0.021}_{-0.021}$&&& $^{+0.021}_{-0.020}$ & $^{+0.086}_{-0.051}$ & $^{+0.085}_{-0.050}$ & $^{+0.24}_{-0.15}$}
\noalign{\vskip 1mm}
   \eightcolumns{NLTT 23956}{5.922845}{1007.42}{0.0135}{289.1}{32.469}{56.149}{13.688 & 0.097 & 0.337 & 0.036 & 0.14 & 0.052 & 0.33}
    \eightcolumns{}{$^{+0.000061}_{-0.000059}$}{$^{+0.107}_{-0.094}$}{$^{+0.0012}_{-0.0012}$}{$^{+6.4}_{-5.6}$}{$^{+0.055}_{-0.056}$}{$^{+0.091}_{-0.095}$}{$^{+0.033}_{-0.028}$& & &$^{+0.040}_{-0.026}$& $^{+0.108}_{-0.090}$ & $^{+0.060}_{-0.037}$ & $^{+0.14}_{-0.14}$}
\noalign{\vskip 1mm}

  \eightcolumns{GJ 3612}{119.411}{718.42}{0.0655}{326.0}{10.638}{20.99}{$-$62.089 & 0.147 & 0.382 & 0.025 & 0.064 & 0.18 & 0.451}
  \eightcolumns{}{$^{+0.035}_{-0.035}$}{$^{+0.57}_{-0.57}$}{$^{+0.0024}_{-0.0024}$}{$^{+1.7}_{-1.7}$}{$^{+0.028}_{-0.029}$}{$^{+0.10}_{-0.10}$}{$^{+0.018}_{-0.018}$&&& $^{+0.026}_{-0.017}$ & $^{+0.064}_{-0.050}$ & $^{+0.106}_{-0.104}$ & $^{+0.121}_{-0.090}$}
 \noalign{\vskip 1mm}
  \eightcolumns{GJ 1182}{154.24}{867.824}{0.5373}{275.76}{11.968}{18.001}{$-$0.683 & 0.139 & 0.370 & 0.033 & 0.17 & 0.064 & 0.500}
  \eightcolumns{}{$^{+0.12}_{-0.12}$}{$^{+0.069}_{-0.070}$}{$^{+0.0016}_{-0.0016}$}{$^{+0.28}_{-0.26}$}{$^{+0.029}_{-0.028}$}{$^{+0.072}_{-0.071}$}{$^{+0.019}_{-0.019}$&&& $^{+0.029}_{-0.023}$ & $^{+0.051}_{-0.041}$ & $^{+0.066}_{-0.044}$ & $^{+0.117}_{-0.088}$}
 \noalign{\vskip 1mm}
  \eightcolumns{UU UMi}{7927}{1118}{0}{0}{2.54}{4.46}{$-$41.17 & 0.257 & 0.254 & 0.0088 & 0.421 & 0.019 & 0.318}
  \eightcolumns{}{$^{+847}_{-637}$}{$^{+64}_{-44}$}{(fixed)}{(fixed)}{$^{+0.23}_{-0.20}$}{$^{+0.22}_{-0.25}$}{$^{+0.20}_{-0.23}$&&& $^{+0.0089}_{-0.0060}$ & $^{+0.090}_{-0.067}$ & $^{+0.023}_{-0.014}$ & $^{+0.084}_{-0.067}$}
\noalign{\vskip 1mm}

 \eightcolumns{LP 395-8}{1.1293392}{620.075}{0.0071}{352}{36.534}{65.29}{$-$26.616 & 0.215 & 0.819 & 0.180 & 0.095 & 0.34 & 1.20}
\eightcolumns{}{$^{+0.0000067}_{-0.0000072}$}{$^{+0.038}_{-0.039}$}{$^{+0.0026}_{-0.0022}$}{$^{+12}_{-12}$}{$^{+0.098}_{-0.088}$}{$^{+0.19}_{-0.20}$}{$^{+0.056}_{-0.060}$&&& $^{+0.094}_{-0.090}$ & $^{+0.135}_{-0.071}$ & $^{+0.23}_{-0.21}$ & $^{+0.38}_{-0.30}$}
 \noalign{\vskip 1mm}
  \eightcolumns{GJ 810A}{812}{822.1}{0.402}{238.5}{5.57}{6.74}{$-$142.098 & 0.452 & 0.246 & 0.053 & 0.63 & 0.035 & 0.262}
  \eightcolumns{}{$^{+58}_{-40}$}{$^{+3.9}_{-3.7}$}{$^{+0.059}_{-0.046}$}{$^{+1.6}_{-1.7}$}{$^{+0.16}_{-0.23}$}{$^{+0.19}_{-0.28}$}{$^{+0.021}_{-0.022}$&&& $^{+0.055}_{-0.037}$ & $^{+0.18}_{-0.11}$ & $^{+0.038}_{-0.024}$ & $^{+0.075}_{-0.056}$}

\noalign{\smallskip}
\hline
\end{tabular}
}
\end{table*}

\begin{table*}[t]
\centering                                                                \caption{Physical parameters derived from the radial velocity fits.}
\label{tab:physical}
\begin{tabular}{lcccccc}
\hline
\hline
\noalign{\smallskip}
\sevencolumns{Name}{$(a_1+a_2)\sin i$}{$\mathcal{M}_1 \sin^3i$}{$\alpha \sin i$}{$\mathcal{M}_2 \sin^3i$ }{$\mathcal{M}_2/\mathcal{M}_1$}{$f(\mathcal{M})$}
\sevencolumns{}{[au]}{[$M_{\odot}$]}{[mas]}{[$M_{\odot}$]}{}{[$10^{-3}M_{\odot}$]}
\noalign{\smallskip}
\hline
\noalign{\smallskip}
\noalign{\vskip 1mm}
  \sevencolumns{EZ Psc}{$0.038623^{+0.000086}_{-0.000090}$}{$0.3632^{+0.0027}_{-0.0028}$}{$2.5393^{+0.0068}_{-0.0070}$}{$0.12778^{+0.00065}_{-0.00069}$}{$0.3518^{+0.0012}_{-0.0013}$}{$8.675^{+0.053}_{-0.056}$}
\noalign{\vskip 1mm}
  \sevencolumns{GJ 1029}{$0.13299^{+0.00091}_{-0.00092}$}{$0.02005^{+0.00048}_{-0.00049}$}{$10.618^{+0.085}_{-0.086}$}{$0.01422^{+0.00025}_{-0.00026}$}{$0.7090^{+0.0085}_{-0.0088}$}{$2.452^{+0.060}_{-0.064}$}
\noalign{\vskip 1mm}
 \sevencolumns{Ross 59}{$0.6656^{+0.0085}_{-0.0085}$}{$0.0567^{+0.0024}_{-0.0024}$}{$34.3^{+3.1}_{-3.2}$}{$0.01887^{+0.00066}_{-0.00066}$}{$0.3327^{+0.0066}_{-0.0067}$}{$1.179^{+0.063}_{-0.063}$}
\noalign{\vskip 1mm}
   \sevencolumns{NLTT 23956}{$0.048242^{+0.000058}_{-0.000060}$}{$0.2705^{+0.0010}_{-0.0011}$}{$1.9048^{+0.0048}_{-0.0048} $}{$0.15643^{+0.00056}_{-0.00057}$}{$0.5783^{+0.0014}_{-0.0014}$}{$21.05^{+0.13}_{-0.13}$}
\noalign{\vskip 1mm}
  \sevencolumns{GJ 3612}{$0.3464^{+0.0011}_{-0.0011}$}{$0.2581^{+0.0029}_{-0.0029}$}{$26.79^{+0.13}_{-0.13}$}{$0.1308^{+0.0010}_{-0.0010}$}{$0.5068^{+0.0028}_{-0.0028}$}{$14.8^{+1.6}_{-1.6}$}
 \noalign{\vskip 1mm}
 \sevencolumns{GJ 1182}{$0.35835^{+0.00100}_{-0.00099}$}{$0.1550^{+0.0015}_{-0.0015}$}{$25.49^{+0.16}_{-0.16}$}{$0.10305^{+0.00077}_{-0.00077}$}{$0.6649^{+0.0031}_{-0.0030}$}{$16.47^{+0.16}_{-0.16}$}
\noalign{\vskip 1mm}
  \sevencolumns{UU UMi}{$5.10^{+0.46}_{-0.38}$}{$0.179^{+0.030}_{-0.029}$}{$348^{+32}_{-27}$}{$0.102^{+0.020}_{-0.018}$}{$0.570^{+0.059}_{-0.055}$}{$13.4^{+3.9}_{-3.4}$}
 \noalign{\vskip 1mm}
  \sevencolumns{LP 395-8}{$0.010570^{+0.000022}_{-0.000023}$}{$0.07920^{+0.00055}_{-0.00057}$}{$ 0.3602^{+0.0011}_{-0.0011}$}{$0.04432^{+0.00026}_{-0.00025}$}{$0.5596^{+0.0022}_{-0.0022}$}{$5.719^{+0.064}_{-0.056}$}
\noalign{\vskip 1mm}
\sevencolumns{GJ 810A}{$0.841^{+0.046}_{-0.038}$}{$0.0660^{+0.0064}_{-0.0071}$}{$64.8^{+3.6}_{-3.0}$}{$0.0545^{+0.0052}_{-0.0056}$}{$0.826^{+0.033}_{-0.048}$}{$11.2^{+1.3}_{-1.5}$}
\noalign{\smallskip}
\hline
\end{tabular}
\end{table*}

\subsection{Photometric analysis}
\label{sec:phot-analysis}

To fully characterize our binary systems, we also carried out a bibliographic search for photometric light curves in public archives from surveys such as the Wide Angle Search for Planets \citep[SuperWASP --][]{Pollacco2006}, The MEarth Project \citep[MEarth --][]{Charbonneau2008, Irwin2011,  Berta2012}, the All-Sky Automated Survey \citep[ASAS --][]{Pojmanski1997} and the Northern Sky Variability Survey \citep[NSVS --][]{Wozniak2004}. The aim was to search for eclipses in the light curves and the estimation of the rotation period of the systems. Before analysing the photometry we removed outliers as explained in \cite{Diez2018}, rejecting iteratively datapoints deviated more than 2.5$\sigma$ from the mean of the full photometric dataset for each targets. However, outliers were further inspected by eye in order to make sure that possible eclipses were not removed.

The rotation period was determined by computing the Lomb-Scargle periodogram \citep{Scargle1981} of the photometric light curves, and then looking for strong signals between 1 and 200 days \citep{Newton2016}. Uncertainties are estimated as half the full width at half maximum of the periodogram peak, as a conservative approach. To evaluate the significance of the signals, we used the False Alarm Probability (hereafter, FAP) as described in \cite{Scargle1982}, which measures the probability that the signal randomly arises from white noise. We defined as significant those periodic signals with FAP~$<0.1 \%$.

We searched for eclipses in the light curves using two different approaches: in the case of binary systems with well determined periods from radial velocities, we folded the light curve of each target in a phase-magnitude diagram using the orbital period found in Section \ref{subsec:rv-analysis}, and checked for a decrease in brightness within a narrow phase region compatible with the radial velocity orbit. We also made use of the Box-fitting Least Squares code \citep[hereafter BLS;][]{Kovacs2002} to identify eclipses with depth similar to the photometric scatter of each curve. BLS was also used in the case of binary systems with poorly constrained periods, although the eclipse probability is very small for long period systems.
Both methodologies yielded negative results in all cases and, thus, we concluded that none of our nine SB2s is an eclipsing binary within the limits of the sampling and measurement accuracy of the photometric data, which are given in Table \ref{tab:photometry}.

Table~\ref{tab:photometry} lists the significant photometric periods found for our targets and the H$\alpha$ line pseudo-equivalent width resulting from the CARMENES pipeline \citep{Zechmeister2017} as an indicator of stellar activity \citep{Reid1995,Hawley1996}. Figure~\ref{fig:plot-photometry} shows the available photometry, the periodogram, and the light curves phase-folded to the best period found. Significant signals for seven of the systems studied here were found, which we identify as corresponding to the rotation period of the main component (the brightest star) of the systems, assuming that both components may have similar activity levels. Using the same data but with a more conservative approach, \cite{Diez2018} reported photometric periods for four of our stars (EZ Psc, GJ 3612, UU UMi and LP 395-8). In all cases, the measured $P_{\rm rot}$ are identical within
uncertainties.

Interestingly, the three short-period systems (EZ~Psc, NLTT~23956, LP~395-8) all have rotation periods below 10~days, and they are active systems showing the H$\alpha$ line in emission. Besides, in these cases the broadening of the spectral lines, which depends on the rotation period of the components and the instrumental resolution, is also larger (see Table~\ref{tab:TODMOR}). However, only in the case of LP~395-8, the binary system with the shortest orbital period, rotation seems to be pseudo-synchronized with the orbital motion at periastron, although an alias period of $\sim$9~days can not be excluded with the present data.

EZ~Psc and NLTT~23956 seem to be sub-synchronous systems, with rotation periods larger than their orbital period. This may indicate that these systems could be young binary systems still in the process of reaching synchronization. However, synchronization timescales are relatively short and even pre-main sequence stars are synchronized for orbital periods below 8-10 days \citep{Mazeh2008}. By statistically analysing the Kepler eclipsing binary candidates, \cite{Lurie2017} suggested that differential rotation could also cause an apparent non-synchronization of orbital and rotational periods if photospheric active regions are located at higher latitudes, as expected for fast rotation systems \citep{Strassmeier2002}. In their study, they found that 13\,\% of the FGK-type primaries with periods between 2 and 10 days, and with small expected mass ratios, are sub-synchronous, showing a ratio between orbital and rotational period of $P_{\rm orb}/P_{\rm rot} \sim 0.87$. This is not far from the values we would expect for EZ~Psc and NLTT~23956 assuming pseudo-synchronization, $P_{\rm orb}/P_{\rm rot} \sim 0.82$, although smaller.

Altough synchronization is not expected for long period binary systems, GJ~3612 also shows significant variability with a semi-amplitude of $\sim14$~mmag and a period of $\sim$123 days, consistent with the orbital period within uncertainties. However, the rotation period determination may be affected by poor photometric sampling and the narrow time interval covered by the observations and, therefore, interaction between the components cannot be confirmed with the present data.

\begin{table*}[t]
\caption{Available photometry for the spectroscopic binaries analysed in this work. Number of observations after 2.5-$\sigma$ clipping, standard deviation, best period and variability semi-amplitude are listed for each target\tablefootmark{a}.}
\label{tab:photometry}
\centering
\begin{tabular}{lccccccc}
\hline\hline
\noalign{\smallskip}

\ninecolumns{Karmn}{pEW(H$\alpha$)}{Survey}{$N_{\rm obs}$ ($N_{\rm used}$)}{$\sigma$}{$P_{\rm rot}$}{A$_{\lambda}$}{FAP}{$\Delta$t}
\ninecolumns{}{[$\AA$]}{}{[\#]}{[mmag]}{[d]}{[mmag]}{[\%]}{[d]}

\noalign{\smallskip}
\hline
\noalign{\smallskip}
\ninecolumns{EZ Psc}{$-4.16\pm0.06$}{MEarth}{1660 (1581)}{7.5}{4.8063$\pm$0.0023}{3.9}{$<10^{-4}$}{2841}
\ninecolumns{GJ 1029}{$-0.12\pm0.02$}{MEarth}{862 (833)}{6.1}{16.32$\pm$0.24}{4.3}{$<10^{-4}$}{2195}
\ninecolumns{Ross 59}{$-0.034\pm0.001$}{ASAS}{318 (296)}{22.4}{$\cdots$}{$\cdots$}{$\cdots$}{2544}
\ninecolumns{NLTT 23956}{$-7.63\pm0.09$}{SuperWASP}{14814 (12450)}{49.9}{7.314$\pm$0.040}{13.3}{$<10^{-4}$}{504}
\ninecolumns{GJ 3612}{$0.070\pm0.008$}{NSVS}{163 (151)}{19.1}{123$\pm$15}{10.8}{$<10^{-4}$}{359}
\ninecolumns{GJ 1182}{$0.10\pm0.10$}{MEarth}{1203 (1145)}{4.8}{$8.92\pm0.24$}{1.7}{$<10^{-4}$}{901}
\ninecolumns{UU UMi}{$-0.11\pm0.02$}{MEarth}{1432 (1405)}{6.6}{90$\pm$26}{2.6}{$<10^{-4}$}{747}
\ninecolumns{LP 395-8}{$-3.03\pm0.16$}{SuperWASP}{986 (880)}{21.2}{1.125$\pm$0.011}{15.1}{$<10^{-4}$}{67}
\ninecolumns{GJ 810A}{$-0.02\pm0.07$}{ASAS}{417 (394)}{65.8}{$\cdots$}{$\cdots$}{$\cdots$}{3170}
\noalign{\smallskip}
\hline
\end{tabular}
\tablefoot{
\tablefoottext{a}{Only periods with FAP$<0.1\%$ are given. The pseudo-equivalent width of the H$\alpha$ line is taken from \cite{Jeffers2017}, and is also reported as an activity indicator .}
}
\end{table*}

\section{Results and discussion}
\label{sec:discussion}

\subsection{Individual masses and radii}

The analysis of the radial velocities of SB2 systems only yields the minimum masses of the components. However, it is possible to estimate absolute values using additional constraints such as mass-luminosity calibrations and mass ratios. We made use of the empirical mass-luminosity relationship ($\mathcal{M}$-M$_{K_s}$) in \cite{Benedict2016}, which is based on mass measurements of astrometric M dwarf binaries. We assumed uncertainties of 0.02 M$_{\odot}$ according to the scatter of the residuals of this relationship.

To estimate the individual masses, we used the systems' $K_{s}$ band magnitude of each binary system given in Table~\ref{tab:basicprop}. From this magnitude and the flux ratio of the system, which we iteratively change between 0 and 1 in steps of 0.01, we computed a set of $K_{s,1}$ and $K_{s,2}$ values corresponding to each component of the system, and converted them to absolute magnitudes M$_{K_{s,1}}$ and M$_{K_{s,2}}$ using the distance of each system. Then, we determined $\mathcal{M}_1$ and $\mathcal{M}_2$ for each absolute magnitude using the $\mathcal{M}$-M$_{K_s}$ relationship. From the set of possible values, we chose as individual masses those reproducing the mass ratio obtained from the radial velocity analysis, which are listed in  Table \ref{tab:physical}. Alternatively, it is also possible to use the flux ratios derived from the spectral analysis with {\sc todmor}, but in our case, they correspond to the flux ratio at the effective wavelengths of the VIS and NIR CARMENES channels, which have wavelengths shortwards of $K_{s}$ used for the mass calibration. The second and third columns in Table~\ref{MSBOP} show the calculated individual masses for each system, with their uncertainties estimated as the standard deviation of 10000 Monte Carlo realizations of the input parameter distribution.

\begin{figure}
\includegraphics[width=88 mm]{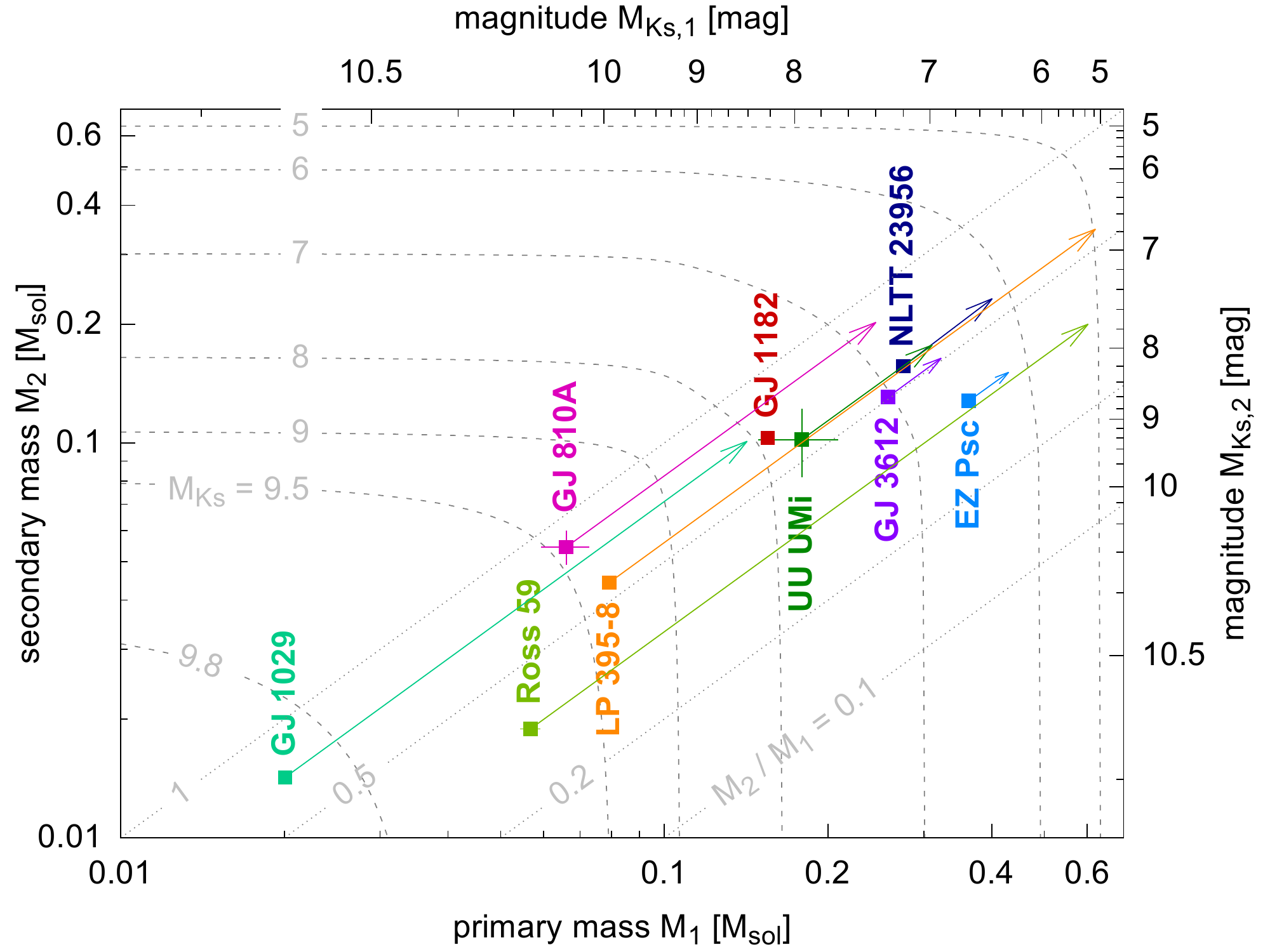}
\caption{Minimum masses, $\mathcal{M} \sin^3i$, for the primary and secondary components of the SB2 binaries (squares). The magnitudes on the top and right axes are computed according to the $\mathcal{M}$-M$_{K_s}$ relation in \cite{Benedict2016}. Lines of constant mass ratio values are shown as dotted diagonals. Dashed contours correspond to the same total flux for pairs M$_{K_s,1}$ and M$_{K_s,2}$. The arrows point to the estimated absolute masses and magnitude M$_{K_s}$. Long arrows are indicative of low orbital inclinations.
}

\label{fig:benedict}
\end{figure}

To check the consistency of individual masses, we compared them with the minimum masses reported in Table \ref{tab:physical}, finding no discrepant values. Figure \ref{fig:benedict} shows the minimum masses found for our systems (see Table~\ref{tab:physical}) compared with the $\mathcal{M}$-$M_{K_s}$ calibration from \cite{Benedict2016}. The arrows point towards the absolute masses derived from each component (see Table~\ref{MSBOP}). Vertical arrows indicate large inclination angles between the visual and the normal to the orbital plane, while long horizontal arrows indicate systems with low inclination (i.e., small $\sin{i}$).

Since none of the binary systems presented here are eclipsing, we computed individual radii from individual masses derived in this section and the empirical mass-radius relation in Schweitzer et al. (in prep.), $R = a \mathcal{M} + b$, where $a$~=~0.934$\pm$0.015, $b$~=~0.0286$\pm$0.066, and $R$ and $M$ are in solar units.
This relation is based on masses and radii of eclipsing binaries, and is valid on a mass range from 0.092 M$_{\odot}$ to 0.73 M$_{\odot}$. The last two columns in Table \ref{MSBOP} provide the individual radii of the components. We have used the radii of the primary components of the binary systems to estimate the inclination of the targets showing rotation periods below 10 days, for which the spectral broadening may be close to the rotation velocity, $v \sin i$. However, only a consistent value of ~38 deg was found for LP~395-8, compatible with the lack of eclipses.

\begin{table*}[t]
\centering
\caption{Individual masses and absolute magnitudes computed with the mass-luminosity relation in \cite{Benedict2016} and the mass ratio in Table \ref{tab:physical}, and individual radii computed with the empirical mass-radius relation in Schweitzer et al. (in prep.).}

\label{MSBOP}

\begin{tabular}[t]{lcccccc}
\hline
\hline
\noalign{\smallskip}
{Name} & $\mathcal{M}_1 ~[M_{\odot}]$ & $\mathcal{M}_2 ~[M_{\odot}]$ & $\mathcal{R}_1 ~[R_{\odot}]$ & $\mathcal{R}_2 ~[R_{\odot}]$ & $K_{s,1} ~[mag]$ & $K_{s,2} ~[mag]$\\
\noalign{\smallskip}
 \hline
 \noalign{\smallskip}
EZ Psc & $0.430\pm0.021$ & $0.151\pm0.020$ &$0.430\pm0.021$&$0.170\pm0.020$&$6.341\pm0.023$ &$8.271\pm0.023$\\
GJ 1029  & $0.142\pm0.020$ & $0.101\pm0.020$ &$0.161\pm0.020$&$0.123\pm0.020$&$8.407\pm0.022$ &$9.473\pm0.038$\\
Ross 59  &$0.602\pm0.032$&$0.200\pm0.022$&$0.591\pm0.032$&$0.216\pm0.022$&$5.31\pm0.21$ &$7.748\pm0.084$\\
NLTT 23956  &$0.401\pm0.021$&$0.232\pm0.020$&$0.403\pm0.021$&$0.245\pm0.020$&$6.490\pm0.025$ &$7.494\pm0.020$\\
GJ 3612  &$0.323\pm0.020$&$0.164\pm0.020$&$0.330\pm0.021$&$0.181\pm0.020$&$6.914\pm0.021$ &$8.113\pm0.023$\\
GJ 1182  &$0.157\pm0.020$&$0.104\pm0.020$&$0.175\pm0.020$&$0.126\pm0.020$& $8.198\pm0.023$&$9.359\pm0.042$\\
UU UMi  &$0.311\pm0.024$&$0.177\pm0.024$&$0.319\pm0.024$&$0.194\pm0.024$&$6.983\pm0.074$ &$7.98\pm0.15$\\
LP 395-8  &$0.621\pm0.020$&$0.348\pm0.020$ &$0.608\pm0.022$&$0.353\pm0.021$&$5.168\pm0.021$ &$6.7767\pm0.0095$\\
GJ 810A &$0.245\pm0.021$&$0.202\pm0.021$&$0.258\pm0.021$&$0.217\pm0.021$&$7.398\pm0.036$ & $7.730\pm0.042$\\
\noalign{\smallskip}
\hline
\end{tabular}

\end{table*}

\subsection{Parameter distribution}

We compared our M dwarf SB2 systems with those already reported. We considered the SB9 catalogue of spectroscopic binary orbits \citep[][last update April 2018]{Pourbaix2004} which contains the orbital parameters of 3595 spectroscopic binaries, of which 1093 are SB2 systems. Only 18 of these systems correspond to SB2 M dwarf main sequence binary systems.
A bibliographic search also results in 40 further known systems with published orbital parameters not included in SB9. Therefore, the nine systems studied in the present work bring the total number of M dwarf SB2 systems to 67, of which 29 are eclipsing, thus increasing the number of known SB2 systems by 15.5\,\%. Table~\ref{tab:Mdwarfs} compiles the radial velocity parameters for all the 67 spectroscopic binary systems with M dwarf main sequence components we have found in the literature, including the systems analysed in this paper for completeness.

\begin{figure*}[t]

\centering
\includegraphics[scale=0.65]{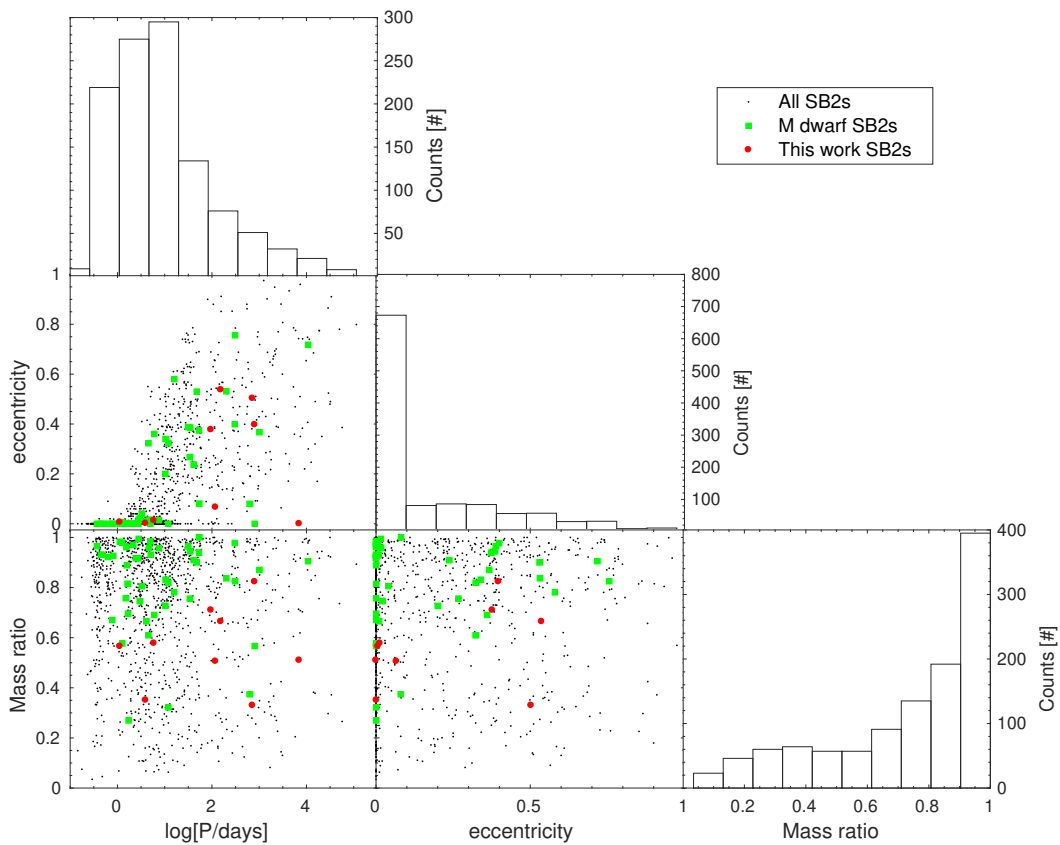}

\caption{Parameter distribution of the SB2 spectroscopic binary systems in SB9 (black dots) and in this work (red circles). M dwarf spectroscopic binary systems in SB9 and coming from the literature are also shown as green squares. The upper panel of each column displays the one-dimensional distribution of each of parameter.}
\label{fig:distribution}
\end{figure*}

Figure~\ref{fig:distribution} shows the parameter distribution of the SB2 systems in SB9, the M dwarf systems coming from both SB9 and the literature, and our reported new systems (red circles). 
The SB2s analysed in this work have typically smaller mass ratios than previously published M dwarf binaries. This results from a combination of several factors, including the high signal-to-noise ratio and resolution of the CARMENES data, the lower semi-amplitudes induced by less massive components, and our previous literature compilation and preparatory observations, which discarded already known binary systems from the CARMENES sample of targets \citep{Caballero2016,CortesContreras2017, Jeffers2017}. This initial cleaning also explains the apparently low binary fraction of the sample, with only 9 of the 342 surveyed stars found to be SB2s. Actually, \cite{CortesContreras2017} analyzed the CARMENES input catalogue and found a multiplicity fraction of 36.5$\pm$2.6\,\%.

Given the orbital periods and separations of the binary systems studied here, it is worth estimating if their orbits could be resolved by Gaia, since this would provide precise individual absolute masses independent from calibrations. Using the individual masses in Table~\ref{MSBOP} and the 1~Gyr stellar models in \cite{Baraffe2015}, we estimated individual $G$-band magnitudes for each component of the systems, which are listed in Table \ref{tab:gaia}. We then computed the semi-major axis of the photocenter motion in the $G$-band, $\alpha_G$, shown in the last column in Table \ref{tab:gaia}. We found values ranging from 0.1 to 65~mas, therefore, given the Gaia astrometric precision of 50 $\mu$as \citep{2018arXiv180409366L}, the astrometric orbits of all systems could be, in principle, resolved. Besides, all the binary systems analysed here show an \textit{astrometric excess noise} \citep{Lindegren2012} parameter above 0.25~mas, always above the median of all the Gaia sources~\citep{2018arXiv180409366L}. This may indicate that individual astrometric measurements are affected by the orbital motion of the system. Moreover, UU UMi and Ross 59, are flagged as duplicate sources in the second data release, indicating that they may be spatially resolved by Gaia.

\begin{table}[h]

\caption{Individual Gaia $G$-band magnitudes estimated using individual masses in Table \ref{MSBOP} and the 1 Gyr stellar models in \cite{Baraffe2015}, and motion of the semi-major axis of the photocenter in the $G$-band.}
\label{tab:gaia}
\centering
\begin{tabular}{lccc}
\hline\hline
\noalign{\smallskip}
Name & G$_1$ & G$_2$ &  $\alpha_G$\\
 & [mag] & [mag] & [mas]\\\noalign{\smallskip}
\hline
\noalign{\smallskip}
EZ Psc      &  11.01     &  14.28   &    0.54 \\
GJ 1029     &  14.60    &  17.00     &    3.34 \\
Ross 59     &  9.24   &  13.29    &    7.73 \\
NLTT 23956  &  11.27    &  12.92    &    0.36 \\
GJ 3612     &  11.94   &  13.97    &   5.44 \\
GJ 1182     &  14.14   &  16.74    &    8.05 \\
UU UMi      &  12.02   &  13.78   &   65.2\\
LP 395-8    &  8.97    &  11.70   &  0.10 \\
GJ 810 A    &  12.77   &  13.27      &  3.97 \\
\hline
\end{tabular}
\end{table}

\section{Conclusions}
\label{sec:conclusions}
In this work we analysed nine new M dwarf SB2 systems found in the context of the CARMENES survey of exoplanets, increasing the number of known MM spectroscopic binaries by over 15\,\%. Orbital parameters derived from the radial velocities, i.e. period, eccentricity, argument of the periastron, radial velocity semi-amplitudes and mass ratios, are provided for these systems for the first time. Among them, 3 systems have periods shorter than 10 days, 3 between 70 and 160 days and 3 have periods longer than around 2 years for which additional observations may help to better constrain their properties.

Publicly available photometry for these targets was also analysed. Significant periodic signals attributed to the rotation period are found for 7 of the systems. Unfortunately, no eclipses are found in any case. However, individual masses and radii were estimated using empirical calibrations for systems with parallactic distances, providing the fundamental properties of the components of the systems.

The comparison of the orbital properties of the systems studied here with those from the literature reveals that our set of low-mass binary systems have smaller mass ratios than more massive systems and that of known M dwarfs SB2s. This trend may arise from the better sensitivity of the CARMENES spectrograph towards longer wavelengths. This could also suggest that low-mass binary systems may have lower mass ratios, but more statistics are needed to confirm this trend.

Further observations of these systems will help to better constrain the properties of the long period systems. Precise astrometric measurements from Gaia may also be very valuable to put additional constraints and derive absolute masses and inclinations. This will increase the sample of low-mass stars that can be used to refine the mass-luminosity relationship of these systems, independently of stellar models.

\begin{acknowledgements}
We are grateful to C. Jordi for useful discussions on Gaia data. We also thank the anonymous referee for a thorough and very helpful review of the paper. CARMENES is an instrument for the Centro Astron\'omico Hispano-Alem\'an de  Calar Alto (CAHA, Almer\'{\i}a, Spain).
  CARMENES is funded by the German Max-Planck-Gesellschaft (MPG),
  the Spanish Consejo Superior de Investigaciones Cient\'{\i}ficas (CSIC),
  the European Union through FEDER/ERF FICTS-2011-02 funds,
  and the members of the CARMENES Consortium
  (Max-Planck-Institut f\"ur Astronomie,
  Instituto de Astrof\'{\i}sica de Andaluc\'{\i}a,
  Landessternwarte K\"onigstuhl,
  Institut de Ci\`encies de l'Espai,
  Insitut f\"ur Astrophysik G\"ottingen,
  Universidad Complutense de Madrid,
  Th\"uringer Landessternwarte Tautenburg,
  Instituto de Astrof\'{\i}sica de Canarias,
  Hamburger Sternwarte,
  Centro de Astrobiolog\'{\i}a and
  Centro Astron\'omico Hispano-Alem\'an),
  with additional contributions by the Spanish Ministry of Economy,
  the German Science Foundation through the Major Research Instrumentation
    Programme and DFG Research Unit FOR2544 ``Blue Planets around Red Stars'',
  the Klaus Tschira Stiftung,
  the states of Baden-W\"urttemberg and Niedersachsen,
  and by the Junta de Andaluc\'{\i}a.
  We acknowledge support from the Spanish Ministry of Economy and Competitiveness
  (MINECO) and the Fondo Europeo de Desarrollo Regional (FEDER) through grants
  ESP2013-48391-C4-1-R, ESP2014-57495-C2-2-R and AYA2015-69350-C3-2-P, AYA2016-79425–C3–1/2/3–P, ESP2016-80435-C2-1-R, as
  well as the support of the Generalitat de Catalunya/CERCA programme. We also acknowledge support
  from the Ag\`encia de Gesti\'o d'Ajuts Universitaris i de Recerca of the Generalitat de Catalunya
  through grant 2018 FI\_B\_00188.
  This work makes use of data from the HARPS-N Project, a collaboration between the Astronomical Observatory of the Geneva University (lead), the CfA in Cambridge, the Universities of St. Andrews and Edinburgh, the Queens University of Belfast, and the TNG-INAF Observatory; from the public release of the WASP data as provided by the WASP consortium and services at the NASA Exoplanet Archive, which is operated by the California Institute of Technology, under contract with the National Aeronautics and Space Administration under the Exoplanet Exploration Program; from the MEarth Project, which is a collaboration between Harvard University and the Smithsonian Astrophysical Observatory; and from the Northern Sky Variability Survey created jointly by the Los Alamos National Laboratory and University of Michigan and funded by the Department of Energy, the National Aeronautics and Space Administration, and the National Science Foundation. This work has made use of data from the European Space Agency (ESA) mission
{\it Gaia} (\url{https://www.cosmos.esa.int/gaia}), processed by the {\it Gaia}
Data Processing and Analysis Consortium (DPAC,
\url{https://www.cosmos.esa.int/web/gaia/dpac/consortium}). Funding for the DPAC
has been provided by national institutions, in particular the institutions
participating in the {\it Gaia} Multilateral Agreement.

\end{acknowledgements}

%
%


\begin{appendix}

\section{Radial velocity data}
\label{sec:appdata}

\begin{table}[h]
\caption{Radial velocity measurements.}
\label{tab:rvs}
\centering
\begin{tabular}{lccc}
\hline\hline
Name& BJD            & RV$_{1}$       & RV$_{2}$\\
&(2,450,000+)   & (km\,s$^{-1}$) & (km\,s$^{-1}$)\\         
\hline                                                              
Ez Psc& 7591.6809 & -18.51 $\pm$ 0.08 & 50.71 $\pm$ 0.38\\
(VIS)& 7595.6767 & -19.85 $\pm$ 0.08 & 54.41 $\pm$ 0.38\\
& 7604.6730 & -17.19 $\pm$ 0.08 & 47.27 $\pm$ 0.35\\
& 7618.5834 & 14.03 $\pm$ 0.09 & -41.43 $\pm$ 0.37\\
& 7634.5438 & 8.71 $\pm$ 0.09 & -26.74 $\pm$ 0.40\\
& 7650.6270 & -2.40 $\pm$ 0.09 & 6.37 $\pm$ 0.34\\
& 7655.5387 & -28.11 $\pm$ 0.08 & 78.12 $\pm$ 0.34\\
& 7676.4600 & 6.47 $\pm$ 0.08 & -20.38 $\pm$ 0.37\\
& 7677.5432 & 25.01 $\pm$ 0.09 & -73.18 $\pm$ 0.39\\
& 8120.2882 & 26.57 $\pm$ 0.08 & -77.31 $\pm$ 0.37\\
\hline
Ez Psc& 7591.6810 & -18.57 $\pm$ 0.14 & 51.51 $\pm$ 0.46\\
(NIR)& 7595.6779 & -20.17 $\pm$ 0.16 & 54.12 $\pm$ 0.41\\
& 7604.6730 & -17.38 $\pm$ 0.15 & 47.52 $\pm$ 0.38\\
& 7618.5845 & 13.84 $\pm$ 0.15 & -42.25 $\pm$ 0.43\\
& 7634.5464 & 8.50 $\pm$ 0.17 & -26.00 $\pm$ 0.61\\
& 7650.6280 & -2.61 $\pm$ 0.15 & 5.69 $\pm$ 0.31\\
& 7655.5384 & -28.40 $\pm$ 0.16 & 76.52 $\pm$ 0.63\\
& 7676.4601 & 6.33 $\pm$ 0.18 & -20.77 $\pm$ 0.56\\
& 7677.5434 & 24.87 $\pm$ 0.14 & -72.71 $\pm$ 0.57\\
& 8120.2883 & 26.14 $\pm$ 0.16 & -76.99 $\pm$ 0.59\\
\hline
GJ 1029& 7611.6660 & -7.11 $\pm$ 0.09 & -17.00 $\pm$ 0.24\\
(VIS)& 7619.6197 & -7.81 $\pm$ 0.10 & -16.46 $\pm$ 0.22\\
& 7620.6000 & -7.95 $\pm$ 0.10 & -16.40 $\pm$ 0.22\\
& 7625.5827 & -8.69 $\pm$ 0.10 & -16.15 $\pm$ 0.23\\
& 7786.3730 & -7.13 $\pm$ 0.12 & -16.86 $\pm$ 0.30\\
& 7936.6348 & -15.13 $\pm$ 0.09 & -5.92 $\pm$ 0.23\\
& 7951.6386 & -20.17 $\pm$ 0.10 & 0.88 $\pm$ 0.29\\
& 7981.6498 & -6.86 $\pm$ 0.09 & -17.15 $\pm$ 0.25\\
& 7999.6108 & -7.59 $\pm$ 0.09 & -16.62 $\pm$ 0.22\\
& 8031.5371 & -14.89 $\pm$ 0.09 & -6.34 $\pm$ 0.22\\
& 8033.5194 & -15.66 $\pm$ 0.10 & -5.34 $\pm$ 0.26\\
& 8040.4951 & -19.03 $\pm$ 0.10 & -0.66 $\pm$ 0.27\\
& 8047.4815 & -20.11 $\pm$ 0.09 & 0.87 $\pm$ 0.27\\
& 8051.4670 & -17.52 $\pm$ 0.09 & -2.52 $\pm$ 0.25\\
& 8055.4708 & -14.22 $\pm$ 0.10 & -6.74 $\pm$ 0.26\\
\hline
GJ 1029& 7611.6661 & -7.42 $\pm$ 0.16 & -17.18 $\pm$ 0.34\\
(NIR)& 7619.6204 & -8.41 $\pm$ 0.22 & -15.70 $\pm$ 0.44\\
& 7620.6000 & -8.51 $\pm$ 0.19 & -15.36 $\pm$ 0.34\\
& 7625.5824 & -9.18 $\pm$ 0.16 & -14.90 $\pm$ 0.28\\
& 7786.3723 & -7.58 $\pm$ 0.22 & -17.18 $\pm$ 0.43\\
& 7936.6348 & -14.16 $\pm$ 0.14 & -5.81 $\pm$ 0.27\\
& 7951.6386 & -19.49 $\pm$ 0.13 & 1.81 $\pm$ 0.28\\
& 7981.6496 & -7.11 $\pm$ 0.15 & -17.40 $\pm$ 0.31\\
& 7999.6106 & -8.19 $\pm$ 0.16 & -16.49 $\pm$ 0.34\\
& 8031.5359 & -14.23 $\pm$ 0.14 & -6.94 $\pm$ 0.30\\
& 8033.5196 & -15.33 $\pm$ 0.17 & -5.38 $\pm$ 0.35\\
& 8040.4955 & -19.04 $\pm$ 0.14 & -0.40 $\pm$ 0.30\\
& 8047.4813 & -20.13 $\pm$ 0.13 & 1.12 $\pm$ 0.29\\
& 8051.4666 & -17.38 $\pm$ 0.14 & -2.42 $\pm$ 0.30\\
& 8055.4693 & -13.64 $\pm$ 0.16 & -7.96 $\pm$ 0.31\\
\hline
Ross 59& 6255.6871 & 28.32 $\pm$ 0.07 & 23.97 $\pm$ 0.62\\
(HARPS-N)& 6255.7830 & 28.33 $\pm$ 0.08 & 24.12 $\pm$ 0.65\\
& 6604.6893 & 24.04 $\pm$ 0.07 & 37.34 $\pm$ 1.09\\
& 6605.6889 & 24.04 $\pm$ 0.07 & 37.50 $\pm$ 0.82\\
\hline

\end{tabular}
\end{table}

\addtocounter{table}{-1}

\begin{table}[h]
\caption{Continued.}
\centering
\begin{tabular}{cccc}
\hline\hline
Name& BJD            & RV$_{1}$       & RV$_{2}$\\  
&(2,450,000+)   & (km\,s$^{-1}$) & (km\,s$^{-1}$)\\         
\hline
& 6606.6984 & 24.04 $\pm$ 0.07 & 37.08 $\pm$ 0.95\\
& 6607.6778 & 24.04 $\pm$ 0.07 & 37.33 $\pm$ 0.98\\
\hline

Ross 59& 7652.6555 & 28.07 $\pm$ 0.03 & 25.59 $\pm$ 0.24\\
(VIS)& 7656.6492 & 28.08 $\pm$ 0.03 & 25.52 $\pm$ 0.24\\
& 7676.6923 & 28.23 $\pm$ 0.03 & 25.03 $\pm$ 0.20\\
& 7689.6558 & 28.35 $\pm$ 0.04 & 24.63 $\pm$ 0.31\\
& 7691.6467 & 28.36 $\pm$ 0.03 & 24.65 $\pm$ 0.18\\
& 7692.6448 & 28.34 $\pm$ 0.03 & 24.48 $\pm$ 0.19\\
& 7693.6418 & 28.35 $\pm$ 0.03 & 24.49 $\pm$ 0.22\\
& 7701.6213 & 28.46 $\pm$ 0.04 & 24.49 $\pm$ 0.26\\
& 7766.5483 & 29.01 $\pm$ 0.03 & 23.04 $\pm$ 0.19\\
& 7815.3666 & 29.46 $\pm$ 0.02 & 21.91 $\pm$ 0.16\\
& 7830.4308 & 29.58 $\pm$ 0.03 & 21.50 $\pm$ 0.18\\
& 7851.3390 & 29.80 $\pm$ 0.03 & 21.03 $\pm$ 0.21\\
& 8017.6101 & 25.18 $\pm$ 0.03 & 34.67 $\pm$ 0.20\\
& 8088.6191 & 24.54 $\pm$ 0.03 & 36.61 $\pm$ 0.20\\
& 8118.4902 & 24.98 $\pm$ 0.03 & 35.07 $\pm$ 0.16\\
& 8149.4038 & 25.51 $\pm$ 0.03 & 33.71 $\pm$ 0.18\\
\hline
Ross 59& 7766.5484 & 28.84 $\pm$ 0.06 & 22.14 $\pm$ 0.17\\
(NIR)& 7815.3670 & 29.33 $\pm$ 0.05 & 21.84 $\pm$ 0.17\\
& 7830.4318 & 29.44 $\pm$ 0.06 & 21.73 $\pm$ 0.19\\
& 7851.3392 & 29.71 $\pm$ 0.07 & 21.29 $\pm$ 0.33\\
& 8017.6116 & 25.59 $\pm$ 0.05 & 34.40 $\pm$ 0.28\\
& 8088.6191 & 24.51 $\pm$ 0.10 & 36.47 $\pm$ 0.41\\
& 8118.4902 & 25.06 $\pm$ 0.07 & 34.13 $\pm$ 0.33\\
& 8149.4025 & 25.57 $\pm$ 0.07 & 32.92 $\pm$ 0.19\\
\hline
NLTT 23956& 7735.7087 & 43.19 $\pm$ 0.13 & -37.45 $\pm$ 0.19\\
(VIS)& 7799.5416 & 4.10 $\pm$ 0.13 & 30.29 $\pm$ 0.21\\
& 7814.4799 & 18.19 $\pm$ 0.13 & 6.49 $\pm$ 0.21\\
& 7821.4617 & -12.76 $\pm$ 0.13 & 59.86 $\pm$ 0.19\\
& 7830.4822 & 43.27 $\pm$ 0.14 & -37.63 $\pm$ 0.20\\
& 7832.4844 & 10.38 $\pm$ 0.13 & 19.63 $\pm$ 0.19\\
& 7848.4245 & 45.39 $\pm$ 0.13 & -40.90 $\pm$ 0.21\\
& 7850.3877 & 5.73 $\pm$ 0.14 & 27.13 $\pm$ 0.21\\
& 7856.3647 & 4.15 $\pm$ 0.14 & 30.29 $\pm$ 0.22\\
& 7861.3722 & 33.45 $\pm$ 0.14 & -20.22 $\pm$ 0.20\\
& 7866.3890 & 46.20 $\pm$ 0.14 & -42.50 $\pm$ 0.21\\
& 8080.7339 & 26.90 $\pm$ 0.14 & -9.02 $\pm$ 0.20\\
& 8092.7091 & 22.71 $\pm$ 0.13 & -2.18 $\pm$ 0.19\\
& 8094.7477 & -17.47 $\pm$ 0.13 & 67.61 $\pm$ 0.18\\
\hline
NLTT 23956& 7735.7080 & 43.30 $\pm$ 0.19 & -37.46 $\pm$ 0.30\\
(NIR)& 7799.5416 & 4.07 $\pm$ 0.20 & 30.74 $\pm$ 0.30\\
& 7814.4798 & 18.15 $\pm$ 0.20 & 6.67 $\pm$ 0.31\\
& 7821.4619 & -12.90 $\pm$ 0.18 & 59.38 $\pm$ 0.29\\
& 7830.4824 & 43.35 $\pm$ 0.21 & -37.85 $\pm$ 0.33\\
& 7832.4845 & 10.44 $\pm$ 0.22 & 18.57 $\pm$ 0.30\\
& 7848.4241 & 45.42 $\pm$ 0.20 & -41.07 $\pm$ 0.30\\
& 7850.3878 & 5.64 $\pm$ 0.31 & 26.98 $\pm$ 0.45\\
& 7856.3646 & 3.94 $\pm$ 0.24 & 30.00 $\pm$ 0.35\\
& 7861.3722 & 33.36 $\pm$ 0.22 & -20.42 $\pm$ 0.34\\
& 7866.3885 & 46.05 $\pm$ 0.28 & -42.43 $\pm$ 0.46\\
& 8080.7338 & 27.00 $\pm$ 0.33 & -9.47 $\pm$ 0.50\\
& 8092.7084 & 22.77 $\pm$ 0.21 & -2.13 $\pm$ 0.32\\
& 8094.7472 & -17.60 $\pm$ 0.24 & 67.59 $\pm$ 0.31\\
\hline
GJ 3612& 7419.6610 & -70.23 $\pm$ 0.07 & -46.10 $\pm$ 0.25\\
(VIS)& 7474.4705 & -54.91 $\pm$ 0.07 & -76.35 $\pm$ 0.26\\
& 7494.4779 & -51.12 $\pm$ 0.07 & -83.26 $\pm$ 0.27\\

\hline
\end{tabular}
\end{table}

\addtocounter{table}{-1}

\begin{table}[t]
\caption{Continued.}
\centering
\begin{tabular}{c c c c c}
\hline\hline
Name& BJD            & RV$_{1}$       & RV$_{2}$\\  
&(2,450,000+)   & (km\,s$^{-1}$) & (km\,s$^{-1}$)\\
\hline
& 7672.7357 & -72.14 $\pm$ 0.08 & -42.13 $\pm$ 0.27\\
& 7690.6364 & -67.61 $\pm$ 0.07 & -51.27 $\pm$ 0.23\\
& 7709.5717 & -56.88 $\pm$ 0.07 & -72.03 $\pm$ 0.25\\
& 7759.5694 & -62.78 $\pm$ 0.10 & -60.47 $\pm$ 0.33\\
& 7761.6902 & -63.70 $\pm$ 0.08 & -58.34 $\pm$ 0.30\\
& 7762.5343 & -64.08 $\pm$ 0.07 & -57.64 $\pm$ 0.25\\
& 7763.6222 & -64.59 $\pm$ 0.07 & -56.92 $\pm$ 0.23\\
& 7766.5875 & -65.93 $\pm$ 0.07 & -54.60 $\pm$ 0.26\\
& 7787.5680 & -72.05 $\pm$ 0.07 & -42.33 $\pm$ 0.25\\
& 7802.7532 & -70.42 $\pm$ 0.07 & -45.70 $\pm$ 0.25\\
& 7815.7454 & -64.75 $\pm$ 0.08 & -56.77 $\pm$ 0.25\\
& 7819.7395 & -62.71 $\pm$ 0.11 & -60.59 $\pm$ 0.36\\
& 7823.4053 & -60.12 $\pm$ 0.08 & -66.54 $\pm$ 0.26\\
& 7830.6792 & -55.93 $\pm$ 0.07 & -73.98 $\pm$ 0.27\\
& 7833.3296 & -54.62 $\pm$ 0.07 & -76.98 $\pm$ 0.26\\
& 7850.5303 & -51.01 $\pm$ 0.07 & -84.04 $\pm$ 0.26\\
& 8088.7086 & -50.96 $\pm$ 0.08 & -84.04 $\pm$ 0.29\\
& 8105.7093 & -56.30 $\pm$ 0.07 & -73.22 $\pm$ 0.27\\
\hline
GJ 3612& 7474.4704 & -55.37 $\pm$ 0.07 & -76.48 $\pm$ 0.18\\
(NIR)& 7672.7355 & -72.35 $\pm$ 0.07 & -42.68 $\pm$ 0.17\\
& 7690.6361 & -67.82 $\pm$ 0.09 & -51.42 $\pm$ 0.26\\
& 7709.5699 & -56.95 $\pm$ 0.08 & -72.16 $\pm$ 0.22\\
& 7763.6248 & -64.56 $\pm$ 0.08 & -57.30 $\pm$ 0.17\\
& 7766.5882 & -66.06 $\pm$ 0.07 & -54.88 $\pm$ 0.18\\
& 7787.5640 & -72.21 $\pm$ 0.08 & -42.67 $\pm$ 0.22\\
& 7802.7533 & -70.49 $\pm$ 0.07 & -45.81 $\pm$ 0.19\\
& 7815.7448 & -64.65 $\pm$ 0.11 & -57.03 $\pm$ 0.22\\
& 7823.4070 & -60.66 $\pm$ 0.07 & -67.05 $\pm$ 0.14\\
& 7830.6779 & -56.01 $\pm$ 0.08 & -74.42 $\pm$ 0.22\\
& 7833.3285 & -55.00 $\pm$ 0.07 & -76.86 $\pm$ 0.17\\
& 8088.7084 & -51.06 $\pm$ 0.14 & -84.19 $\pm$ 0.36\\
& 8105.7097 & -56.46 $\pm$ 0.10 & -73.54 $\pm$ 0.23\\
\hline
GJ 1182& 7488.5925 & -2.53 $\pm$ 0.08 & 2.18 $\pm$ 0.23\\
(VIS)& 7494.5471 & -3.28 $\pm$ 0.08 & 3.34 $\pm$ 0.19\\
& 7505.5087 & -5.12 $\pm$ 0.08 & 6.02 $\pm$ 0.22\\
& 7802.6332 & -3.34 $\pm$ 0.07 & 3.30 $\pm$ 0.18\\
& 7822.6421 & -6.80 $\pm$ 0.07 & 8.54 $\pm$ 0.19\\
& 7833.6210 & -9.04 $\pm$ 0.07 & 11.87 $\pm$ 0.20\\
& 7848.5930 & -11.69 $\pm$ 0.07 & 15.97 $\pm$ 0.21\\
& 7850.5784 & -11.86 $\pm$ 0.07 & 16.31 $\pm$ 0.20\\
& 7854.6329 & -11.88 $\pm$ 0.07 & 16.31 $\pm$ 0.20\\
& 7858.5516 & -10.87 $\pm$ 0.08 & 14.52 $\pm$ 0.21\\
& 7860.5355 & -9.69 $\pm$ 0.08 & 12.68 $\pm$ 0.21\\
& 7862.5431 & -7.58 $\pm$ 0.08 & 9.57 $\pm$ 0.21\\
& 7864.5170 & -4.87 $\pm$ 0.09 & 5.52 $\pm$ 0.25\\
& 7875.5984 & 11.00 $\pm$ 0.08 & -18.30 $\pm$ 0.20\\
& 7888.5064 & 10.53 $\pm$ 0.07 & -17.71 $\pm$ 0.20\\
& 7889.5389 & 10.28 $\pm$ 0.08 & -17.34 $\pm$ 0.22\\
& 7894.4574 & 9.11 $\pm$ 0.07 & -15.34 $\pm$ 0.20\\
& 7901.4802 & 7.34 $\pm$ 0.08 & -13.06 $\pm$ 0.22\\
& 7907.4334 & 5.91 $\pm$ 0.07 & -10.27 $\pm$ 0.22\\
& 7912.4396 & 4.86 $\pm$ 0.08 & -8.71 $\pm$ 0.21\\
& 7917.4318 & 3.85 $\pm$ 0.08 & -7.08 $\pm$ 0.23\\
\hline
GJ 1182& 7488.6016 & -2.23 $\pm$ 0.08 & 3.59 $\pm$ 0.17\\
(NIR)& 7802.6334 & -2.95 $\pm$ 0.12 & 3.87 $\pm$ 0.18\\
& 7822.6418 & -7.00 $\pm$ 0.09 & 8.43 $\pm$ 0.18\\
& 7833.6206 & -9.18 $\pm$ 0.09 & 11.59 $\pm$ 0.17\\
& 7848.5927 & -11.81 $\pm$ 0.11 & 15.89 $\pm$ 0.22\\
& 7852.5741 & -12.19 $\pm$ 0.08 & 16.13 $\pm$ 0.17\\
& 7854.6322 & -12.07 $\pm$ 0.13 & 15.97 $\pm$ 0.24\\

\hline
\end{tabular}
\end{table}

\addtocounter{table}{-1}

\begin{table}[t]
\caption{Continued.}
\centering
\begin{tabular}{c c c c c}
\hline\hline
Name& BJD            & RV$_{1}$       & RV$_{2}$\\  
&(2,450,000+)   & (km\,s$^{-1}$) & (km\,s$^{-1}$)\\
\hline
& 7858.5532 & -11.04 $\pm$ 0.09 & 14.38 $\pm$ 0.17\\
& 7860.5360 & -9.85 $\pm$ 0.11 & 12.57 $\pm$ 0.21\\
& 7862.5435 & -7.89 $\pm$ 0.08 & 9.67 $\pm$ 0.18\\
& 7875.5984 & 10.82 $\pm$ 0.11 & -18.71 $\pm$ 0.22\\
& 7888.5058 & 10.57 $\pm$ 0.10 & -17.53 $\pm$ 0.20\\
& 7889.5374 & 10.31 $\pm$ 0.12 & -17.14 $\pm$ 0.23\\
& 7894.4546 & 9.22 $\pm$ 0.10 & -15.19 $\pm$ 0.19\\
& 7901.4794 & 6.93 $\pm$ 0.13 & -13.19 $\pm$ 0.25\\
& 7907.4328 & 5.82 $\pm$ 0.12 & -10.84 $\pm$ 0.24\\
& 7912.4389 & 4.75 $\pm$ 0.09 & -8.97 $\pm$ 0.19\\
& 7917.4175 & 3.63 $\pm$ 0.11 & -7.63 $\pm$ 0.21\\
\hline
UU UMi & 7472.6413 & -38.92 $\pm$ 0.02 & -44.98 $\pm$ 0.09\\
(VIS)& 7504.5670 & -38.94 $\pm$ 0.03 & -45.06 $\pm$ 0.09\\
& 7529.4912 & -38.90 $\pm$ 0.03 & -45.12 $\pm$ 0.10\\
& 7556.4629 & -38.87 $\pm$ 0.03 & -45.15 $\pm$ 0.11\\
& 7559.5434 & -38.91 $\pm$ 0.03 & -45.18 $\pm$ 0.11\\
& 7763.6439 & -38.75 $\pm$ 0.03 & -45.46 $\pm$ 0.10\\
& 7800.7530 & -38.71 $\pm$ 0.02 & -45.52 $\pm$ 0.09\\
& 7815.5392 & -38.68 $\pm$ 0.03 & -45.51 $\pm$ 0.09\\
& 7832.5606 & -38.67 $\pm$ 0.03 & -45.50 $\pm$ 0.10\\
& 7848.6126 & -38.71 $\pm$ 0.03 & -45.53 $\pm$ 0.09\\
& 7867.5390 & -38.65 $\pm$ 0.03 & -45.53 $\pm$ 0.11\\
& 7897.4662 & -38.67 $\pm$ 0.03 & -45.52 $\pm$ 0.11\\
& 7931.5387 & -38.64 $\pm$ 0.03 & -45.57 $\pm$ 0.13\\
& 7961.3985 & -38.62 $\pm$ 0.03 & -45.61 $\pm$ 0.12\\
& 7993.3479 & -38.61 $\pm$ 0.03 & -45.64 $\pm$ 0.15\\
& 8054.3032 & -38.65 $\pm$ 0.03 & -45.65 $\pm$ 0.10\\
& 8117.7403 & -38.67 $\pm$ 0.03 & -45.66 $\pm$ 0.09\\
& 8161.7033 & -38.65 $\pm$ 0.03 & -45.64 $\pm$ 0.10\\
& 8200.5577 & -38.60 $\pm$ 0.03 & -45.60 $\pm$ 0.11\\
\hline
UU UMi& 7472.6401 & -39.33 $\pm$ 0.06 & -45.45 $\pm$ 0.12\\
(NIR)& 7504.5679 & -39.19 $\pm$ 0.07 & -45.33 $\pm$ 0.12\\
& 7529.4903 & -39.29 $\pm$ 0.06 & -45.42 $\pm$ 0.10\\
& 7556.4628 & -39.70 $\pm$ 0.08 & -45.90 $\pm$ 0.14\\
& 7559.5432 & -38.92 $\pm$ 0.07 & -45.12 $\pm$ 0.11\\
& 7763.6438 & -39.06 $\pm$ 0.08 & -45.60 $\pm$ 0.16\\
& 7800.7531 & -39.11 $\pm$ 0.06 & -45.70 $\pm$ 0.13\\
& 7815.5387 & -39.02 $\pm$ 0.06 & -45.65 $\pm$ 0.13\\
& 7832.5591 & -39.07 $\pm$ 0.06 & -45.65 $\pm$ 0.13\\
& 7848.6135 & -39.03 $\pm$ 0.08 & -45.61 $\pm$ 0.16\\
& 7867.5386 & -38.97 $\pm$ 0.06 & -45.60 $\pm$ 0.14\\
& 7897.4628 & -38.67 $\pm$ 0.07 & -45.29 $\pm$ 0.14\\
& 7931.5379 & -38.73 $\pm$ 0.06 & -45.36 $\pm$ 0.16\\
& 7961.3982 & -38.03 $\pm$ 0.09 & -44.70 $\pm$ 0.20\\
& 7993.3477 & -38.56 $\pm$ 0.11 & -45.08 $\pm$ 0.24\\
& 8054.3030 & -39.33 $\pm$ 0.09 & -45.97 $\pm$ 0.25\\
& 8117.7404 & -39.05 $\pm$ 0.09 & -45.66 $\pm$ 0.17\\
& 8200.5573 & -39.06 $\pm$ 0.13 & -45.34 $\pm$ 0.24\\
\hline
LP 395-8& 7545.6530 & 5.91 $\pm$ 0.16 & -84.53 $\pm$ 0.31\\
(VIS)& 7566.6209 & -49.26 $\pm$ 0.18 & 12.96 $\pm$ 0.36\\
& 7573.5523 & -20.07 $\pm$ 0.20 & -39.89 $\pm$ 0.41\\
& 7593.4966 & -60.67 $\pm$ 0.18 & 34.48 $\pm$ 0.32\\
& 7633.4363 & -14.10 $\pm$ 0.25 & -49.58 $\pm$ 0.58\\
& 7643.4056 & -50.69 $\pm$ 0.17 & 16.70 $\pm$ 0.31\\
& 7652.3889 & -57.38 $\pm$ 0.22 & 28.04 $\pm$ 0.43\\
& 7652.4062 & -55.33 $\pm$ 0.25 & 25.16 $\pm$ 0.47\\
& 7652.4215 & -53.40 $\pm$ 0.18 & 21.52 $\pm$ 0.49\\
& 7652.4367 & -51.37 $\pm$ 0.17 & 17.13 $\pm$ 0.31\\
& 7654.3796 & -48.09 $\pm$ 0.20 & 13.25 $\pm$ 0.35\\
& 8089.2778 & -60.97 $\pm$ 0.28 & 34.54 $\pm$ 0.87\\
\hline

\end{tabular}
\end{table}

\addtocounter{table}{-1}

\begin{table}[t]
\caption{Continued.}
\centering
\begin{tabular}{c c c c c}
\hline\hline
Name& BJD            & RV$_{1}$       & RV$_{2}$\\  
&(2,450,000+)   & (km\,s$^{-1}$) & (km\,s$^{-1}$)\\
\hline
& 8093.2694 & 10.16 $\pm$ 0.13 & -91.82 $\pm$ 0.26\\
& 8102.2775 & 8.98 $\pm$ 0.17 & -90.36 $\pm$ 0.32\\
\hline
LP 395-8& 7545.6508 & 5.83 $\pm$ 0.19 & -84.06 $\pm$ 0.35\\
(NIR)& 7593.4976 & -60.46 $\pm$ 0.26 & 33.66 $\pm$ 0.59\\
& 7633.4363 & -14.19 $\pm$ 0.40 & -46.36 $\pm$ 0.55\\

& 7643.4069 & -50.66 $\pm$ 0.24 & 15.68 $\pm$ 0.45\\
& 7652.3889 & -57.74 $\pm$ 0.50 & 27.92 $\pm$ 0.98\\
& 7652.4063 & -55.93 $\pm$ 0.49 & 25.07 $\pm$ 1.36\\
& 7652.4216 & -53.72 $\pm$ 0.29 & 22.93 $\pm$ 0.50\\
& 7652.4353 & -51.50 $\pm$ 0.24 & 18.49 $\pm$ 0.42\\
& 7654.3791 & -48.37 $\pm$ 0.27 & 12.20 $\pm$ 0.49\\
& 8089.2780 & -60.99 $\pm$ 0.76 & 33.26 $\pm$ 1.32\\
& 8093.2686 & 9.75 $\pm$ 0.24 & -90.38 $\pm$ 0.65\\
& 8102.2776 & 9.08 $\pm$ 0.44 & -89.16 $\pm$ 1.46\\
\hline
GJ 810A& 7626.5156 & -144.98 $\pm$ 0.12 & -138.40 $\pm$ 0.14\\
(VIS)& 7630.4562 & -145.09 $\pm$ 0.12 & -138.25 $\pm$ 0.14\\
& 7642.4209 & -145.54 $\pm$ 0.10 & -137.85 $\pm$ 0.12\\
& 7652.3733 & -146.00 $\pm$ 0.13 & -137.62 $\pm$ 0.15\\
& 7655.4534 & -145.94 $\pm$ 0.11 & -137.40 $\pm$ 0.12\\
& 7673.3313 & -146.57 $\pm$ 0.11 & -136.80 $\pm$ 0.12\\
& 7677.3211 & -146.70 $\pm$ 0.11 & -136.59 $\pm$ 0.13\\
& 7704.2706 & -147.54 $\pm$ 0.10 & -135.56 $\pm$ 0.12\\
& 7911.6195 & -138.92 $\pm$ 0.12 & -146.13 $\pm$ 0.12\\
& 7928.5978 & -138.42 $\pm$ 0.11 & -146.59 $\pm$ 0.12\\
& 7943.5924 & -138.08 $\pm$ 0.12 & -146.90 $\pm$ 0.13\\
& 7958.5804 & -137.85 $\pm$ 0.10 & -147.18 $\pm$ 0.12\\
& 7975.5164 & -137.60 $\pm$ 0.11 & -147.31 $\pm$ 0.12\\
& 7990.4604 & -137.54 $\pm$ 0.11 & -147.36 $\pm$ 0.13\\
& 8020.3798 & -137.62 $\pm$ 0.11 & -147.36 $\pm$ 0.12\\
& 8048.3226 & -137.83 $\pm$ 0.11 & -147.13 $\pm$ 0.12\\
& 8089.2567 & -138.58 $\pm$ 0.12 & -146.74 $\pm$ 0.13\\
& 8093.2450 & -138.62 $\pm$ 0.11 & -146.66 $\pm$ 0.12\\
\hline
GJ 810A& 7626.5157 & -143.92 $\pm$ 0.14 & -138.14 $\pm$ 0.16\\
(NIR)& 7630.4562 & -144.10 $\pm$ 0.15 & -138.16 $\pm$ 0.15\\
& 7652.3734 & -145.31 $\pm$ 0.32 & -137.77 $\pm$ 0.33\\
& 7655.4529 & -145.62 $\pm$ 0.15 & -137.95 $\pm$ 0.14\\
& 7673.3310 & -146.40 $\pm$ 0.13 & -136.75 $\pm$ 0.14\\
& 7677.3211 & -146.72 $\pm$ 0.17 & -136.72 $\pm$ 0.20\\
& 7704.2692 & -147.59 $\pm$ 0.16 & -135.49 $\pm$ 0.17\\
& 7928.5971 & -139.18 $\pm$ 0.13 & -146.65 $\pm$ 0.13\\
& 7943.5931 & -138.37 $\pm$ 0.18 & -146.70 $\pm$ 0.17\\
& 7958.5790 & -137.46 $\pm$ 0.14 & -146.58 $\pm$ 0.15\\
& 7975.5159 & -137.72 $\pm$ 0.15 & -147.29 $\pm$ 0.16\\
& 7990.4607 & -137.68 $\pm$ 0.15 & -147.29 $\pm$ 0.15\\
& 8020.3803 & -137.54 $\pm$ 0.14 & -147.12 $\pm$ 0.14\\
& 8048.3221 & -138.46 $\pm$ 0.14 & -147.44 $\pm$ 0.15\\
& 7911.6058 & -139.92 $\pm$ 0.12 & -146.24 $\pm$ 0.12\\
& 8074.2568 & -139.03 $\pm$ 0.16 & -147.09 $\pm$ 0.17\\
& 8089.2560 & -139.25 $\pm$ 0.45 & -147.44 $\pm$ 0.50\\
& 8093.2439 & -139.10 $\pm$ 0.15 & -146.51 $\pm$ 0.16\\

\hline                                                                           
\end{tabular}
\end{table}

\clearpage
\onecolumn
\section{Radial velocity data fits}
\label{sec:appRVfit}

\begin{figure*}[h]
\centering
\includegraphics[width=85 mm]{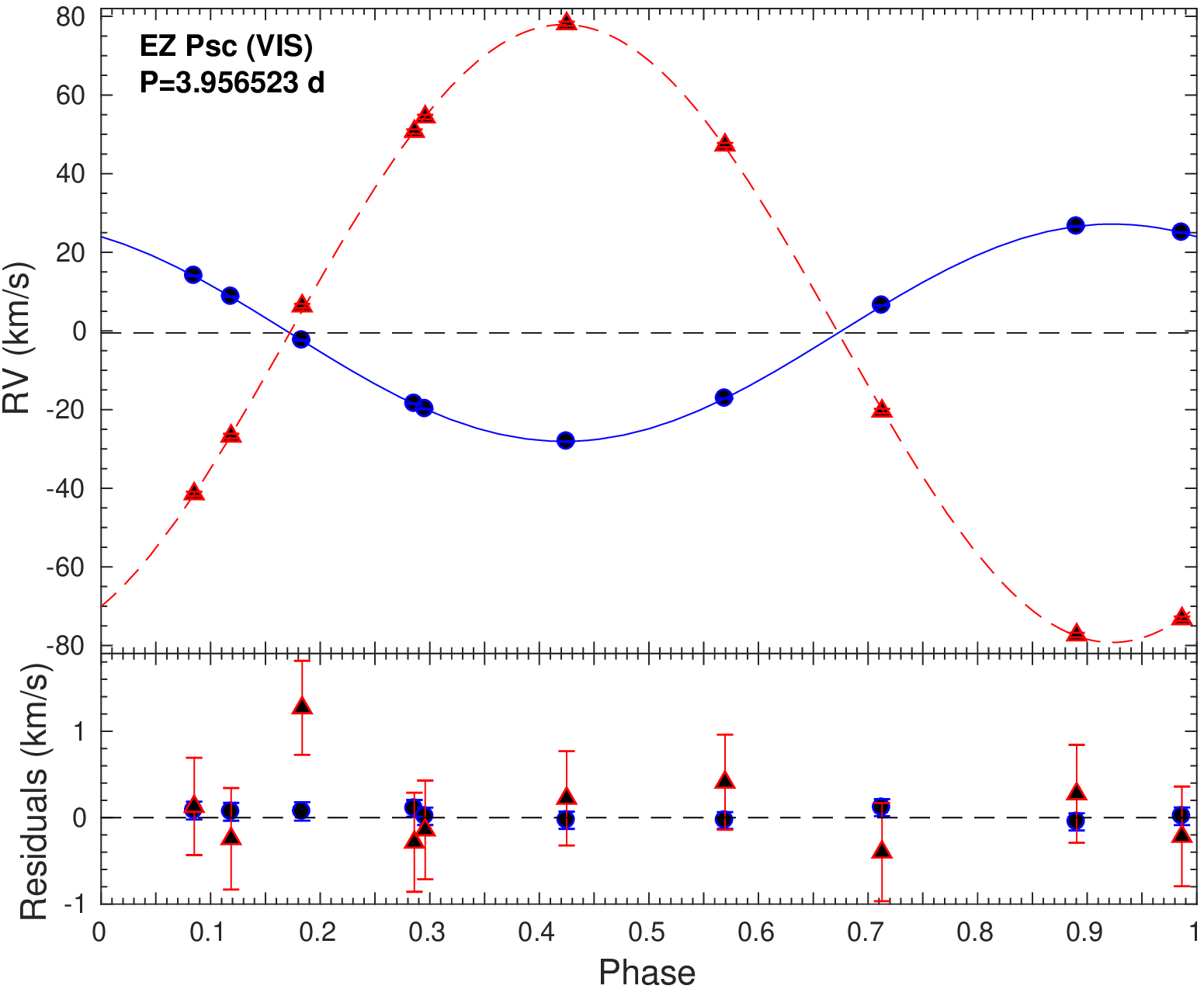}\includegraphics[width=85 mm]{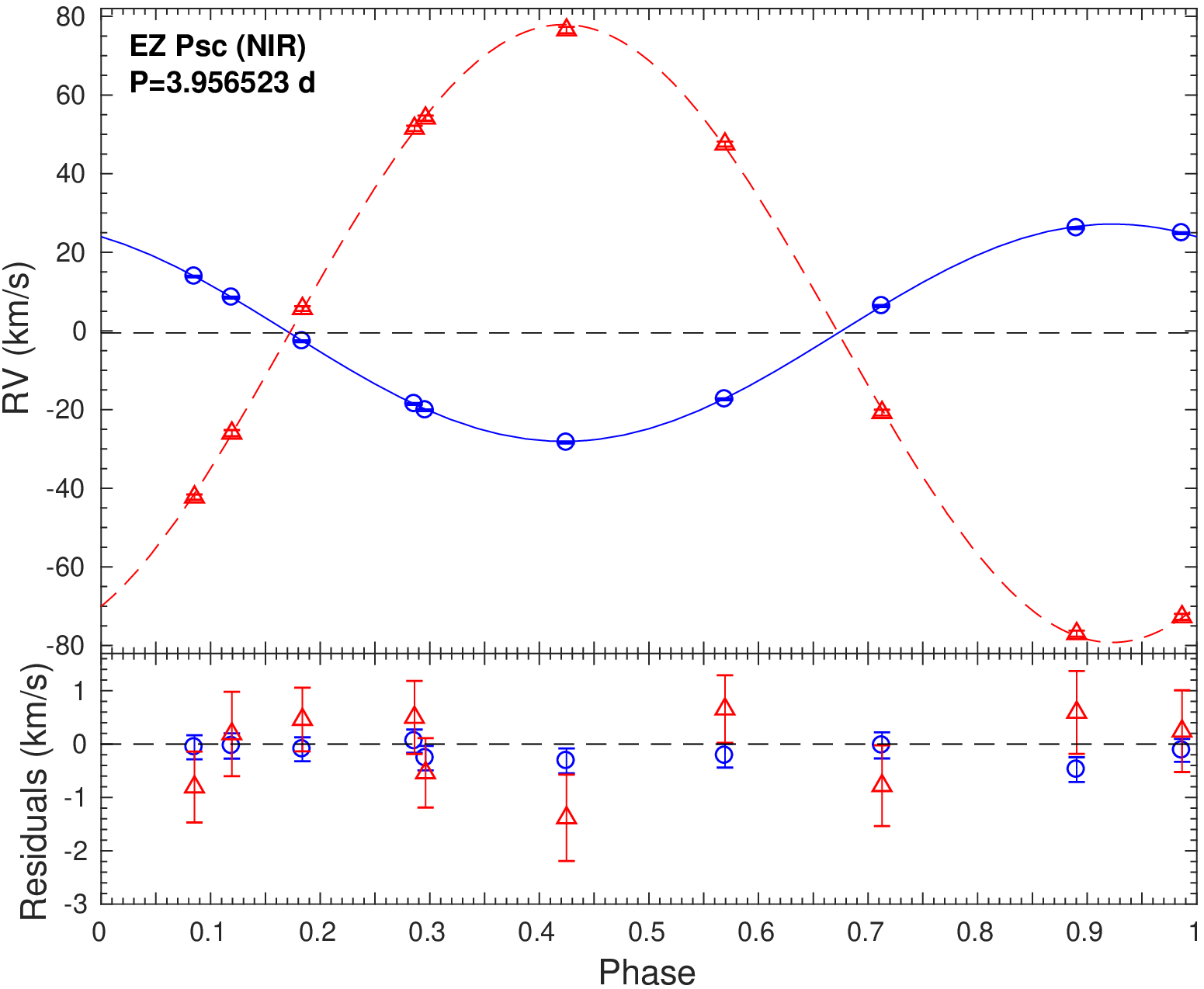}
\includegraphics[width=85 mm]{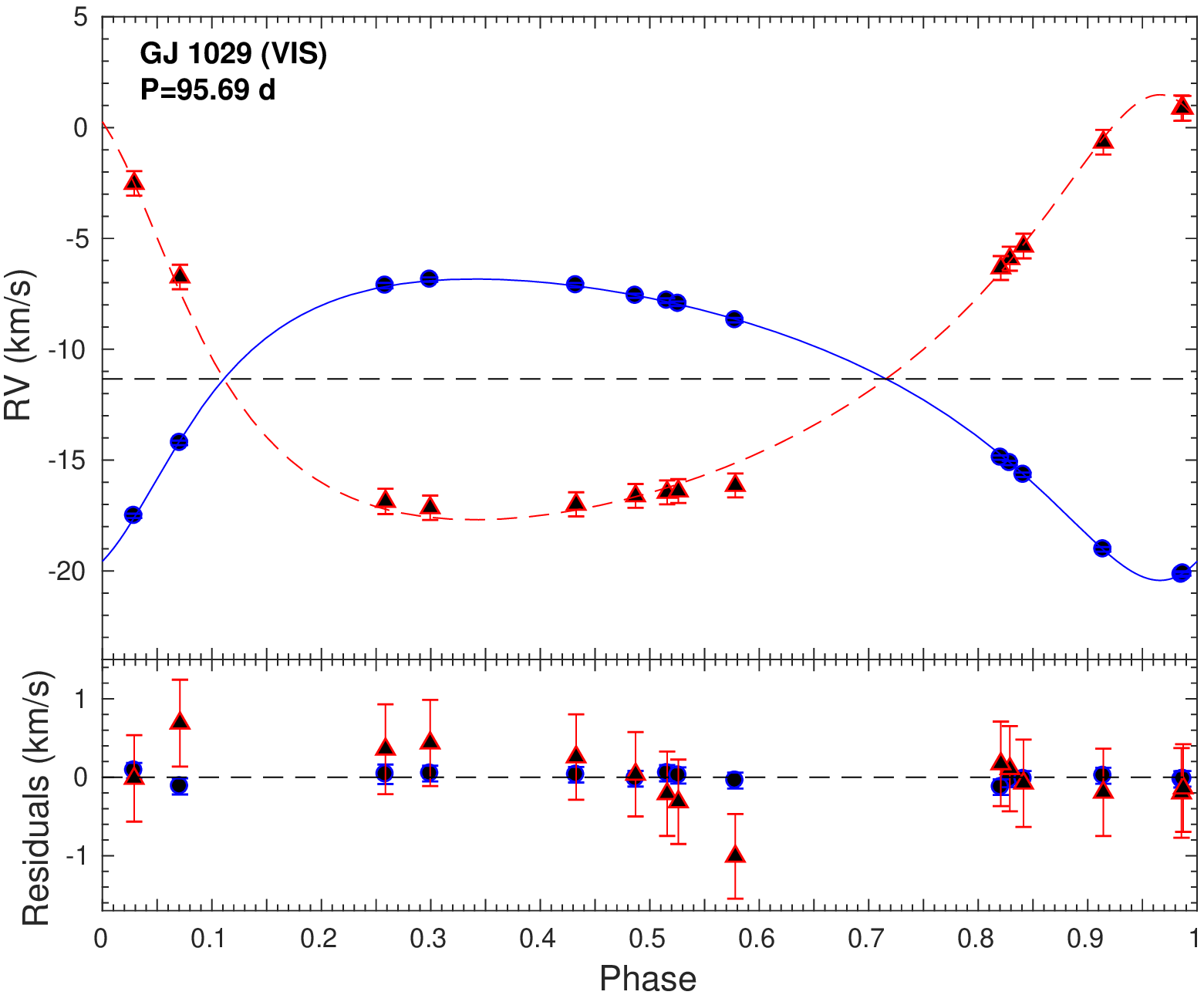}
\includegraphics[width=85 mm]{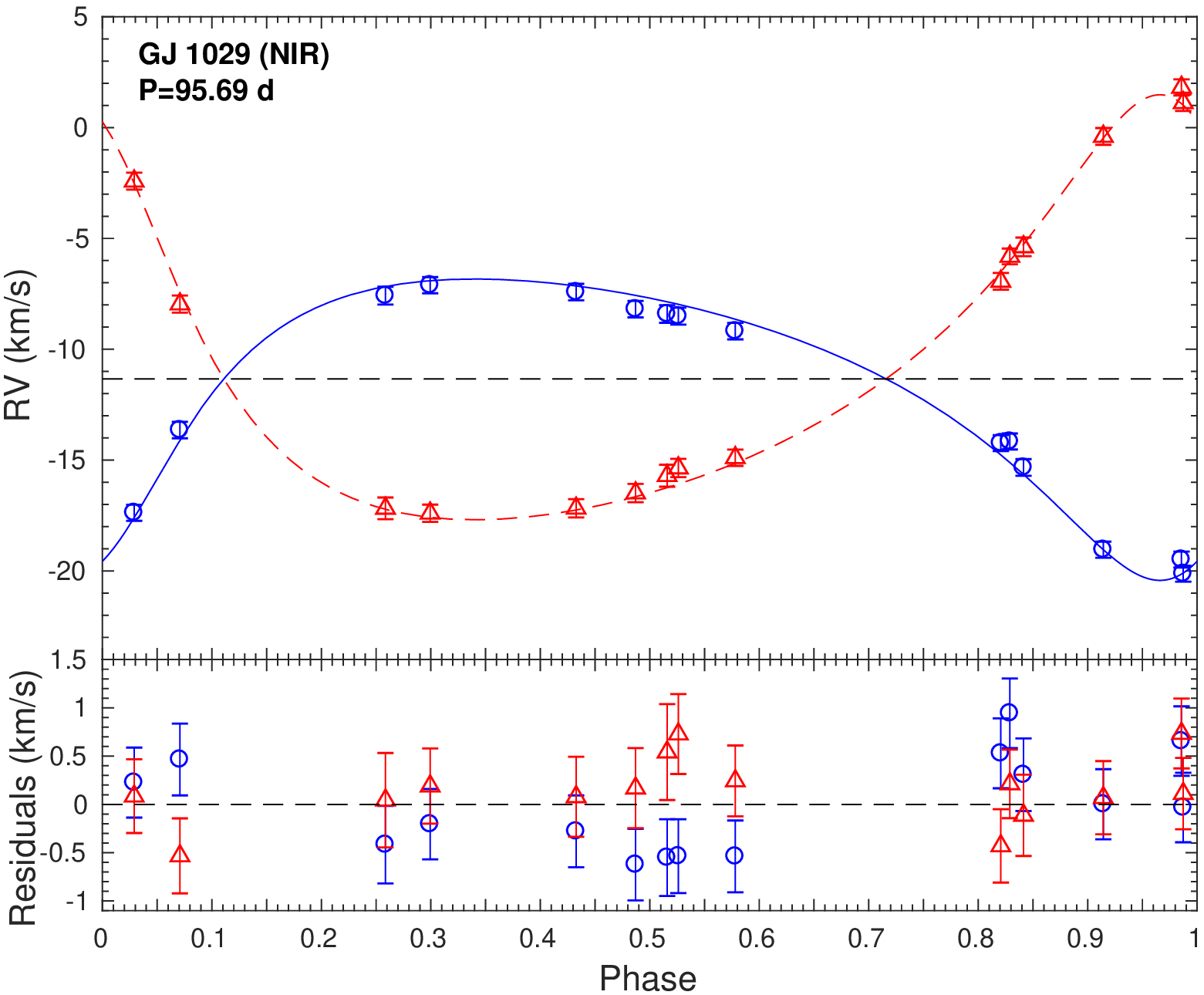}
\includegraphics[width=85 mm]{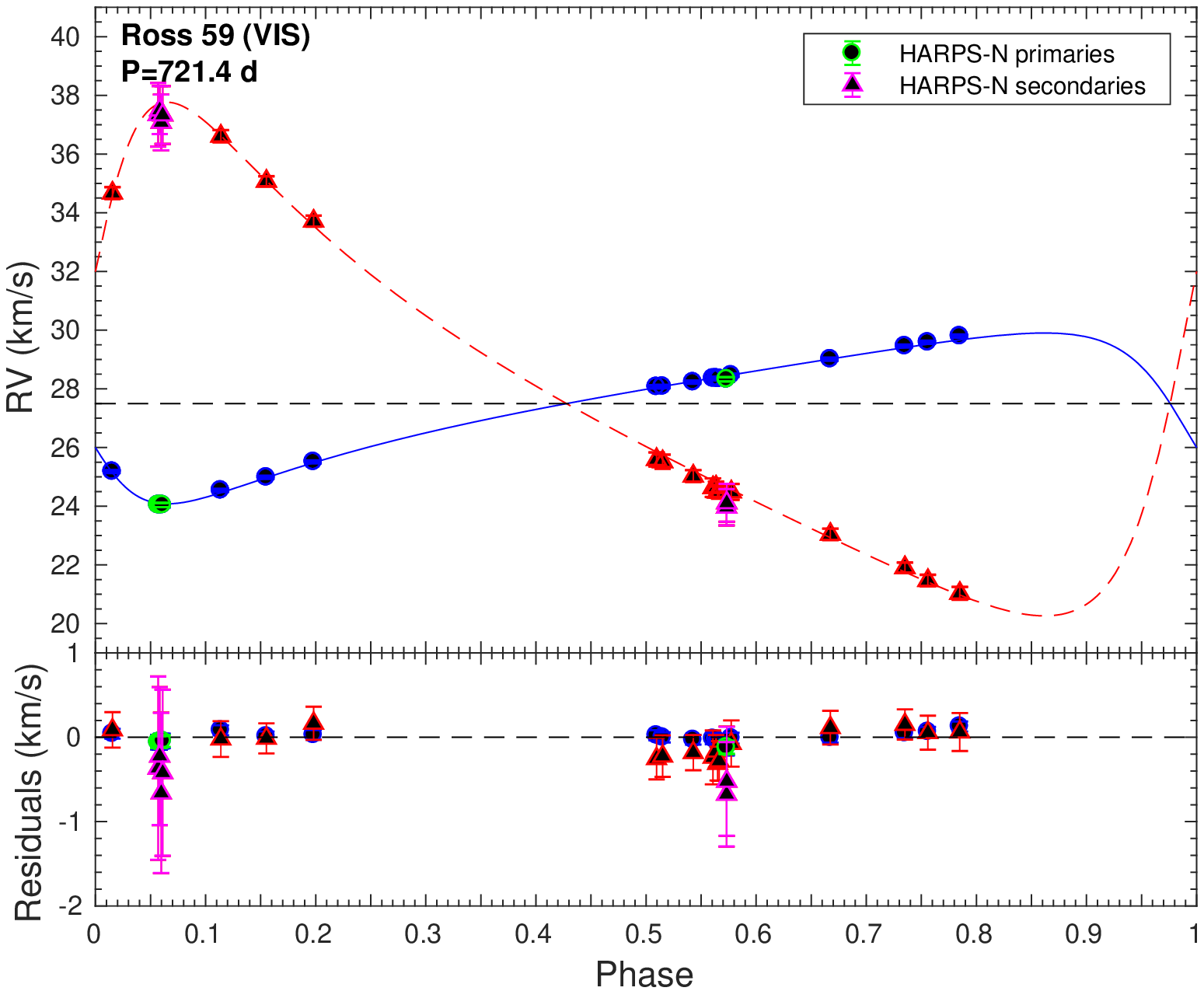}
\includegraphics[width=85 mm]{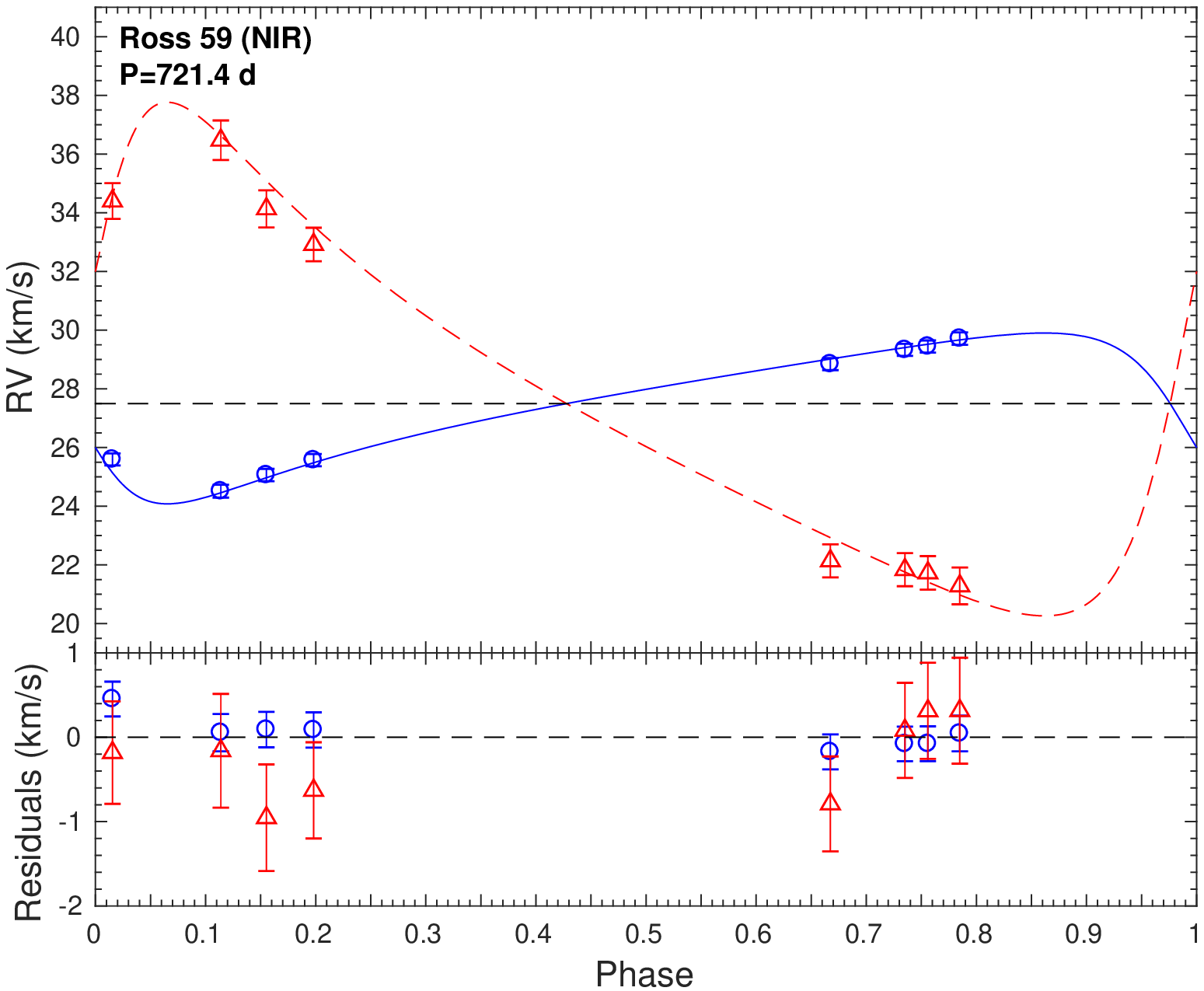}
\caption{Radial velocity curves of our targets as a function of the orbital phase. VIS and NIR CARMENES data are shown in the left and right panels, respectively, for each target as labeled. The top plot in each panel displays the radial velocity data of the primary (blue circle) and secondary (red triangle) components, along with their best-fitting models (blue solid and red dashed lines, respectively). The bottom plot on each panel shows the residuals of the best fit. For Ross 59, HARPS data for the primary (green circles) and secondary (violet triangles) are also shown.}
\label{fig:plot-rv}
\end{figure*}

\addtocounter{figure}{-1}
\begin{figure*}[t]
\centering
\includegraphics[width=85 mm]{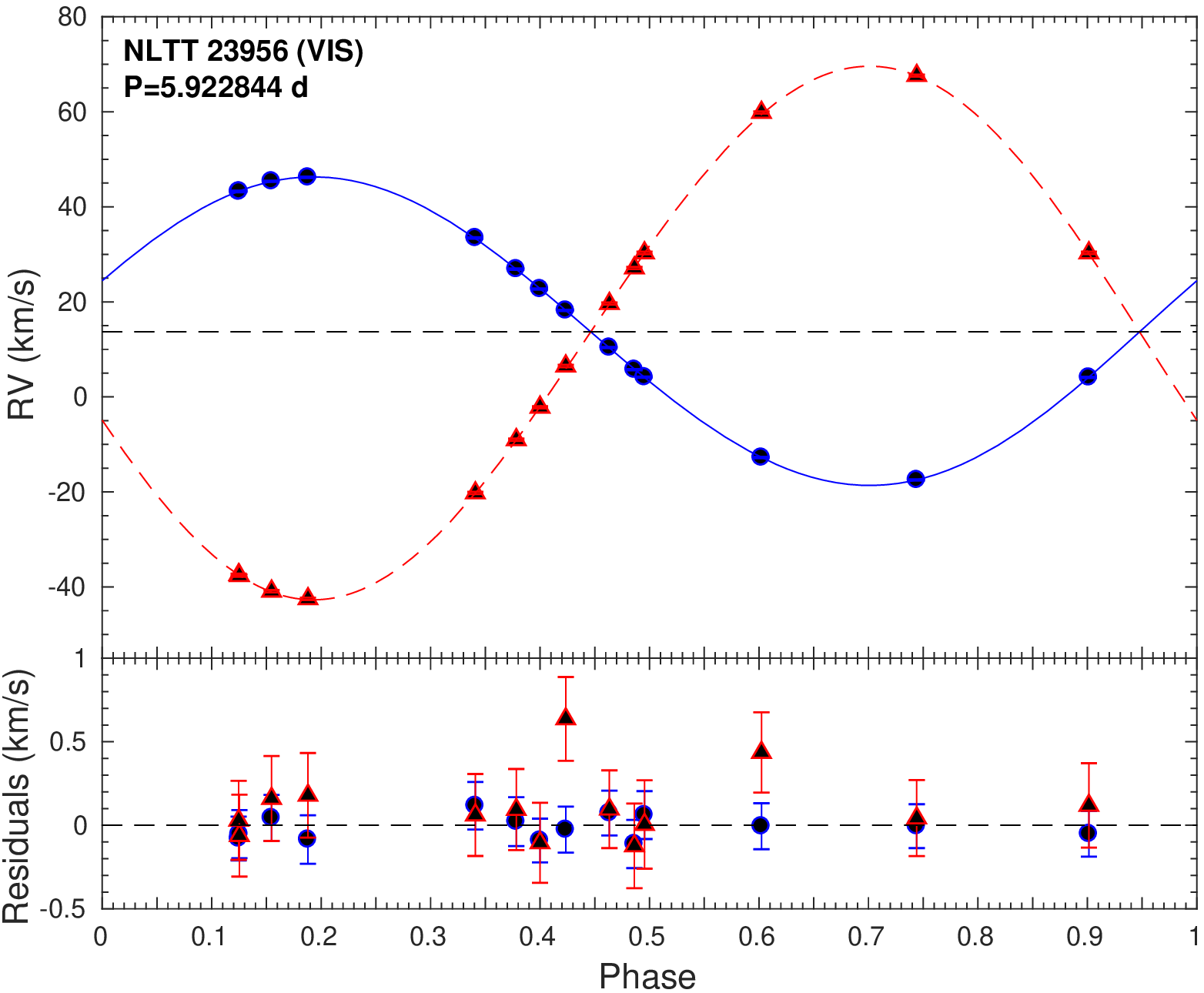}
\includegraphics[width=85 mm]{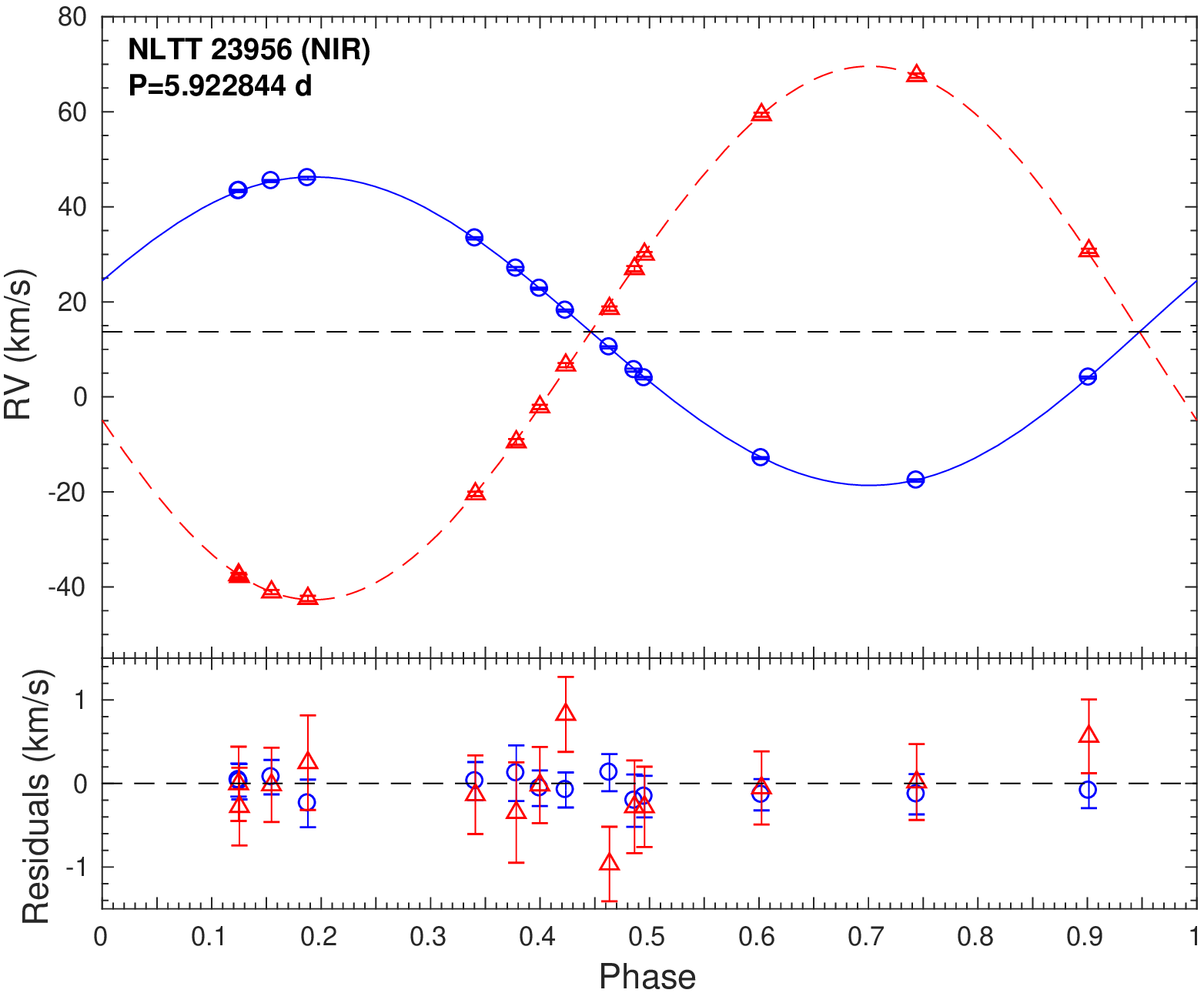}
\includegraphics[width=85 mm]{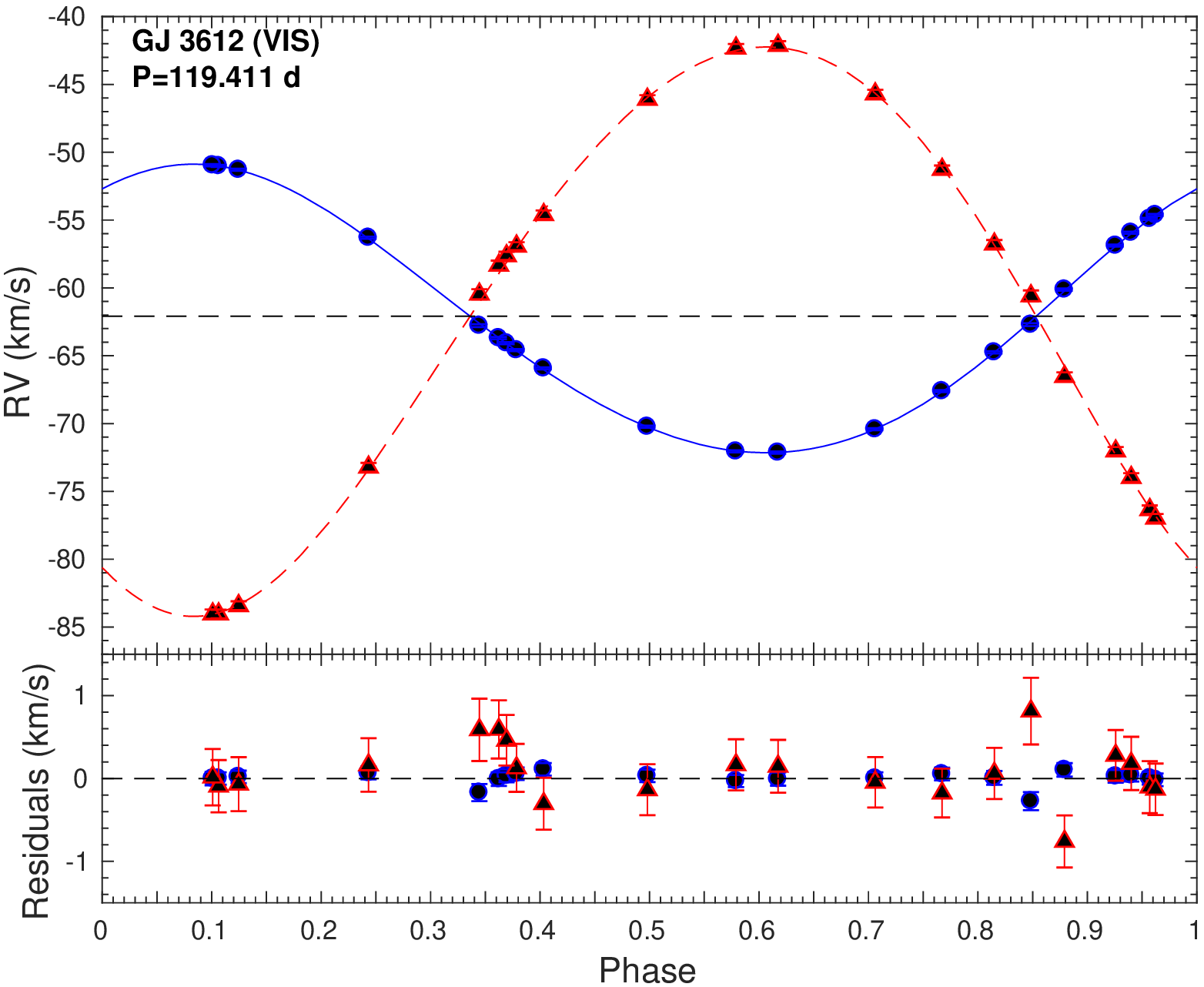}
\includegraphics[width=85 mm]{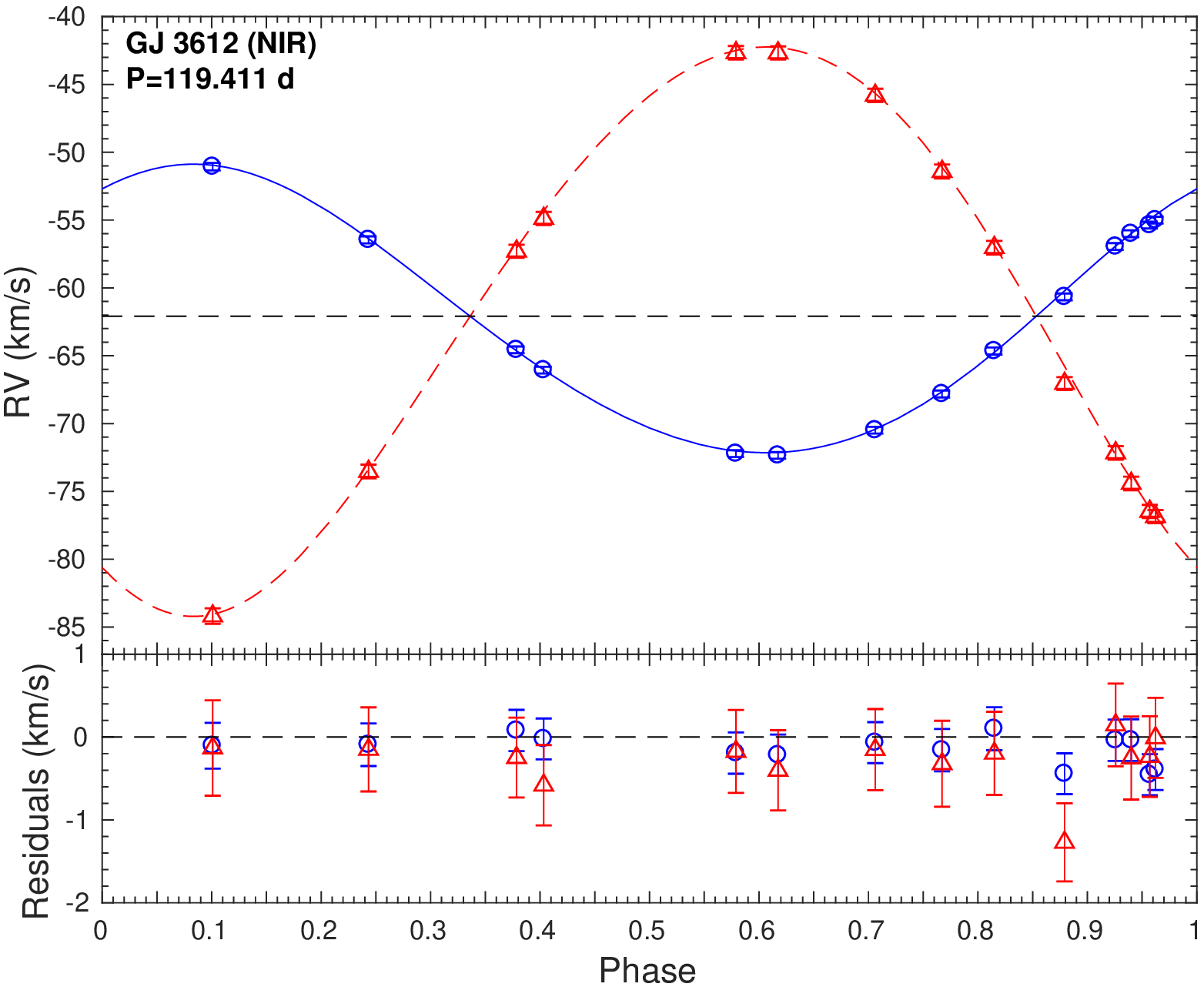}
\includegraphics[width=85 mm]{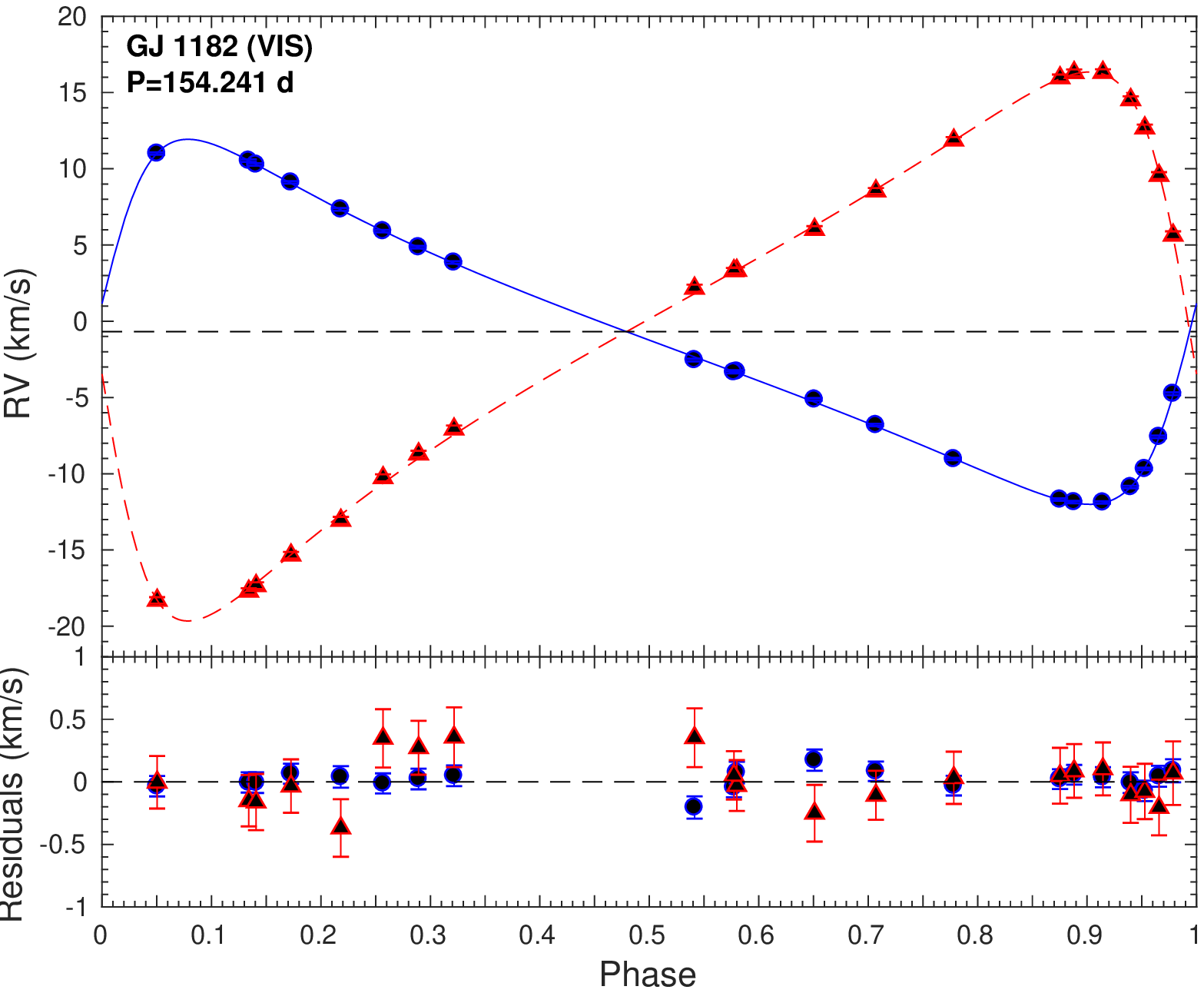}
\includegraphics[width=85 mm]{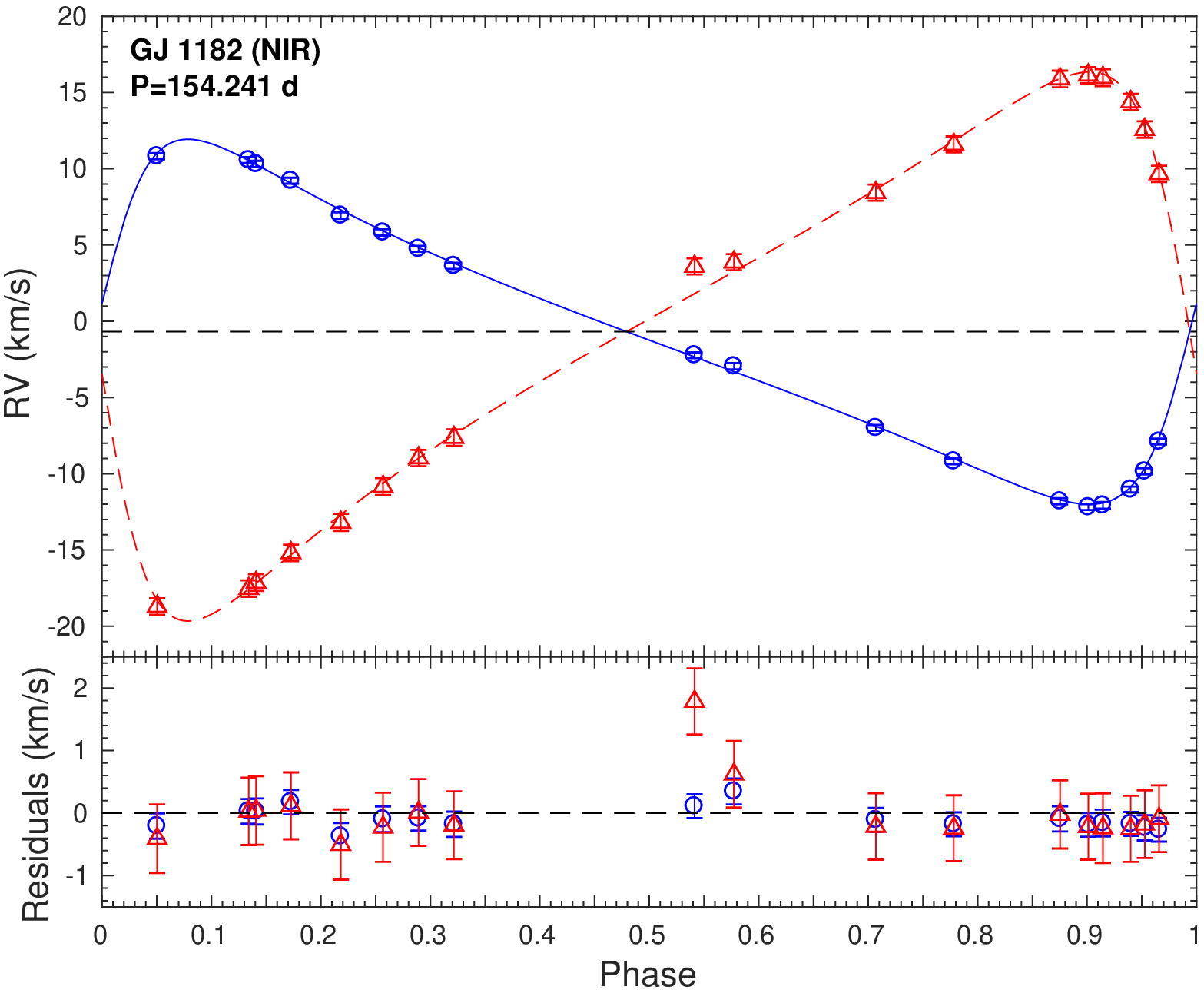}
\caption{Continued.}

\end{figure*}

\addtocounter{figure}{-1}
\begin{figure*}[t]
\centering

\includegraphics[width=85 mm]{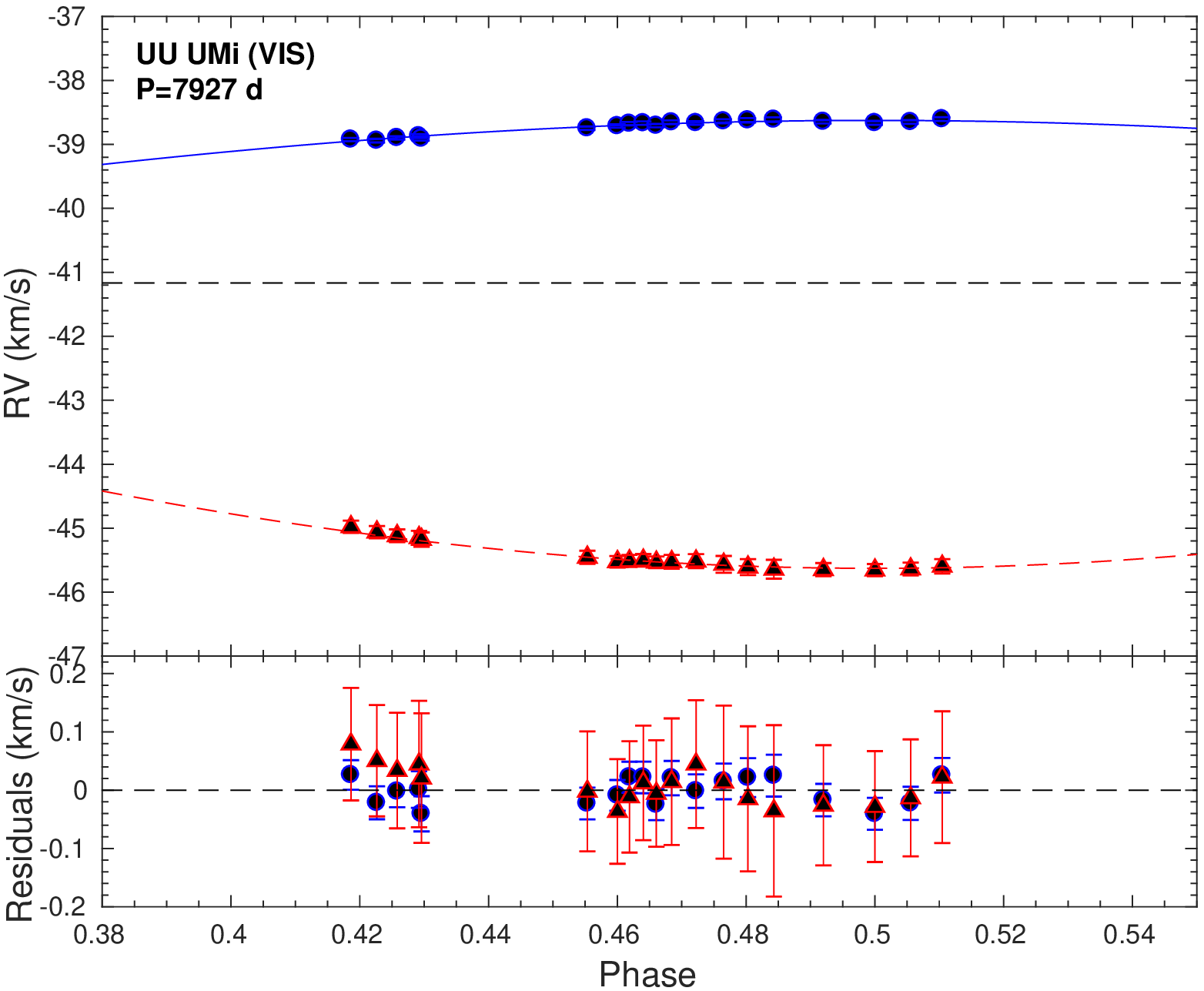}
\includegraphics[width=85 mm]{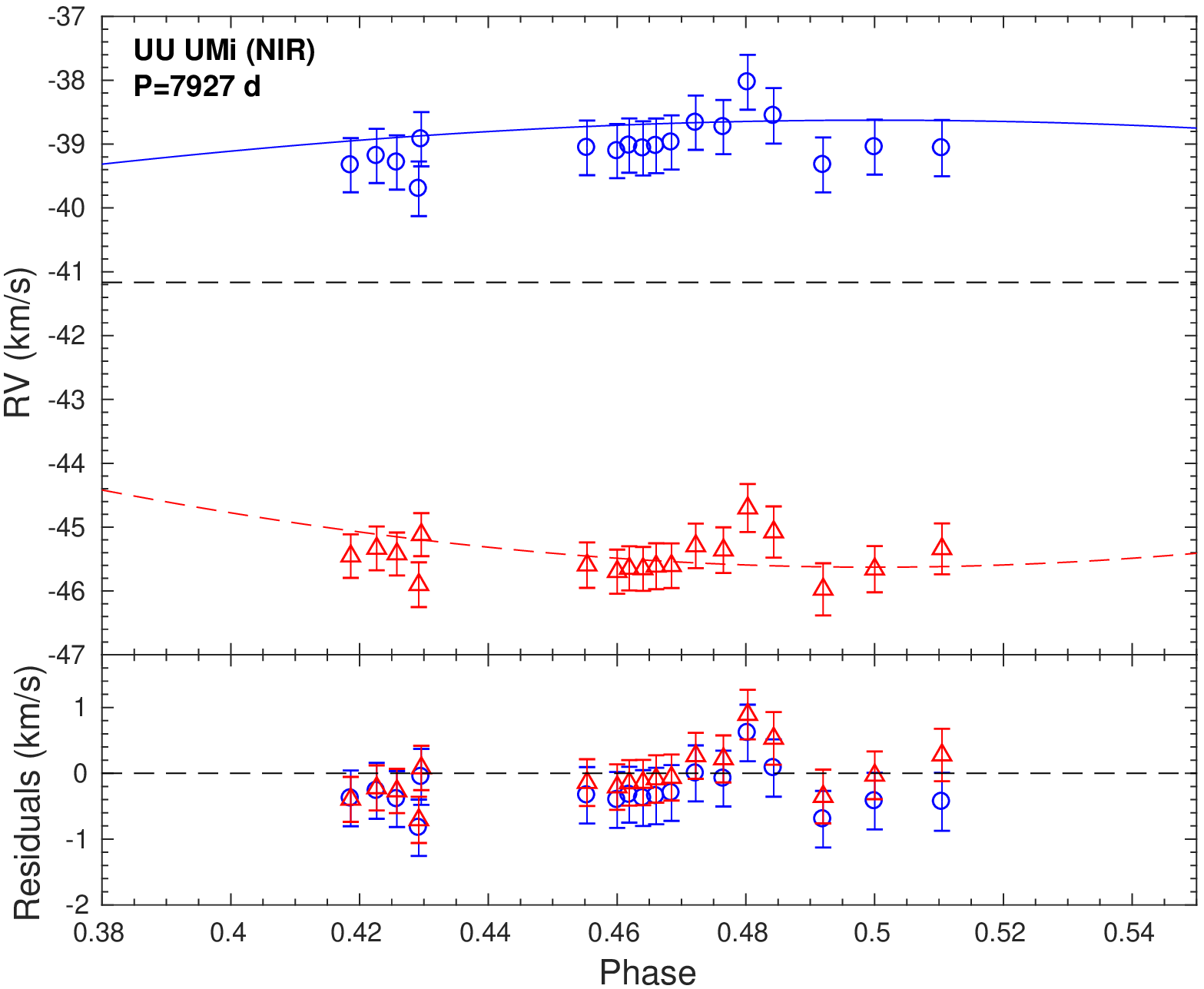}
\includegraphics[width=85 mm]{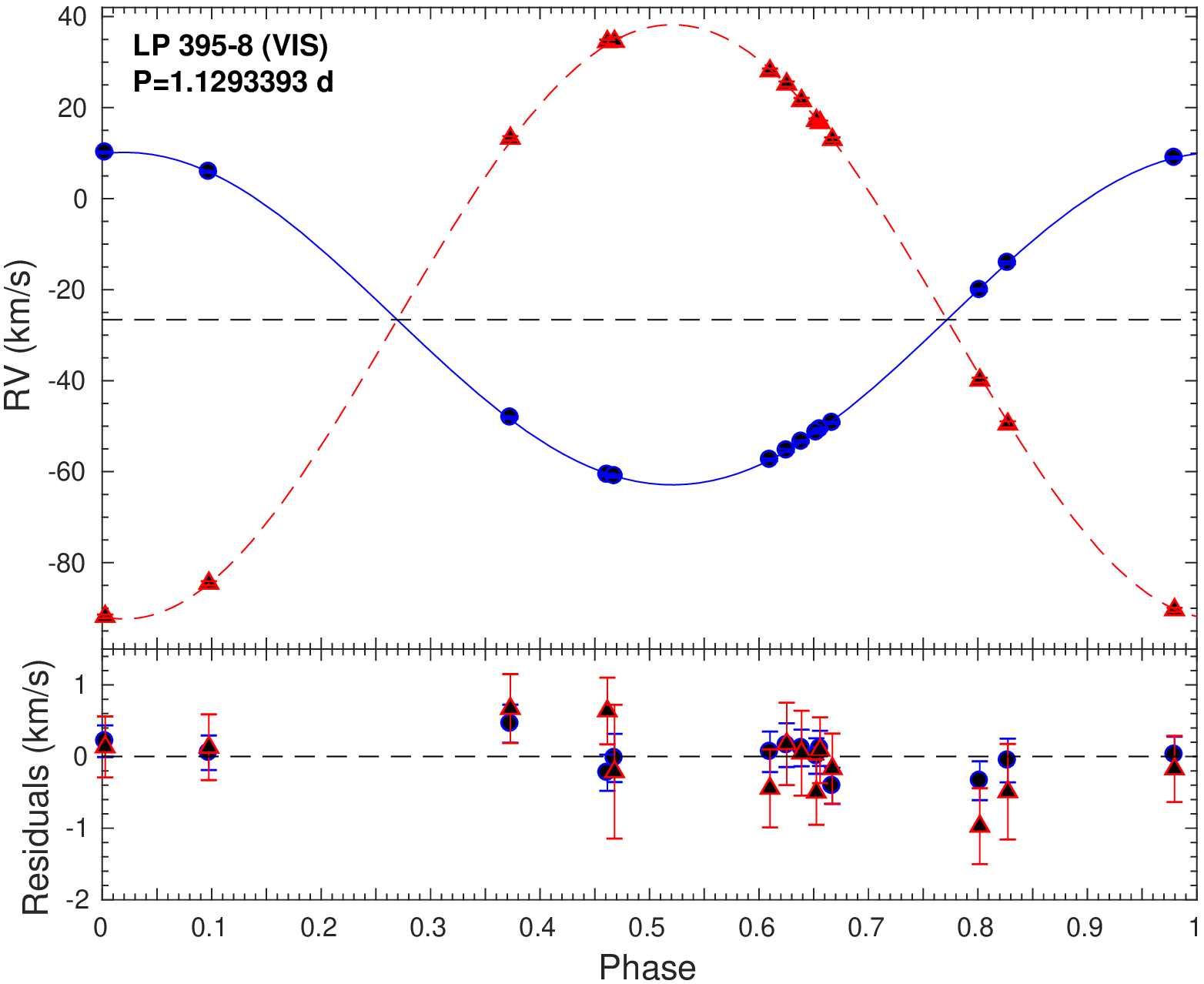}
\includegraphics[width=85 mm]{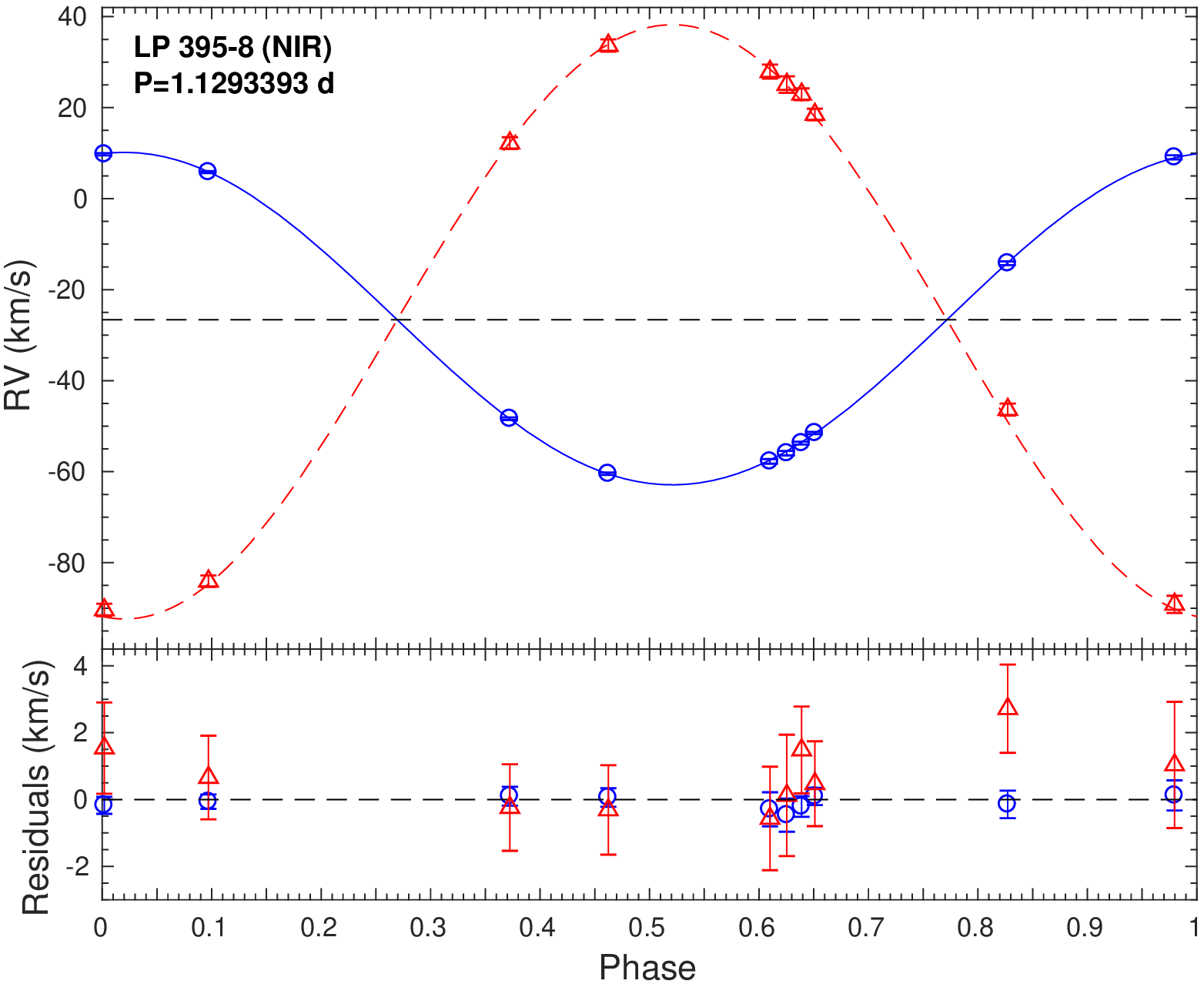}
\includegraphics[width=85 mm]{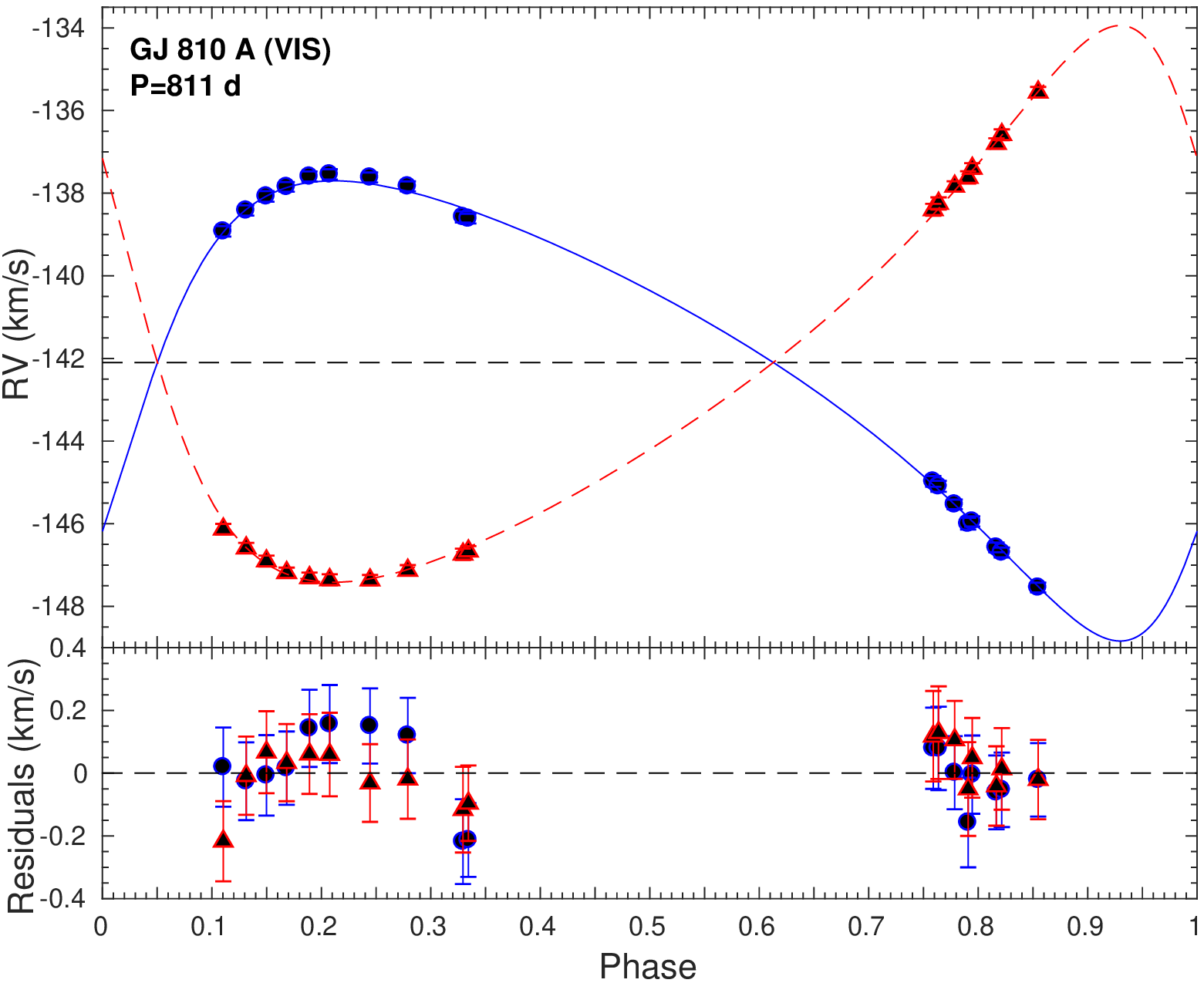}
\includegraphics[width=85 mm]{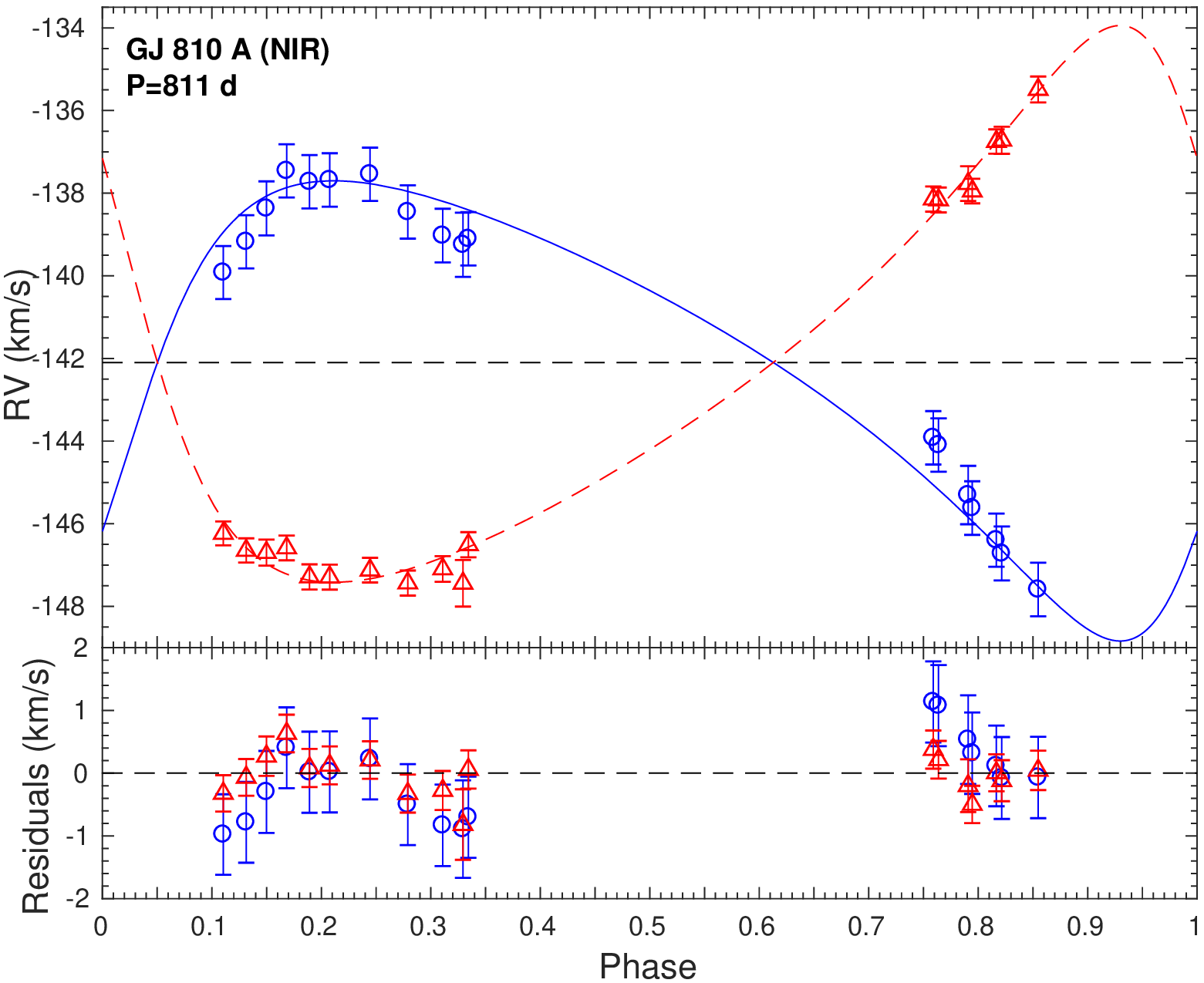}
\caption{Continued.}
\end{figure*}

\clearpage

\section{Photometric data and periodogram analysis}
\label{sec:appPHOT}

\begin{figure*}[h]
\centering
\includegraphics[width=80 mm]{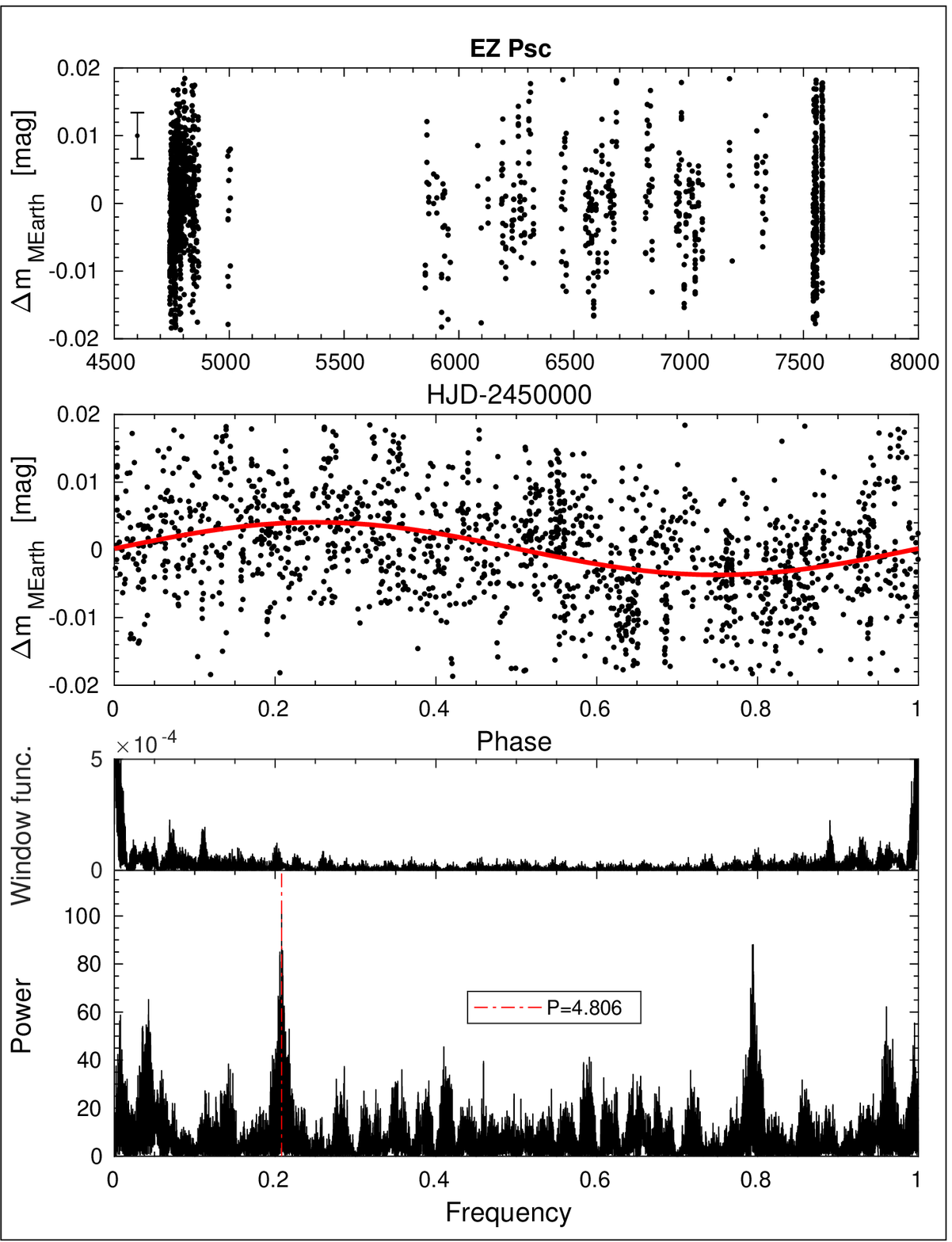}
\includegraphics[width=80 mm]{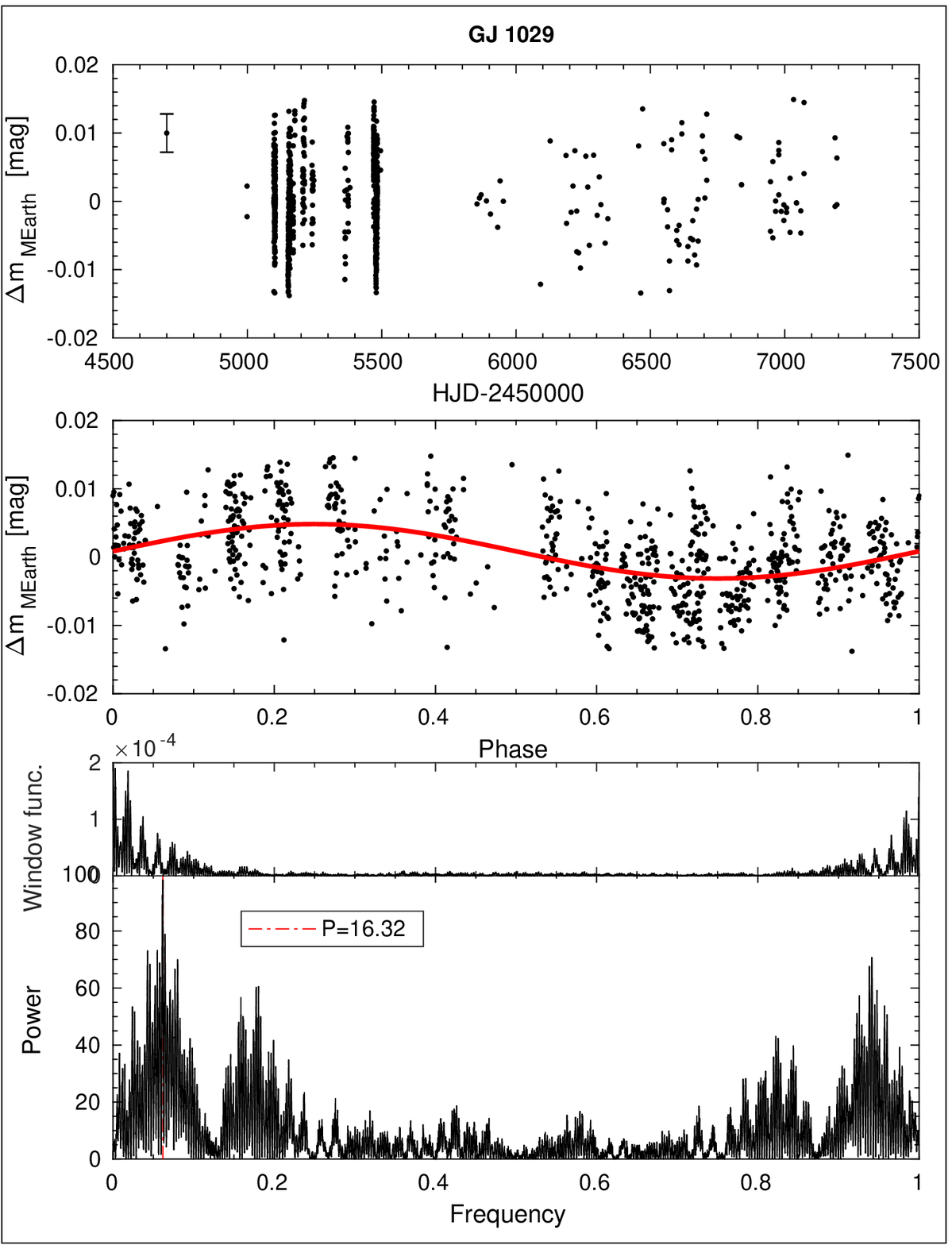}
\includegraphics[width=80 mm]{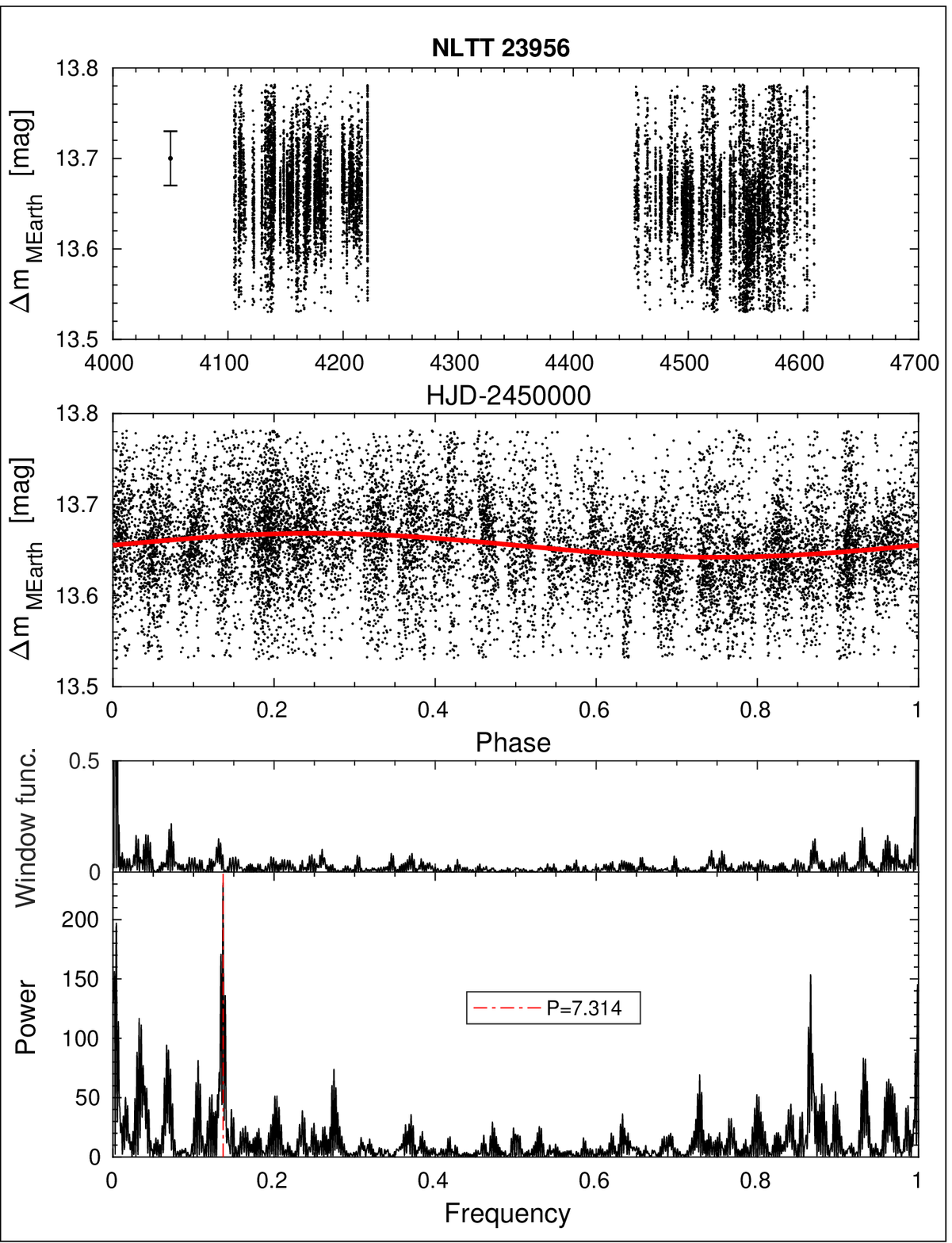}
\includegraphics[width=80 mm]{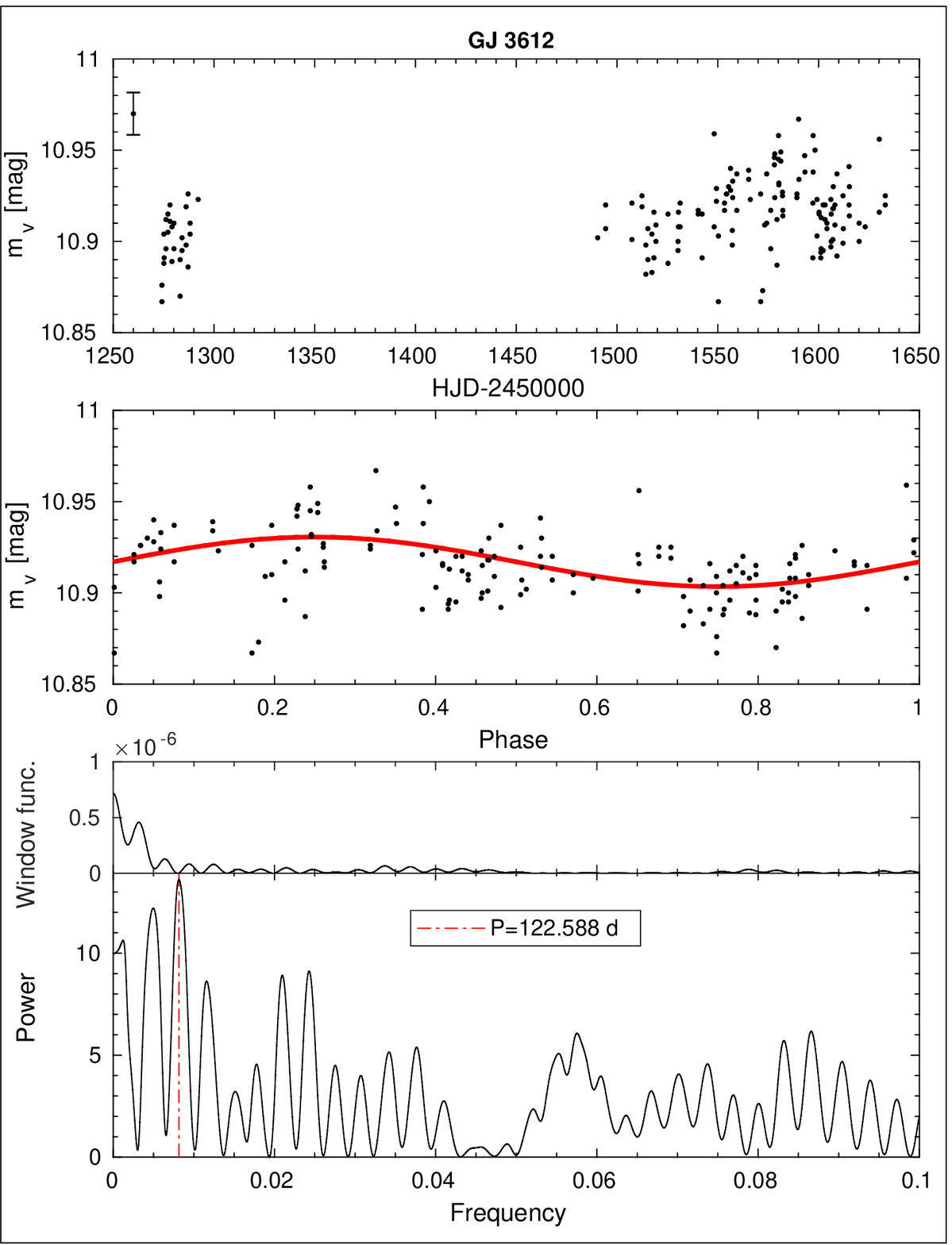}
\caption{Photometry data and analysis for the stellar systems analysed in this work. Each panel corresponds to a binary system as labeled. For each system, the top panel shows the light curve as a function of time and the mean value of the uncertainty of the observations. The middle panel displays the light curve phase to the photometric period found and the best sinusoidal fit. The bottom panel shows the periodogram and window function of the data and the best period found (red dot-dashed line).}
\label{fig:plot-photometry}

\end{figure*}

\addtocounter{figure}{-1}
\begin{figure*}[t]
\centering

\includegraphics[width=85 mm]{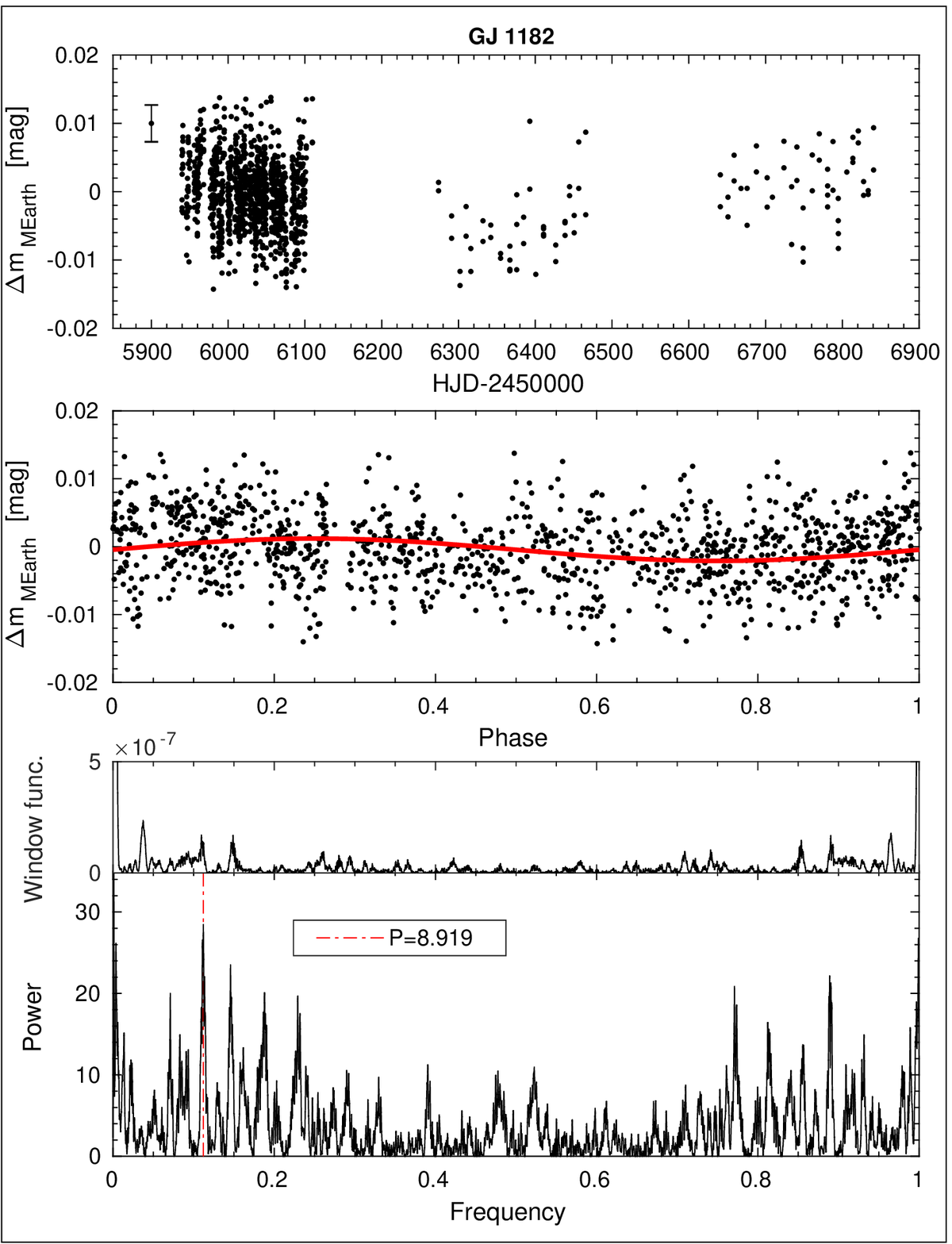}
\includegraphics[width=85 mm]{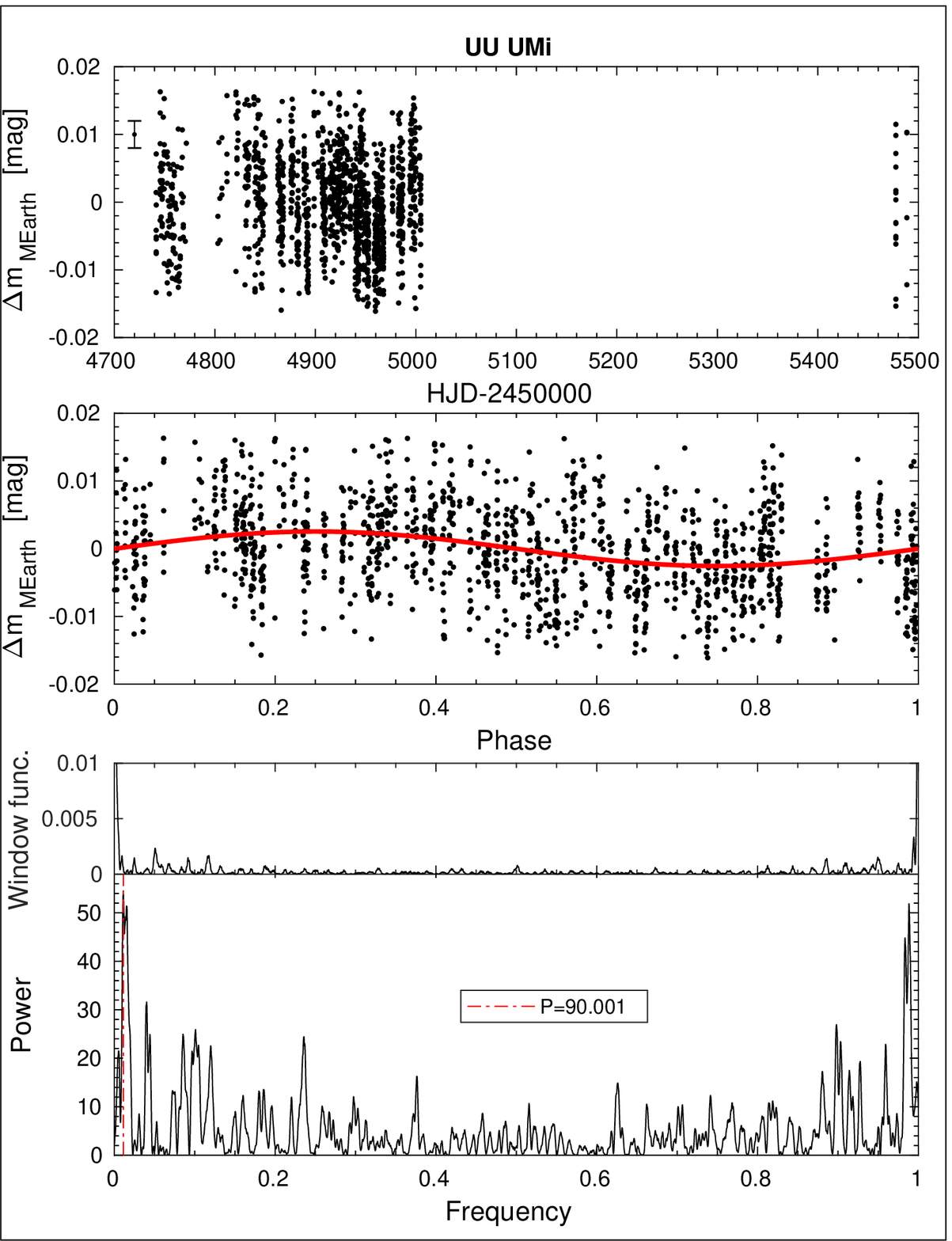}
\includegraphics[width=85 mm]{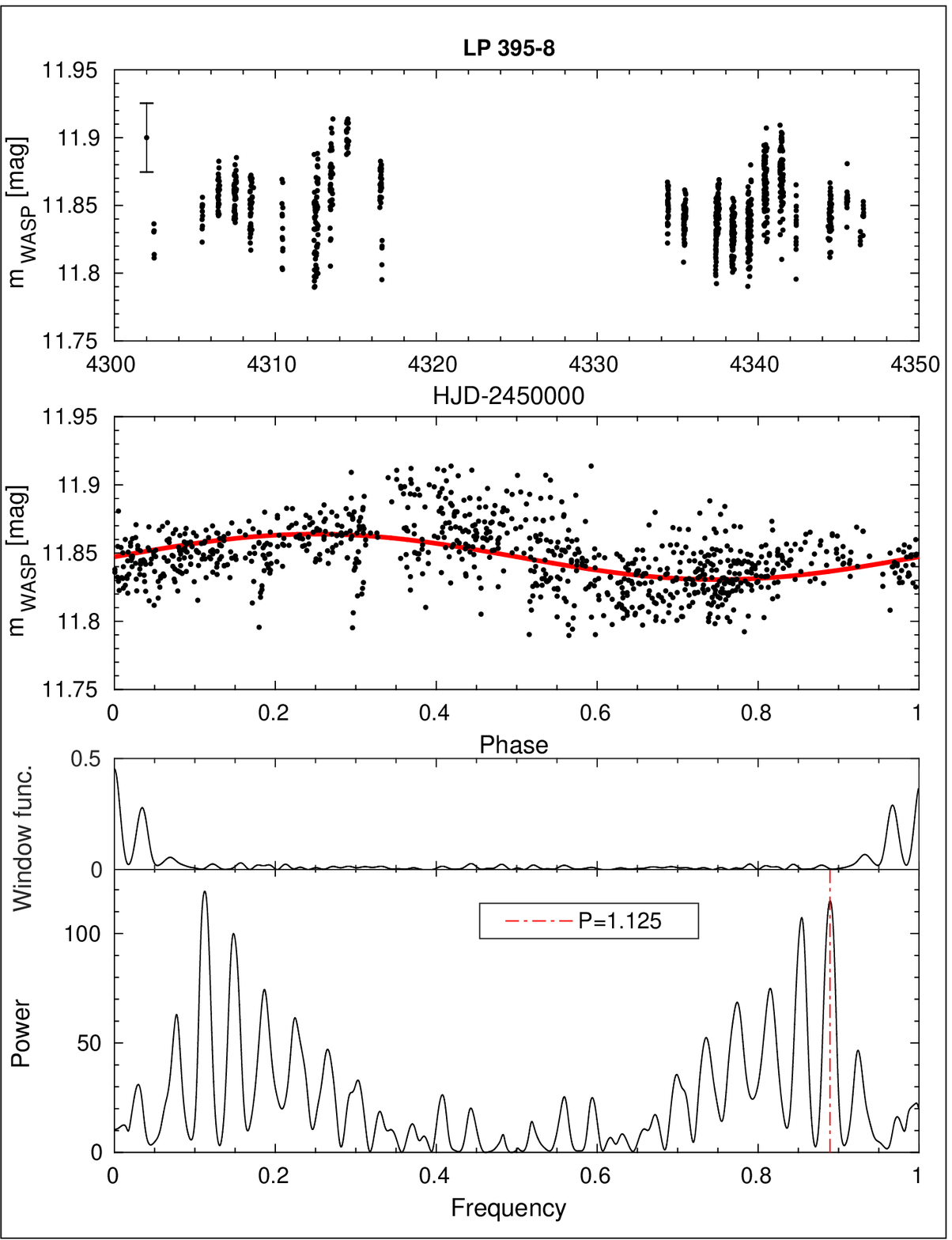}
\caption{Continued.}
\label{fig:plot-photometry-cont1}
\end{figure*}

\clearpage

\section{Known double-line spectroscopic binaries}
\label{sec:appSB2}

\begin{table*}[h]
\centering                                                                \caption{Known M dwarf SB2 systems with published orbital parameters. Systems are sorted by their right ascension. This table will be made available via CDS.}
\label{tab:Mdwarfs}
\newcommand{\abcdresults}[8]{ #1 & #2 & #3 & #4 & #5 & #6 & #7 & #8 \\}
\resizebox{0.95\textwidth}{!}{
\begin{tabular}{lcccccccc}
\hline\hline

\abcdresults{Name&$P_{orb}$}{$T$}{$e$}{$\omega$}{$K_1$}{$K_2$}{$\mathcal{M}_2/\mathcal{M}_2$}{Ref.\tablefootmark{a}}
\abcdresults{&[d]}{[JD-2400000]}{}{[deg]}{[km~s$^{-1}$]}{[km~s$^{-1}$]}{}{}

\hline
\noalign{\smallskip}
\abcdresults{EZ Psc &3.95652$\pm$0.00008}{57709.2$\pm$0.2}{0.0022$\pm$0.0009}{28$\pm$20}{27.64$\pm$0.05}{78.6$\pm$0.2}{0.352$\pm$0.001}{$\star$}
\abcdresults{FF And &2.1703}{42708.359}{$\cdots$}{$\cdots$}{72.1}{74.3}{0.970}{BF77}
\abcdresults{GJ 1029 &95.7$\pm$0.1}{57857.3$\pm$0.4}{0.386$\pm$0.005}{209$\pm$2}{6.79$\pm$0.04}{9.59$\pm$0.10}{0.709$\pm$0.009}{$\star$}
\abcdresults{CD-39 325\tablefootmark{b}&0.4455960 $\pm$ 0.0000002}{51868.8393 $\pm$ 0.0003}{$\cdots$}{$\cdots$}{118 $\pm$ 2}{163 $\pm$ 3}{0.727 $\pm$ 0.019}{Hel12}
\abcdresults{2MASS J02545247-0709255 &11.7951$\pm$ 0.0006}{50509.94 $\pm$ 0.02}{$\cdots$}{$\cdots$}{24.6 $\pm$ 0.3}{43 $\pm$ 2}{0.58 $\pm$ 0.02}{Tor02}
\abcdresults{2MASS J03182386-0100183\tablefootmark{b}&0.407037 $\pm$ 0.000014}{53988.7993 $\pm$ 0.0006}{$\cdots$}{$\cdots$}{108 $\pm$ 5}{122 $\pm$ 4}{0.88 $\pm$ 0.05}{Bla08}
\abcdresults{2MASS J03262072+0312362\tablefootmark{b}&1.5862046 $\pm$ 0.0000008}{5478.9163 $\pm$ 0.0001}{$\cdots$}{$\cdots$}{88.4 $\pm$ 0.2}{94.9 $\pm$ 0.2}{0.931 $\pm$ 0.002}{Kra11}
\abcdresults{GJ 3236\tablefootmark{b}&0.771260 $\pm$ 0.000002}{54734.9950 $\pm$ 0.0001}{$\cdots$}{$\cdots$}{86 $\pm$ 2}{115 $\pm$ 2}{0.746 $\pm$ 0.022}{Irw09}
\abcdresults{BD+03 515&31.16 $\pm$ 0.02}{47778.1 $\pm$ 0.4}{0.39 $\pm$ 0.04}{182 $\pm$ 5}{19 $\pm$ 1}{22 $\pm$ 2}{0.87 $\pm$ 0.08}{Tok91}
\abcdresults{2MASS J04463285+1901432\tablefootmark{b}&0.618790$\pm$ 0.000005}{52530.2622 $\pm$ 0.0004}{$\cdots$}{$\cdots$}{63 $\pm$ 1}{152 $\pm$ 7}{0.41 $\pm$ 0.02}{Heb06}
\abcdresults{UCAC3 56-8629\tablefootmark{b}& 1.1112861$\pm$0.0000004}{55255.9633 $\pm$ 0.0007}{0.03$\pm$ 0.01}{271.2 $\pm$0.2}{91 $\pm$ 5}{103 $\pm$ 5}{0.88 $\pm$ 0.06}{Lub17}
\abcdresults{DQ Tau&15.810 $\pm$ 0.006}{49582.78 $\pm$ 0.23}{0.58 $\pm$ 0.07}{228 $\pm$ 5}{22 $\pm$ 2}{22 $\pm$ 2}{0.96 $\pm$ 0.14}{Mat97}
\abcdresults{LP 476-207&11.9623 $\pm$ 0.0005}{49799.47 $\pm$ 0.04}{0.323 $\pm$ 0.006}{212.0 $\pm$ 0.6}{9.96 $\pm$ 0.03}{17.57 $\pm$ 0.03}{0.567 $\pm$ 0.002}{Del99}
\abcdresults{V1236 Tau\tablefootmark{b}&2.58791$\pm$0.00001}{52251.512$\pm$0.005}{$\cdots$}{$\cdots$}{88.4$\pm$0.5}{95.5$\pm$0.6}{0.98 $\pm$ 0.02}{BO06}
\abcdresults{V2212 Ori\tablefootmark{b}&4.67390 $\pm$ 0.00006}{54849.9008 $\pm$ 0.0005}{0.017 $\pm$ 0.003}{1.48 $\pm$ 0.01}{57 $\pm$ 2}{58 $\pm$ 3}{0.98 $\pm$ 0.06}{Gom12}
\abcdresults{2MASS J05445791-2456095\tablefootmark{b}&4.077017$\pm$0.000001}{56374.0171 $\pm$ 0.0001}{0.002 $\pm$ 0.002}{274 $\pm$ 42}{42 $\pm$ 2}{58 $\pm$ 4}{0.72 $\pm$ 0.06}{Zho15}
\abcdresults{Ross 59&721$\pm$2}{58007 $\pm$ 2}{0.509 $\pm$ 0.009}{109 $\pm$ 2}{2.91 $\pm$ 0.04}{8.8 $\pm$ 0.1}{0.333 $\pm$ 0.007}{$\star$}
\abcdresults{LHS 6100&2.63}{56261.747$\pm$0.006}{0.010$\pm$0.003}{128.4$\pm$35}{32.29$\pm$0.14}{38.44$\pm$0.32}{0.840$\pm$0.006}{Ski18}
\abcdresults{QY Aur&10.428}{45770.8}{0.34}{217.5}{33.1}{39.8}{0.831}{TP86}
\abcdresults{G 109-55 &304.4 $\pm$ 0.3}{48826 $\pm$ 2}{0.400$\pm$ 0.008}{273.8 $\pm$ 0.9}{12.47 $\pm$ 0.08}{18.6 $\pm$ 0.2}{0.670 $\pm$ 0.008}{Del99}
\abcdresults{YY Gem\tablefootmark{b}&0.81428}{49345.112}{$\cdots$}{$\cdots$}{121.2 $\pm$ 0.4}{120.5 $\pm$ 0.4}{0.994 $\pm$ 0.005}{TR02}
\abcdresults{Ross 775\tablefootmark{b}&1.5484492 $\pm$ 0.0000006}{5473.73166 $\pm$ 0.00003}{$\cdots$}{$\cdots$}{92.3 $\pm$ 0.2}{99.2 $\pm$ 0.2}{0.931 $\pm$ 0.003}{Kra11}
\abcdresults{CU Cnc\tablefootmark{b}& 2.771472$\pm$0.000004 }{50207.8128 $\pm$ 0.0009}{$\cdots$}{$\cdots$}{68.03 $\pm$ 0.09}{73.06 $\pm$ 0.09}{0.931 $\pm$ 0.002}{Del99}
\abcdresults{2MASS J08503296+1208239\tablefootmark{b}& 3.3439539$\pm$0.0000002 }{57214.94178 $\pm$ 0.00003}{ 0.014$\pm$0.03}{269.80$\pm$0.03}{48.2 $\pm$ 0.2}{79.4 $\pm$ 0.9}{0.607 $\pm$ 0.007}{Har18}

\abcdresults{2MASS J08504984+1948364\tablefootmark{b}& 6.015742$\pm$ 0.000002}{57148.9041 $\pm$ 0.0001}{0.0017$\pm$0.0006}{38$\pm$17}{34.3 $\pm$ 0.2}{64.7 $\pm$ 0.7}{0.531 $\pm$ 0.005}{Kra17}

\abcdresults{G 41-14&7.555 $\pm$ 0.002}{50471.2 $\pm$ 0.2}{0.014 $\pm$ 0.002}{7 $\pm$ 9}{30.15 $\pm$ 0.05}{36.79 $\pm$ 0.09}{0.820 $\pm$ 0.002}{Del99}
\abcdresults{BD-02 3000&47.709 $\pm$ 0.053}{49345.44 $\pm$ 0.52}{0.53 $\pm$ 0.02}{81 $\pm$ 2}{30.2 $\pm$ 0.6}{40 $\pm$ 2}{0.76 $\pm$ 0.03}{Har96}
\abcdresults{NLTT 23956&5.92285 $\pm$ 0.00006}{58007.4 $\pm$ 0.1}{0.014 $\pm$ 0.001}{289 $\pm$ 6}{32.47 $\pm$ 0.06}{56.15 $\pm$ 0.09}{0.578 $\pm$ 0.001}{$\star$}
\abcdresults{2MASS J10305521+0334265\tablefootmark{b}&1.637530 $\pm$ 0.000002}{54547.83444 $\pm$ 0.00008}{$\cdots$}{$\cdots$}{83.3 $\pm$ 0.2}{93.6 $\pm$ 0.2}{0.89 $\pm$ 0.003}{Kra11}
\abcdresults{GJ 3612& 119.41$\pm$ 0.04}{57718.4 $\pm$ 0.6}{0.066 $\pm$ 0.002}{326 $\pm$ 2}{10.64 $\pm$ 0.03}{21.0 $\pm$ 0.1}{0.507 $\pm$ 0.003}{$\star$}
\abcdresults{LSPM J1112+7626\tablefootmark{b} &41.03237 $\pm$ 0.00002}{55290.0462 $\pm$ 0.0002}{0.238 $\pm$ 0.006}{50.1 $\pm$ 0.2}{22.81 $\pm$ 0.06}{32.79 $\pm$ 0.06}{0.696 $\pm$ 0.002}{Irw11}
\abcdresults{DP Dra&54.075 $\pm$ 0.006}{50506.2 $\pm$ 0.5}{0.081 $\pm$ 0.005}{137 $\pm$ 3}{22.0 $\pm$ 0.5}{22.5 $\pm$ 0.5}{0.98 $\pm$ 0.03}{Del99}
\abcdresults{BD+35 2436&200.26 $\pm$ 0.09}{48780.3 $\pm$ 0.4}{0.53 $\pm$ 0.01}{349 $\pm$ 2}{17.1 $\pm$ 0.2}{20.7 $\pm$ 0.4}{0.83 $\pm$ 0.02}{Tok97}
\abcdresults{UCAC4 847-011196\tablefootmark{b}&0.3681414 $\pm$ 0.0000003}{53473.98266 $\pm$ 0.00002}{$\cdots$}{$\cdots$}{143.9 $\pm$ 0.4}{156.1 $\pm$ 0.9}{0.922 $\pm$ 0.006}{Lop06}
\abcdresults{GJ 1182&154.2 $\pm$ 0.1}{57867.82 $\pm$ 0.07}{0.537 $\pm$ 0.002}{275.8 $\pm$ 0.3}{11.97 $\pm$ 0.03}{18.00 $\pm$ 0.07}{0.665 $\pm$ 0.003}{$\star$}
\abcdresults{HD 131976&308.884 $\pm$ 0.004}{50270.22 $\pm$ 0.01}{0.7559 $\pm$ 0.0002}{127.56 $\pm$ 0.05}{18.19 $\pm$ 0.01}{27.32 $\pm$ 0.03}{0.6656 $\pm$ 0.0007}{For99}
\abcdresults{GU Boo\tablefootmark{b}&0.488728 $\pm$ 0.000002}{52723.9811 $\pm$ 0.0003}{$\cdots$}{$\cdots$}{142.7 $\pm$ 0.7}{145.1 $\pm$ 0.7}{0.983 $\pm$ 0.007}{Lop05}
\abcdresults{UU UMi&7927 $\pm$ 700}{58118 $\pm$ 50}{$\cdots$}{$\cdots$}{2.5 $\pm$ 0.2}{4.5 $\pm$ 0.2}{0.57 $\pm$ 0.06}{$\star$}
\abcdresults{G 179-55\tablefootmark{b}&3.550018 $\pm$ 0.000002}{51232.8953 $\pm$ 0.0009}{$\cdots$}{$\cdots$}{56.0 $\pm$ 0.8}{55.8 $\pm$ 0.8}{1.00 $\pm$ 0.02}{Har11}
\abcdresults{BD+11 2874&1015$\pm$ 3}{50828$\pm$ 11}{0.37 $\pm$ 0.02}{340 $\pm$ 5}{6.6 $\pm$ 0.3}{7.0 $\pm$ 0.3}{0.94 $\pm$ 0.05}{Tok00}
\abcdresults{2MASS J15595050-1944373\tablefootmark{b}&34.00070 $\pm$ 0.00009}{56909.2511 $\pm$ 0.0009}{0.2673 $\pm$ 0.0002}{175.9 $\pm$ 0.7}{28.96 $\pm$ 0.09}{30.19 $\pm$ 0.09}{0.959 $\pm$ 0.004}{Dav16}
\abcdresults{UGCS J161630.67-251220.2\tablefootmark{b}&2.80885 $\pm$ 0.00002}{56894.7139 $\pm$ 0.0005}{0.016 $\pm$ 0.009}{259 $\pm$ 9}{43.4 $\pm$ 0.6}{47.8 $\pm$ 0.4}{0.91 $\pm$ 0.01}{Dav16}
\abcdresults{2MASS J16502074+4639013\tablefootmark{b}&1.12079 $\pm$ 0.00001}{53139.7495 $\pm$ 0.0008}{$\cdots$}{$\cdots$}{100.5$\pm$ 0.3}{101.3 $\pm$ 0.3}{0.9921 $\pm$ 0.0004}{Cre05}
\abcdresults{CM Dra\tablefootmark{b}&1.268390 $\pm$ 0.000001}{46058.5640 $\pm$ 0.0003}{0.005 $\pm$ 0.001}{129 $\pm$ 16}{72.2 $\pm$ 0.1}{78.0 $\pm$ 0.1}{0.93 $\pm$ 0.002}{Mor09}
\abcdresults{G 203-60&3.29}{5649.38$\pm$ 0.30}{0.002 $\pm$ 0.002}{54 $\pm$ 33}{43.87 $\pm$ 0.08}{66.47 $\pm$ 0.16}{0.660 $\pm$ 0.001}{Ski18}
\abcdresults{GJ 644 A&626 $\pm$ 2}{47185$\pm$ 28}{0.08 $\pm$ 0.02}{285 $\pm$ 17}{4.4 $\pm$ 0.1}{2.7 $\pm$ 0.1}{0.61 $\pm$ 0.08}{Maz01}
\abcdresults{GJ 644 B&2.96552 $\pm$ 0.00002}{47337.2 $\pm$ 0.2}{0.021 $\pm$ 0.007}{162 $\pm$ 21}{17.2 $\pm$ 0.2}{19.0 $\pm$ 0.2}{0.91 $\pm$ 0.01}{Maz01}
\abcdresults{BD+04 3562&34.50 $\pm$ 0.01}{49178.4 $\pm$ 0.2}{0.39 $\pm$ 0.02}{132 $\pm$ 2}{29.7 $\pm$ 0.4}{35$\pm$ 1}{0.84 $\pm$ 0.03}{Tok94}
\abcdresults{BY Dra&5.9751}{41147.09}{0.36}{220}{28.2}{28.8}{0.979}{VF79}
\abcdresults{GJ 1230 A&5.06880 $\pm$ 0.00005}{50643.7$\pm$ 0.2}{0.009 $\pm$ 0.001}{230 $\pm$ 10}{46.9 $\pm$ 0.1}{49.0 $\pm$ 0.1}{0.957 $\pm$ 0.003}{Del99}
\abcdresults{MCC 188&10.319 $\pm$ 0.008}{46188.981 $\pm$ 0.085}{0.20 $\pm$ 0.01}{177 $\pm$ 3}{21.9 $\pm$ 0.4}{23.3 $\pm$ 0.4}{0.94 $\pm$ 0.02}{DM88}
\abcdresults{2MASS J19324321+3636534\tablefootmark{b} &1.6734372 $\pm$ 0.0000005}{54374.8082 $\pm$ 0.0002}{$\cdots$}{$\cdots$}{72 $\pm$ 2}{95 $\pm$ 3}{0.76 $\pm$ 0.03}{Bir12}
\abcdresults{2MASS J19341550+3628271\tablefootmark{b}&1.4985177 $\pm$ 0.0000004}{54332.88980 $\pm$ 0.00008}{$\cdots$}{$\cdots$}{91 $\pm$ 2}{94 $\pm$ 2}{0.96 $\pm$ 0.03}{Bir12}
\abcdresults{2MASS J19350355+3631165\tablefootmark{b}&2.44178$\pm$ 0.00003}{54319.83270 $\pm$ 0.00002}{$\cdots$}{$\cdots$}{29.4 $\pm$ 0.5}{109 $\pm$ 2}{0.270 $\pm$ 0.006}{Nef13}
\abcdresults{2MASS J19364065+3642460\tablefootmark{b}&4.939095 $\pm$ 0.000002}{54393.8079 $\pm$ 0.0002}{$\cdots$}{$\cdots$}{55 $\pm$ 2}{60$\pm$ 1}{0.92 $\pm$ 0.04}{Bir12}
\abcdresults{2MASS J20115132+0337194\tablefootmark{b}&0.6303135 $\pm$ 0.0000002}{54738.74970 $\pm$ 0.00004}{$\cdots$}{$\cdots$}{124.8 $\pm$ 0.1}{129.9 $\pm$ 0.1}{0.961 $\pm$ 0.001}{Kra11}
\abcdresults{LP 395-8&1.129339 $\pm$ 0.000007}{57620.08 $\pm$ 0.04}{0.007 $\pm$ 0.002}{352 $\pm$ 12}{36.5 $\pm$ 0.1}{65.3 $\pm$ 0.2}{0.560 $\pm$ 0.002}{$\star$}
\abcdresults{GJ 810 A &812 $\pm$ 51}{57822 $\pm$ 4}{0.40 $\pm$ 0.05}{239 $\pm$ 2}{5.6 $\pm$ 0.2}{6.7 $\pm$ 0.2}{0.83 $\pm$ 0.04}{$\star$}
\abcdresults{Wolf 1084&0.795340 $\pm$ 0.000003}{54140.530$\pm$0.001}{$\cdots$}{$\cdots$}{73.8 $\pm$ 1.4}{75.4 $\pm$ 1.4}{0.98 $\pm$ 0.03}{Ire08}
\abcdresults{BD+40 883 A&3.2756}{43270.94}{0.04}{3.9}{39.5}{57.2}{0.691}{Fek78}
\abcdresults{BD+40 883 B&10777.10 $\pm$ 241.06}{43099$\pm$ 17}{0.72 $\pm$ 0.01}{309 $\pm$ 3}{3.7 $\pm$ 0.2}{12 $\pm$ 2}{0.32 $\pm$ 0.05}{DM88}
\abcdresults{G 212-34&8.17}{56488.08$\pm$ 0.20}{0.062 $\pm$ 0.012}{127.8 $\pm$ 8.8}{58.51 $\pm$ 0.59}{64.7 $\pm$ 1.3}{0.905 $\pm$ 0.016}{Ski18}
\abcdresults{Ross 775&53.221 $\pm$ 0.004}{48980.2$\pm$ 0.2}{0.374 $\pm$ 0.004}{300 $\pm$ 1}{18.7 $\pm$ 0.1}{18.7 $\pm$ 0.1}{1.000 $\pm$ 0.008}{Del99}
\abcdresults{2MASS J21442066+4211363&3.30}{56205.3381$\pm$0.0057}{$\cdots$}{$\cdots$}{61.16$\pm$0.46}{64.6$\pm$2.6}{0.947$\pm$0.037}{Ski18}
\abcdresults{CD-51 13128&1.123 $\pm$ 0.007}{48467.07 $\pm$ 0.02}{$\cdots$}{$\cdots$}{36 $\pm$ 2}{40 $\pm$ 2}{0.92 $\pm$ 0.08}{JB93}
\abcdresults{FK Aqr&4.0832}{37144.123}{0.01}{356}{46.8}{58.1}{0.806}{HM65}
\abcdresults{2MASS J23143816+0339493\tablefootmark{b} &1.722821 $\pm$ 0.000004}{54730.78778 $\pm$ 0.00004}{$\cdots$}{$\cdots$}{75.4 $\pm$ 0.2}{92.5 $\pm$ 0.2}{0.815 $\pm$ 0.002}{Kra11}

\hline

\end{tabular}

}
\tablefoot{
\tablefoottext{a}{$\star$: This work; Bir12: \cite{2012MNRAS.426.1507B}; BF77: \cite{1977PASP...89...65B}; Bla08: \cite{2008ApJ...684..635B};
BO06: \cite{2006ApJ...651.1155B}; Cre05: \cite{2005ApJ...625L.127C}; Dav16: \cite{2016ApJ...816...21D}; Del99: \cite{1999A&A...344..897D}; DM88: \cite{1988A&A...200..135D}; Fek78: \cite{1978AJ.....83.1445F}; For99: \cite{1999A&A...351..619F}; Gom12: \cite{2012ApJ...745...58G}; Har96: \cite{1996AJ....112.2222H}; Har11: \cite{2011AJ....141..166H};
Har18: \citep{2018AJ....155..114H}; Heb06: \cite{2006AJ....131..555H};  Hel12: \cite{2012MNRAS.425.1245H}; HM65: \cite{1965ApJ...141..649H}; Ire08: \cite{2008ApJ...678..463I}; Irw09: \cite{2009ApJ...701.1436I}; Irw11: \cite{2011ApJ...742..123I}; JB93: \cite{1993MNRAS.260..132J}; Kra11: \cite{2011ApJ...728...48K};
Kra17: \cite{2017ApJ...845...72K}; Lop05: \cite{2005ApJ...631.1120L}; Lop06: \cite{2006astro.ph.10225L}; Lub17: \cite{2017ApJ...844..134L}; Mat97: \cite{1997AJ....113.1841M}; Maz01: \cite{2001MNRAS.325..343M}; Mor09: \cite{2009ApJ...691.1400M}; Nef13: \cite{2013MNRAS.431.3240N};  Ski18: \cite{2018arXiv180602395S}; Tok91: \cite{1991A&AS...91..497T}; Tok94: \cite{1994AstL...20..309T}; Tok97: \cite{1997A&AS..121...71T}; Tok00: \cite{2000AstL...26..668T}; Tor02: \cite{2002AJ....123.1701T}; TP86: \cite{1986AJ.....92.1424T}; TR02: \cite{2002ApJ...567.1140T}; VF79: \cite{1979ApJ...234..958V}; Zho15: \cite{2015MNRAS.451.2263Z}.}
\tablefoottext{b}{Eclipsing binary systems.}}
\end{table*}
\end{appendix}

\end{document}